\pdfoutput=1
\documentclass[11pt,a4paper]{book}
\usepackage{a4}
\usepackage{amsmath}
\usepackage{graphicx}
\usepackage{amssymb}
\usepackage{epstopdf}
\usepackage{hyperref}
\hypersetup {
pdftitle = {On the Correspondence of Open and Closed Strings},
pdfauthor = {Marco Baumgartl}
}
\usepackage{fancyhdr}
\usepackage{sectsty}
\usepackage[rightcaption]{sidecap}
\renewcommand{\baselinestretch}{1.05}

\allsectionsfont{\sffamily\raggedright}

\renewcommand{\chaptermark}[1]%
         {\markboth{\thechapter.\ #1}{}}
\renewcommand{\sectionmark}[1]%
         {\markright{\thesection\ #1}}
\newcommand{\LMUTitle}[9]{{\sf

  \thispagestyle{empty}

  \vspace*{\stretch{1}}
  {\parindent0cm
  \rule{\linewidth}{.7ex}}
  \begin{flushright}
    \vspace*{\stretch{1}}
    \sffamily\bfseries\Huge
    #1\\
    \vspace*{\stretch{1}}
    \sffamily\bfseries\large
    #2
    \vspace*{\stretch{1}}
  \end{flushright}
  \rule{\linewidth}{.7ex}

  \vspace*{\stretch{3}}
  \begin{center}
    \Large Dissertation\\
    \Large an der #4\\
    \Large der Ludwig--Maximilians--Universit\"at\\
    \Large M\"unchen\\
    \vspace*{\stretch{1}}
    \Large vorgelegt von\\
    \Large #2\\
    \vspace*{\stretch{2}}
  \end{center}

  \newpage
  \thispagestyle{empty}

  \vspace*{\stretch{1}}

  \begin{flushleft}
    \large Erstgutachter:  #7 \\[1mm]
    \large Zweitgutachter: #8 \\[1mm]
    \large Tag der m\"undlichen Pr\"ufung: #9\\
  \end{flushleft}

  \cleardoublepage
  }
}

\def\be{\begin{equation}}
\def\beq{\begin{equation}}
\def\ee{\end{equation}}
\def\eeq{\end{equation}}
\def\ba{\begin{eqnarray}}
\def\ea{\end{eqnarray}}
\def\eqn#1{\begin{equation}\begin{split}#1\end{split}\end{equation}}

\def\P{\partial}
\def\PR{{\overrightarrow\partial}}
\def\PL{{\overleftarrow\partial}}

\def\one{1\! \vrule height 7.5pt width 0.6pt depth 0pt \vrule height  
0.3pt width 1.1pt depth 0.1pt}

\def\Z{{\cal Z}}
\def\S{{\cal S}}
\def\X{{\cal X}}
\def\bdry{{\rm bdry}}
\def\a{\alpha}
\def\bP{{\bar\partial}}
\def\t{\theta}
\def\bz{{\bar z}}
\def\tr{{\rm tr}}
\def\Tr{{\rm Tr}}
\def\Det{{\rm Det}}
\def\Ad{{\rm Ad}}

\def\text#1{{\hbox{#1}}}
\def\1{{-1}}
\def\2{\frac{1}{2}}
\def\<{\langle}
\def\>{\rangle}
\def\b{\beta}
\def\e{{\epsilon}}
\def\g{\gamma}
\def\l{\lambda}
\def\r{\rho}
\def\k{\kappa}
\def\d{\delta}
\def\m{\mu}
\def\n{\nu}
\def\s{\sigma}
\def\w{\omega}
\def\vev#1{\left\<#1\right\>}
\def\VEv#1{\Bigl\<#1\Bigr\>}

\def\hb{\hat{b}}
\def\hc{\hat{c}}

\newcommand{\mfa}{\mathfrak a}
\newcommand{\mfb}{\mathfrak b}
\newcommand{\mfc}{\mathfrak c}

\def\IP{\relax{\rm I\kern-.18em P}}
\def\CC{{\mathbb C}}

\begin{document}

  \frontmatter

  \LMUTitle
      {On the Correspondence of \\
       Open and Closed Strings}               
      {Marco Baumgartl}                       
      {
		}                             
      {Fakult\"at f\"ur Physik}                         
      {Munich 2007}                          
      {10 October 2007}                            
      {Prof.\ Dr.\ I.\ Sachs}                          
      {Prof.\ Dr.\ D.\ L\"ust}                         
      {03. Dezember 2007}                         

\setcounter{tocdepth}{1}
\tableofcontents

\contentsline {chapter}{\numberline {13}Bibliography}{157}{chapter.13}


\parskip = 0.1in
\renewcommand{\baselinestretch}{1.1}

  \markboth{Content}{Content}


  \cleardoublepage

\markboth{Abstract}{Abstract}

The topic of this thesis is the correspondence of open and closed strings. Already for a long time it has been evident that those both sectors of string theory do not only couple to each other, but it is also possible to identify excitations of the closed string in the open string sector. This correspondence has been shown in a multitude of examples, which indicates a deep connection. This thesis tries to understand this from the viewpoint of coupled open-closed moduli spaces and finally from a string field theoretic point of view. Implications of this conjectured correspondence have gained great importance, among them gauge-gravity correspondence, AdS/CFT correspondence as well as non-perturbative effects in open string field theory.

A new approach to bosonic boundary string field theory on curved target spaces is developed, which allows to demonstrate techniques to identify closed string excitations in the open string sector. Certain factorisation properties of path integrals over WZW-models are derived, which lead to a adequate re-formulation of boundary string field theory. It is shown for the first time that this setting can reproduce curved D-branes by tachyon condensation starting from flat space as soon as non-local interactions terms are permitted. The results coincide with expectations from conformal field theory.

Additionally first steps are taken to study more complex supersymmetric string theories on Calabi-Yau manifolds. This includes an exemplary investigation of coupled open-closed moduli spaces. These results are derived in the framework of topological string theory, which constitutes a projection to a finite subspace of the full theory. The recent formalism of matrix factorisations for describing B-type branes is used in order to show how open string moduli spaces can be constructed exactly. Moreover the influence of closed string perturbations appears in form of an effective open-closed superpotential, which also is explicitly computed.

	\cleardoublepage

  \markboth{Acknowledgements}{Acknowledgements}

Primarily I want to thank my supervisor Ivo Sachs for introducing me to string theory, starting with basic lectures as an undergraduate student, revealing to me profound insights into the mathematical description of nature at the most fundamental level, accepting me as his student and guiding me in innumerable discussions. He always took his time to talk about problems, ideas and speculations, providing me with much freedom to study many aspects of string theory and theoretical physics in general. I also want to thank him for the excellent collaboration in organising lectures and tutorials. 

It is my pleasure to thank Samson Shatashvili for support of my research work, for many interesting discussions and the suggestion of research problems.
My special thanks go to Ilka Brunner and Matthias Gaberdiel, who gave me the opportunity to work with them on new and interesting developments in topological string theory.
Cordially I want to thank Andreas Recknagel for making it possible to visit him as postgraduate fellow, but in particular for all those most inspiring and motivating discussions at day and night time.

Also many thanks go to Slava Mukhanov as well as to my various colleges and office mates, among them
Sebastian Guttenberg,
Ingo Kirsch,
Dominique L\"ange,
Steffen Metzger,
Giuseppe Policastro,
and Sergey Solodukhin. 

Finally, the most indispensable thanks go to my family for their constant support, and most importantly to my wife Hengameh who encouraged me and was always extremely patient with me.

\cleardoublepage
\markboth{}{}
$\;$
\clearpage
$\;$
\vfill
    \hfill\includegraphics[width=10mm]{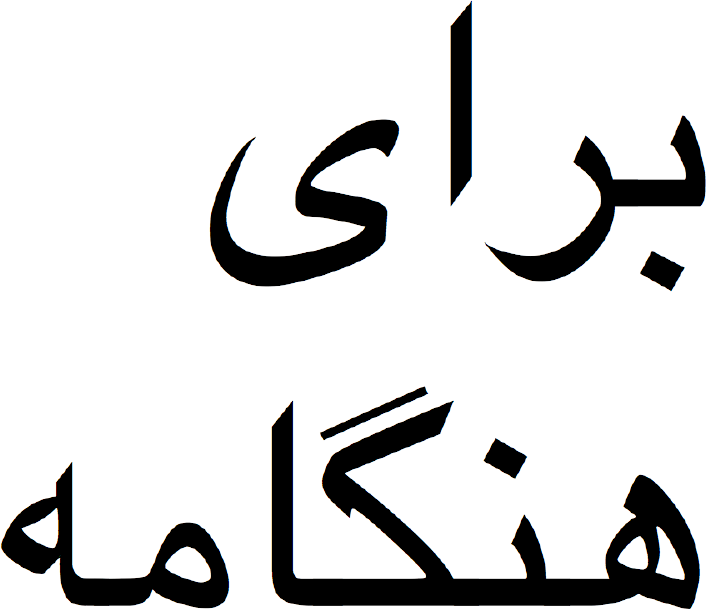}
    
    \hfill{\em for  Hengameh}

  \mainmatter\setcounter{page}{1}

%
%
%
%

%
%
%

\chapter{Introduction}

From a theoretical as well as from an experimental point of view quantum field theory has emerged over many years of research as the correct approach to describe particle interactions at low scales and high energies, up to the GUT scale where the unification of the known forces is predicted. Based upon this theory the standard model of particle physics has been developed, which has been successful in unifying the known forces and particles in a consistent mathematical framework. It provides a scheme where all observed particles can be gathered, classified according to mass, charge, spin etc. This extremely successful model has provided deep insights into the fundamental laws of nature, also from a conceptional point of view.

Despite its success, in our present understanding it still leaves several issues untouched. One of them is the fact that many properties of the particles in the standard model must be determined by experiment. This raises the question if there is some mechanism that fixes for example the particle masses, and if not, then why is there a discrete spectrum observed. Re-prashed differently, there is still a consistent framework missing which explains the basic origin and properties of elementary particles. 

Another flaw in the quantum field theory approach are the mathematical problems which arise when trying to include gravity in the theory. The quantisation of gravity cannot be done by employing usual methods of quantum theory. Despite many attempts, it has not been possible to conduct the quantisation correctly, and it is also not clear if more advanced methods will finally lead to a positive result. One the other hand, it is evident that a thorough understanding of gravity at quantum scales is important.
An example for this is black hole physics, where it has been known for a long time that classical concepts break down at the central singularity while even semi-classical phenomena like Hawking radiation show that black holes must be treated as quantum objects.  Another example is cosmology, where in inflationary scenarios initial perturbations caused by vacuum fluctuations are quantum effects that are presumed to be responsible for the present distribution of matter in the universe.

The lesson to learn is that gravity makes probably a modification of conventional quantum field theory necessary. Due to the different nature of gravity as compared to the other known forces this modification must be rather fundamental. It is therefore unlikely that it is possible to derive such a theory in a bottom-up approach by mildly extending the standard model description. Rather, major shifts in viewpoints must be expected in order to make it possible to understand a theory of quantum gravity as part of a greater framework which also incorporates all other known particles and forces.

Such a modification has been proposed by string theory (see e.g.\ \cite{Green:1987sp, Green:1987mn, Polchinski:1998rq, Polchinski:1998rr}). While it shares many concepts with conventional quantum field theories, in principle it is still able to reproduce the known forces and particles, as well as perturbative gravity. All this is achieved in a first quantised framework, so that string theory in fact provides a perturbative approach to quantum gravity. Generally, perturbative string theory has large `moduli spaces', which means that there are again free parameters in the model. But is has been observed that there are mechanisms at work which tend to fix these parameters and it is hoped, that this fixation is complete. This would then yield a model free of (continuous) parameters, and one would obtain a description of all possible vacua that can arise within string theory. Unfortunately this happens often in the non-perturbative regime of string theory, which is often hard to control as one usually deals with a perturbative theory. Therefore it is of major interest to develop an off-shell version of string theory, referred to as string field theory. Large parts of this thesis are dedicated to contribute to a better understanding of such a string field theory. 

In the following we describe briefly the underlying ideas behind string theory from a conceptual point of view and explain some important properties and mechanisms that appear generically.

The basic construction of string theory, which has its roots originally in the attempt to invent an effective theory for quark interactions, is very simple. The one-dimensional worldline, which would be viewed as a particle in spacetime, is replaced by a two-dimensional worldsheet, so that in each time slice it appears as a `string' in space rather than a point. The string's position in spacetime is then described by the string map $X^\mu(\sigma^\alpha)$ which maps the worldsheet to the target space. It is by no ways clear that such a theory can be quantised, and indeed it does only work for two-dimensional worldsheets, but not for higher-dimensional worldvolumes. Even the string must satisfy certain conditions so that a quantisation is possible, namely the Weyl anomaly cancellation condition which essentially fixes the number of spacetime dimensions to 10 or 26 in the supersymmetric and bosonic case, respectively. Similar conditions cannot be easily satisfied in higher-dimensional settings, so that string theory is in fact singled out by the simplicity of its construction.

The transition from a worldline to a worldsheet introduces new degrees of freedom into the theory which can be imagined as the spacelike modes of the string map on the worldsheet. These are not present for point particles, since for them the worldline is always timelike. The existence of these modes are finally responsible for infinitely many new excitable states of the string. In addition, strings can appear in open and closed form, each version leading to states with characteristic properties. While open strings describe scalar and vector bosons in the massless sector, most remarkably one can find excitations among the closed strings can be identified as gravitons. This discovery has amplified the interest in string theory.

A theory formulated in more than four spacetime dimensions has to explain why only four dimensions are actually observed. Usually extra directions are treated as `internal' dimensions, which are compactified on a scale compatible with experimental observations. This results in a construction where four-dimensional spacetime appears as usual flat Minkowski space, whereas the internal dimensions constitute manifolds with possibly complicated geometry. The spectrum and properties of fields appearing in the extended dimensions depends on the details of the compactification. Of special interest is here the supersymmetric case. In this case the compact manifold must be a six-dimensional Calabi-Yau-manifold. While it is believed that this version of string theory is suitable to build up an extension of the (supersymmetric) standard model, it is also attractive because mathematical methods apply which allowed for an important progress in the past.

It turns out that in string theory typically also tachyons appear in the spectrum. In quantum field theory this is usually an indication for a perturbative instability of the chosen vacuum. Hence it is important to look for tachyon-free vacua, which has lead to a thorough and successful study of supersymmetric string theories. In some situations one can accept the tachyon rather as a feature than a flaw: At least in the open string case, it has been shown that the tachyon creates a potential whose minima describe stable vacua. In these minima also D-branes can appear, which are hypersurfaces on which the end-points of open strings are fixed. The number and configuration of D-branes is an important ingredient for the investigation of consistent string vacua. Given the importance of tachyon condensation for the open string case, it is plausible to believe that similar techniques also apply to closed strings. However, for them new complications arise, which tremendously increase the technical difficulties.

With the study of tachyon potentials and the associated condensation processes from instable to stable vacua one enters already the regime of string field theory, since the tachyon potential can be considered as the static approximation of a string field theory action which is also well-defined off-shell, i.e.\ away from classical solutions. For the open string the tachyon condensation is in principle under control as long as only massless and tachyonic modes are included. But string theory usually comes with infinitely many massive fields, too, which can contribute non-perturbatively. While this on the one hand adds to the technical difficulties, it on the other hand can be shown that open and closed strings might not be so different than originally suspected.

Over the years many examples have been collected which show that open strings sometimes are capable of describing closed string interactions. 
Since the early days of string theory it has been presumed that the distinction between 
open strings and closed strings is not fundamental. This follows already from the 
observation that closed string poles occur as intermediate states in open string scattering 
amplitudes. From the point of view of open string field theory these poles seem to violate 
unitarity unless closed string states are present in the classical open string field theory. 
One possibility is to accept that open string field theory is not unitary and to add extra 
closed string degrees of freedom by hand \cite{Zwiebach:1997fe}.
However, in this approach one has to address 
the problem of overcounting since now the same diagram can be obtained from the open 
and closed string sector of the field theory Lagrangian.

An alternative approach is to try to identify closed string states directly in open 
string field theory \cite{Strominger:1986zd, SAMSONunpub}.
This idea receives further motivation from Sen's work on non-BPS 
branes \cite{Sen:1998rg, Sen:1999mg} which resulted in a very active study of open string field theory in different 
formulations and some progress in understanding the vacuum structure of open strings has 
been achieved \cite{Sen:2004nf}.
The correspondence goes so far that it has been conjectured that generally on-shell closed strings can be described by open strings. It is tempting to conjecture that there is maybe only a single fundamental object in string theory, and that open and closed strings are just different ways of describing the same theory, which leads to easier formulations in one or the other regime.

The purpose of this thesis is to investigate the correspondence between open and closed strings on the level of boundary string field theory (BSFT). 
Roughly, there 
are two types of input data necessary for the construction of the action: first, the bulk 
conformal field theory must be specified, which corresponds to the choice of a closed string 
background. As there is no complete classification of conformal field theories, the space of 
Ôclosed string backgroundsÕ is not well-defined. Equally poorly understood is the space of 
boundary interaction terms, which is used to deform the boundary conformal field theory. 
This space is intimately related to the configuration space of open string field theory.

On the other hand, these two spaces are certainly not independent of each other, 
because some examples of dualities between open and closed strings are known 
\cite{Gross:1970eg, Lovelace:1971fa, Strominger:1986zd, Bonora:2006tm, Katsumata:2004cc, Garousi:2004pi, Sen:2003iv, DiVecchia:2003ae, Shatashvili:2001ux, Mayr:2001xk, Kapustin:2004df}.
Such correspondences appear already on the level of moduli spaces of classical solutions of string theory. In particular, examples are known of coupled open-closed moduli spaces, the most simple realisation being a constant antisymmetric tensor field, a Kalb-Ramond field, in the closed string background which can in the same way be viewed as a gauge field in open string background. The further development of this idea led to Kontsevich's theory of deformation quantisation \cite{Kontsevich:1997vb}. Also AdS/CFT-correspondence, holography and gauge/gravity duality are research directions in string theory which strongly suggest an underlying profound connection between open and closed strings \cite{Maldacena:1997re, Aharony:1999ti}.

In order to make progress on the issue of open-closed correspondence, our strategy will be to consider deformations in the open string sector and compare them to deformations in the closed string sector, starting form a $\s$-model point of view. While this is already a difficult problem on the level of open-closed moduli spaces (i.e.\ for on-shell string theories), we manage to go beyond that classical niveau, at least in the bosonic case, and apply these ideas to open string field theory \cite{Baumgartl:2004iy}. We will see that it is indeed possible to relate deformations of the closed string background to an infinite collection of open string excitations. In order to arrive at this statement it will be necessary to develop a version of boundary string field theory in curved target spaces. On the way to this, a conjecture about factorisation properties of path-integrals on curved target spaces will be made and proven for a large class of models.

The generally obtained statements are supported by calculations in explicit models, where results on tachyon condensation on D-branes are obtained, which are consistent with our knowledge of string theory in curved target spaces, as well as with expectations from open-closed correspondence \cite{Baumgartl:2006xb}. In particular we observe strong hints that tachyons in flat space can condense to curved branes, which are stabilised by the presence of non-local couplings. This leads to the speculation that, roughly speaking, non-local open string couplings are related to closed strings in open string field theory.

Once the bosonic case is understood, it is desirable to apply the lessons learned to supersymmetric string theory on Calabi-Yau manifolds. In this case one encounters complicated moduli spaces and therefore also  the interaction of open and closed string moduli is expected to be difficult. We were able to make progress in the study of a topological version of this model. The outcome presented in the last part of this thesis shall be understood a preparation for further investigations. The study of open-closed moduli spaces is an important topic in itself, therefore the results given are still very interesting from this point of view. The investigation of non-local couplings in supersymmetric settings is an ongoing research project. 

The framework for the construction of open-closed moduli spaces is the topologically B-twisted ${\cal N}=(2,2)$ theory. 
The topological version is very attractive because it contains less degrees or freedom, and these are under better control than in a non-topological theory. This is mainly due to the fact that the renormalisation group flow in the models under investigation is strongly constrained. Yet these theories describe certain quantities of physical importance like for example Yukawa couplings.

Since we work with a topological theory, the appearance of non-local couplings is not expected. Hence this approach gives in a way the reduction to constant modes of the procedure we developed for bosonic string field theory. The results are considered then from a slightly different point of view: we manage to relate the topological calculations to the renormalisation group by conformal field theory methods \cite{Baumgartl:2007an}. This allows us to obtain concrete expressions for effective superpotentials on the quintic which are exact in the open string coupling and first order in the closed string coupling. While these results open up the way to many other intriguing questions, it would be very interesting to extend the calculations to the non-topological case. In particular the role of non-local couplings in supersymmetric theories should be clarified. These issues will be subject of further studies.


This thesis is organised in three parts. The first part gives some general background information on string theory and its mathematical description. Conformal field theory in general is introduced briefly, as well as Wess-Zumino-Witten models, which will be used to formulate open string field theory on curved target spaces. The fourth chapter collects some information on effective theories and the spacetime interpretation of string theory as well as its relation to renormalisation group flow. This will be useful as preparation for the investigation of string field theory.

The second part deals with closed string deformations in open string field theory. Chapter five provides a basic introduction to bosonic boundary string field theory as well as to an associated generalised boundary state formalism. The sixth chapter reviews central aspects of tachyon condensation, since we will see later that the tachyon is the driving force which causes a localisation on branes curved space. This chapter also addresses some issues of open-closed correspondence which appear in this context. Chapter seven contains the main results on the construction of boundary string field theory on curved manifolds. It contains the factorisation conjecture as well as its proof for group manifolds. In the eighth chapter a concrete example is worked out, which demonstrates the condensation of a flat 3-brane to a spherical 2-brane, triggered by the presence of no-local couplings. Issues of stability under tachyonic perturbations is discussed, and perturbative $\b$-functions are calculated explicitly.

The third part contains our result on the deformations of the topologically twisted B-model. Chapter nine provides the technical background by a brief introduction to ${\cal N}=(2,2)$ string theory on Calabi-Yaus, from the conformal field theory point of view as well as from the Landau-Ginzburg point of view. The tenth chapter explains how D-branes arise in this model and how they can be described by so-called matrix factorisations. A few very basic examples are provided and the connection to conformal field theory is discussed. In chapter eleven, the matrix factorisation technique is used to determine the moduli space of 2-branes.\footnote{As the four external dimensions are ignored we speak of 2-branes with respect to the internal Calabi-Yau threefold. Hence when one takes all dimensions into account one should speak of D5-branes rather.} The effect of closed string deformations on the open string moduli space is investigated. By using the connection to conformal field theory, it is possible to explicitly calculate the effective superpotential. Although this method has been developed for a specific example, it is generally applicable. 

\vskip5mm

Main new results are presented in chapters 7, 8 and 11, while many less significant calculations and insights are distributed over other chapter, too. In the course of this research project the main results have been published already in \cite{Baumgartl:2004iy, Baumgartl:2006xb, Baumgartl:2007an}.

\part{String theory}

%
%
%

\chapter{Conformal field theory}
\label{ch-cft}

This chapter gives a short introduction to conformal field theory (CFT) \cite{Belavin:1984vu, Ginsparg:1988ui, DiFrancesco:1997nk}. Often string theory is defined in a path-integral approach (see \cite{Green:1987sp, Green:1987mn, Polchinski:1998rq, Polchinski:1998rr}), where the integrals are taken over all possible embeddings of the 2-dimensional string worldsheets in spacetime. Although this approach provides a very intuitive way of thinking about string scattering diagrams, it is difficult to work with this formalism in general. Conformal field theory provides an efficient and well developed way for perturbative string theory calculations around a classical background configuration. In fact, in a configuration where the string $\b$-functions vanish, there is a mapping between string states and operators in an associated 2-dimensional conformal field theory.

\section{Closed strings}

A propagating closed string is geometrically described by a cylindrical worldsheet $\Sigma=S^1\times{\mathbb R}$, where the non-compact direction is temporal. After a suitable Wick-rotation into Euclidean space, the worldsheet can be furnished with complex coordinates $z$ and $\bz$. At the heart of CFT lies its invariance under conformal transformations. These are locally given by
\be
	z\mapsto f(z)\qquad\bz\mapsto\bar f(\bz)\ ,
\ee
where $f(z)$ and $\bar f(\bz)$ are holomorphic and anti-holomorphic. On the other hand, global conformal transformations of the closed string worldsheet are given by the M\"obius group, acting like
\be
	z\mapsto \frac{az+b}{cz+d} \qquad\text{with}
		\begin{pmatrix}a & b\\c & d\end{pmatrix}
		\in\text{SL}(2,{\mathbb R})\ .
\ee
A similar transformation appears for $\bz$. The infinitesimal generators of this transformation are given by $l_n=-z^{n+1}\P_z$. They can be combined into a energy-momentum tensor, which can be considered as the operator generating scale transformations. In a quantum field theoretic treatment the so-called Witt-algebra
\be
	[l_n,l_m]=(n-m)l_{n+m}
\ee
satisfied by $l_n$ is centrally extended, known as the Virasoro algebra. In this case the generators are denoted by $L_n$ and obey
\be
	[L_n,L_m]=(n-m)L_{n+m} + \frac{c}{12}(n^3-n)\d_{n+m,0}\ .
\ee
The constant $c$ is called central charge \cite{Zamolodchikov:1986gt}.

The energy-momentum tensor $T(z,\bz)=T(z)+\bar T(\bz)$ is then given by
\be
	T(z)=\sum_n \frac{L_n}{z^{n+2}}
\ee
and analogously for $\bar L_n$. One should note that the holomorphic (`left-moving') and anti-holomorphic (`right-moving') operators commute for boun\-dary-less worlsheets,
\be
	[L_n,\bar L_m] = 0\ .
\ee
Thus the full algebra is a product of two Virasoro algebras. This will also be true for the Hilbert space of states, which in the same way factors into a left-moving and a right-moving part,
\eqn{
	{\cal H}^c = {\cal H}\otimes{\cal \bar H} \ .
}
This Hilbert space is determined through the action of the Virasoro generators by imposing
\be
	L_0 |h\> = h|h\> \qquad L_n|h\> = 0 \text{ for } n>0
\ee
on physical states $|h\>$ in each of the two sectors. The eigenvalue $h$ of $L_0$ is the conformal weight of $|h\>$, which 
is called primary state when it is eigenstate of $L_0$. 

In conformal field theory every state $|\phi\>$ can be associated to an operator $\phi(z,\bar z)$ acting on the conformal vacuum via
\be
	|\phi\> = \lim_{z,\bz\to 0} \phi(z,\bz) |0\>\ .
\ee
Using this correspondence, the conformal weight of a primary state acquires an interpretation as scaling exponent under local conformal transformations $(f,\bar f)$,
\be
	h(z,\bar z) \mapsto \left(\frac{\P f}{\P z}\right)^h \left(\frac{\P \bar f}{\P \bz}\right)^{\bar h} h(z,\bar z)\ .
\ee
Locally this is encoded in the action of the energy-momentum tensor. If the conformal transformation is given by $f(z) = z+ \e(z)$ the associated charge (for the holomorphic sector) is defined as
\be
	T_\e = \oint\frac{dz}{2\pi i}\e(z)T(z)\ ,
\ee
where the integral goes along a closed contour around the origin. By expansion one finds a commutator for operators at the same radius\footnote{by a suitable transformation quantisation in time direction has been replaced by radial quantisation.} 
\be
	[T_\e,h(w,\bar w)] = \left( \e(w)\P_w+\frac{\P\e(w)}{\P w} h\right)h(w,\bar w)\ .
\ee
A similar relation holds also under the integral,
\be
	T(z)h(w,\bar w) = \frac{h}{(z-w)^2} h(w,\bar w)  + \frac{1}{z-w}\frac{\P}{\P w}h(w,\bar w) + \dots\, 
\ee
where the dots denote holomorphic functions in $z-w$, which are regular as $z\to w$.

This is an example for a operator-product-expansion (OPE), which is a useful technique in CFT. In fact, for all primary fields $h_i$ such a OPE is given generally by\footnote{only the holomorphic sector is considered, for simplicity.}
\be
\label{OPE0}
	h_i(z) h_j(0) = \sum_k z^{h_k-h_i-h_j}C^k_{ij} h_j(0) = \frac{C^0_{ij}}{z^{h_i+h_j}} + \text{ less singular terms}\ .
\ee 
The constants $C^k_{ij}$ appear as structure constants. Their knowledge determines the OPE and therefore the theory completely.

With the OPE at hand it is possible to find expressions for $n$-point functions $\< h_{i_1} h_{i_2}\cdots h_{i_n}\>$. In particular, as by definition $\<h_i(z)\>=0$ one finds
\eqn{
	\< h_i(z) h_j(0)\> = \frac{C^0_{ij}}{z^{h_i+h_j}}\ ,
}
because all less singular terms in (\ref{OPE0}) are linear in the fields.

\section{Open strings}

\subsection{Boundary fields and correlators}

While closed string worldsheets have the topology of a cylinder, open strings are described by strips of topology $I\times{\mathbb R}$. As an open string has boundaries, the treatment of the string end points introduces new structures into the theory.
 
The main complications arising in the open string sector come from the necessity of imposing boundary conditions, as the worldsheet is not closed any more. The presence of a boundary may destroy conformal invariance, although in a less drastic way than a relevant perturbation of the closed string background. The proper setting up of a boundary conformal field theory (BCFT) means the introduction of a boundary in a well-defined way, and the proper choice of boundary conditions.

Local conformal transformations, which played an essential role in the construction of CFTs in previous sections, are now modified by the presence of a boundary. They must respect the boundary in the sense that only transformations tangential to it are valid. Otherwise the local properties of CFTs with and without boundary do not differ. Global properties, such as the spectrum, do differ immensely. 

The restriction of valid conformal transformations results in a break-down of the left- and right-Virasoro algebra to a single one. Still, the condition for conformal invariance, namely the tracelessness of the energy momentum tensor $T$, should also be satisfied at the boundary. Therefore one has to impose
\eqn{
\label{TbarT}
	T(z) = \bar T(\bz)
}
at the boundary $z=\bz^*$, where the form of $z^*$ depends on the geometry of the boundary\footnote{when the boundary is identical to the real line, $\bz^*\equiv \bz$.}. The resulting conditions from this requirement are called gluing conditions. They encode the conformal boundary conditions.

Starting from this expression, the mode expansion of (\ref{TbarT}) translates into a condition on the Virasoro generators,
\eqn{
	{\cal L}_n \equiv L_n + \bar L_{-n} = 0
}
at the boundary and for all $n$. Like before in the closed string case, the new single set generators ${\cal L}_n$ can be used to define the state space of boundary states and the conformal vacuum.

One way to construct operators in BCFT is the use of the method of images. This requires that any operator $\phi(z,\bar z)$ in the bulk is accomplished by a `mirror operator' $\phi^*(z^*,\bar z^*)$ on the other side of the boundary in a way that local conformal symmetries are untouched. I.e.\ an operator $\phi(z,\bz)$ on the open string worldsheet can be represented as $\phi(z,\bar z)\phi^*(z^*,\bar z^*)$ on the closed string worlsheet. 

While this is fine everywhere in the bulk, it produces singularities when the bulk operator is transported to the boundary. There the field interacts with its image, which can be seen from the OPE,
\eqn{
	\phi(z,\bar z)\phi^*(z^*,\bar z^*) = \frac{A}{|z-\bar z^*|^{2h_\phi}} + \dots
}
As $z=\bar z^*$ defines the boundary, the one-point function
\eqn{
	\<\phi(z,\bar z)\>_{\rm disk} = \frac{A}{|z-\bar z^*|^{2h_\phi}}
}
becomes singular.

In addition there are operators in the spectrum, which are genuine to the open string sector.

In the same way as in the bulk, an open string Hilbert space can be built, consisting of states which are compatible with the gluing conditions. Not surprisingly, an operator-state correspondence can be employed here again, which allows the construction of boundary fields. These fields live by construction at the boundary $x=z=z^*$. Again, they posses OPEs, which are this time of the form
\eqn{
	\psi_i(x)\psi_j(y) = \sum_k \frac{{C^{(\a)}}_{ij}^k}{(x-y)^{h_i+h_j-h_k}} \psi_k(y)\ .
}
In complete analogy, the boundary field $\psi_i$ can be assigned a conformal weight $h_i$ which appears in the OPE with $T(x)-\bar T(x)$. A formal difference to the closed string case is the appearance of an additional label $(\a)$ in the defining constants ${C^{(\a)}}_{ij}^k$ of the boundary OPE. This label denotes a certain boundary condition. Generally, one expects that there are more than one possible boundary conditions for a given CFT, so that the open string Hilbert space becomes a direct sum of the different boundary sectors.

Going back to the idea, that boundary fields describe open strings, it seems plausible to introduce operators, which switch between different boundary conditions. The heuristic view behind that is an open string whose one end obeys boundary condition $(\a)$ and whose other end obeys boundary condition $(\b)$. Indeed it makes sense to introduce such operators in a BCFT. The associated Hilbert space will be denoted by ${\cal H}^{(\a\b)}$. For $\a\ne\b$ its elements will be called boundary changing operators, whereas boundary preserving operators are contained as special case for $\a=\b$ in ${\cal H}^{(\a)} \equiv {\cal H}^{(\a\a)}$. In general, the open string Hilbert is a direct sum of Hilbert spaces ${\cal H}^{(\a\b)}$.

\subsection{Boundary states}
\label{sec-bs}

Let us be more explicit about the construction of the boundary states associated to boundary fields.

When an algebra of fields is given, boundary conditions can be viewed as relations that induce a linear map on the field algebra with values in ${\mathbb C}$. Every element of the Hilbert space defines such a map. But since these maps must obey algebraic constraints, only certain linear combinations of these homomorphisms can be chosen. In this way each boundary condition $(\a)$ can be associated to a `boundary state' $||\a\>\!\>$ via
\eqn{\label{BScorr}
	\<\phi_1\cdots\phi_n\>_{(\a)} \equiv \<\phi_1\cdots\phi_n|| \a\>\!\>\ .
}
The boundary state $||\a\>\!\>$ is a coherent state, that means it is no finite energy state in the Fock space. The coherent states that describe boundary conditions 
are characterised by the property that the left- and right-moving fields corresponding to 
unbroken symmetries are related to one another at the boundary. Thus for boundary preserving symmetry generators $S(z)$ and $\bar S(\bar z)$ there are relations of the form
\eqn{\label{symcond}
	S(z) =\rho(\bar S(z^*))
}
at the boundary $z=z^*$ (e.g.\ for the real line $z^*=\bar z$). The automorphism $\rho$ must leave the stress tensor invariant. The symmetry generator has a mode expansion of the form $S(z)=\sum_n S_nz^{-n-h}$, where $h$ is the conformal weight. After a suitable conformal transformation the condition (\ref{symcond}) can be expressed in modes, acting on the boundary state as
\eqn{
	\left(S_n-(-1)^h\rho(\bar S_{-n})\right)||\a\>\!\>=0\qquad{n\in\mathbb Z}\ .
}
These are called gluing conditions and must be obeyed by any symmetry preserved at the boundary. In particular for the conformal generators the conditions are
\eqn{
	\left(L_n-\bar L_{-n}\right)||\a\>\!\> = 0\ .
}
Every boundary conformal field theory must obey these conditions, since they are independent of the choice of $\rho$, as long as $\rho$ leaves the stress tensor invariant.

For a Hilbert space of the form
\eqn{
	{\mathcal H} = \bigoplus_{i,j} N_k^{ij}{\mathcal H}_i\otimes\bar{\mathcal H}_j,\
}
where $i,j$ label irreducible representation of chiral symmetry algebras, and $N_k^{ij}$ are constants,
one can show that the solutions of (\ref{symcond}) lay in the diagonal Hilbert space, denoted by
\eqn{
	|i\>\!\> \in {\mathcal H}_i\otimes\bar{\mathcal H}_i\ .
}
These special coherent states are called Ishibashi states 
\cite{Ishibashi:1988kg}.
Any boundary state can be expressed as a linear combination of Ishibashi states (if there are finitely many, the theory is called rational)
\eqn{
	||\a\>\!\> = \sum_i B^i_{(\a)}|i\>\!\>\ .
}
The constraints we found earlier on the boundary states must translate in some way on the coefficients $B^i$.
There are two types of such constraints, the Cardy 
\cite{Cardy:1989ir}
conditions and the `sewing relations' 
\cite{Lewellen:1991tb, Cardy:1991tv}

The Cardy conditions come from the fact that the left hand side of (\ref{BScorr}) is a correlator in the open string sector, and it is identified with an expression in the closed string sector on the right hand side. Demanding such an equivalence for the complete partition function, i.e.\ an invariance under modular transformations, this leads to the Cardy constraints, relating open string one-loop diagrams to closed string tree-level amplitudes.

The sewing conditions are conditions which arise already on the upper half plane, i.e.\ without taking loop diagrams into account. These conditions are statements about crossing symmetry, which leads finally to an associative bulk-boundary algebra. One can consider different combinations of bulk and boundary fields in three- and four-point functions, which therefore leads to three sewing relations that involve boundary fields (coming from correlators with four boundary fields, with two boundary fields and one bulk field, and with one boundary field and two bulk fields).

One should note that is no general solution to all these constraints known. It is not even clear, if they are consistent in all cases, and if there is always an unique solution. The investigation of relations between bulk and boundary data in conformal field theories is still an interesting branch of research.

\section{Renormalisation group flow}

Conformal field theories are constructed as scale invariant theories and represent therefore fixed points under scale transformations in the space spanned by all possible perturbations. There is a principle difference between the effects of perturbation by bulk operators and by boundary operators. The general understanding is that perturbations can possibly destroy scale invariance, which makes an adjustment of the coupling constants necessary. The task is to find new values for the couplings so that the theory is again scale invariant and consistent in the sense that no singularities with logarithmic behaviour appear in correlation functions. When this procedure is conducted perturbatively around a known conformal point, it is necessary to choose a certain scheme which determines how to deal with non-logarithmic singularities. This procedure is known as renormalisation. Finally this leads to a set of equations called $\b$-functions, which describe how the couplings must be changed under small perturbations of the theory. The dependence of these functions on the chosen renormalisation scheme is a relict of the perturbative expansion.

Perturbations of conformal field theories without boundaries by bulk fields have been studied starting with \cite{Zamolodchikov:1986gt,Zamolodchikov:1987ti}. There the $c$-theorem has been proven, stating that the central charge $c$ can only decrease under bulk deformations. An approach to renoramlisation group flow in the language of conformal field theory can be found in 
\cite{Cardy:1989daX}.

For the boundary sector the behaviour is different. One can show that perturbations with boundary operators are not able to change the central charge (at least not in the presence of finitely many perturbing operators). That means under boundary renormalisation group flow the basic properties of the bulk CFT are preserved, giving valuable information on the possible end points of the flow. Generally speaking, it is therefore often easier to keep the boundary flow under control than the bulk flow.

The influence of bulk perturbations on boundary CFTs has already been investigated in 
\cite{Cherednik:1985vs, Sklyanin:1987bi, Fring:1993mp, Ghoshal:1993tm}.

A treatment of combined bulk and boundary renormalisation group flow has not been available until \cite{Fredenhagen:2006dn}. In the following their arguments will be briefly reviewed.

Starting from a conformal action $S^*$ the perturbed action has the form
\eqn{\label{cftdefac}
	S=S^* + \sum_i \tilde\lambda_i\int d^2z \phi_i(z,\bar z)  + \sum_j\tilde\mu_j\int dx \psi_j(x)\ ,
}
where $\phi_i$ is a set of bulk fields with associated coupling constants $\tilde\l_i$, and $\psi_j$ are boundary fields with couplings $\tilde\mu_j$. The perturbing operators can have different scaling dimensions, thus the couplings themselves are dimensionful. It makes sense to define dimensionless bare couplings $\l_i$ and $\mu_j$ and introduce an explicit length scale $l$ to compensate:
\eqn{\label{cftbscale}
	\tilde\l_i=\l_i l^{h_{\phi_i}-2}\qquad \tilde\mu_j=\m_j l^{h_{\psi_j}-1}\ .
}
The classical scaling behaviour is determined by the conformal weights $h_{\phi_i}$ and $h_{\psi_j}$. One can immediately see that the couplings are scale invariant for $h_{\phi_i}=2$ and $h_{\psi_j}=1$. The associated operators are called marginal then; in case $h_{\phi_i}<2, h_{\psi_j}<1$ they are called relevant, and they are irrelevant for $h_{\phi_i}>2, h_{\psi_j}>1$. Contributions from irrelevant operators are not crucial for the renormalisation and contribute only to subleading order. Relevant perturbations on the other hand can change the conformal field theory fundamentally. Marginal operators correspond to flat directions in moduli space and therefore describe deformations which lead to families of connected theories. Although marginal perturbations do not change the scaling behaviour to lowest order, they may acquire quantum corrections which threaten scale invariance again. When such corrections are absent, the operators are called exactly or truly marginal.

When the path-integral defined by the action (\ref{cftdefac}) is expanded in the couplings, this will generate insertions of (mixed) bulk and boundary fields. When these operators coincide this leads to infinities, which must be regularised. One possibility is the introduction of an UV cutoff by demanding that
\eqn{
	|z_k^i-z_{k'}^{i'}|>l\qquad |x_k^j-x_{k'}^{j'}|>l\qquad d(z)>\frac{l}{2}\ .
}
Here $z^i_k$ denotes the $k$th insertion of the $i$th bulk operator, and $x^i_k$ analogously for the boundary fields. The third condition demands a cutoff for the distance $d(z)$ of bulk operator insertion points from the boundary. For the upper half plane $d(z)=\text{Im }z$.

These three inequalities are responsible for higher order contributions to $\b$-functions in the coupling constants. The reason is that the cutoffs appear as boundaries of integration domains, and variation with respect to $l$ generates additional contributions from these integrals to the obvious scaling behaviour in (\ref{cftbscale}). Therefore is it possible to determine changes $\d\l_i$ and $\d\m_j$ of the couplings under variations of the scale $l$, so that scale invariance of the theory is ensured. One can choose a variable $t$ parametrising the renormalisation flow and express the resulting $\b$-functions as derivatives of the couplings with respect to this parameter
\eqn{\label{cftbeta}
	\dot \l_k &= (2-h_{\phi_k})\l_k + \pi C_{ijk}\l_i\l_j + {\mathcal O}(\l^3) \\
	\dot\mu_k &= (1-h_{\psi_k})\m_k + \frac{1}{2}B_{ik}\l_i + D_{ijk}\m_i\m_j + {\mathcal O}(\m\l,\m^3,\l^2)\ .
}	
Here $C_{ijk}, D_{ijk}$ and $B_{ik}$ denote the structure constants of the bulk-three-point function, the boundary-three-point function and the bulk-boundary-two point function. Renormalisation group fixed points are determined by the vanishing of the right hand sides of (\ref{cftbeta}).

In the second equation in  (\ref{cftbeta}) we see that there is a contribution from closed string couplings to the $\b$-functions of the open string couplings. Therefore it can happen, that a marginal boundary field does not stay marginal when a bulk pertubation is switched on, i.e.\ even for marginal fields it is not always possible to set them to zero constantly. This situation will be crucial in the arguments in chapter \ref{ch-ren} and \ref{ch-superpot}. In particular one understands that, at least for these perturbative $\b$-functions, changes in the bulk fundamentally modify the boundary theory, but changes in the boundary sector have no effect on the bulk. In chapter \ref{ch-ren} the renormalisation group methods will be applied to the factorised boundary string field theory action, where the distinction of the two sectors is not that clear any more.

%
%
%

\chapter{WZW-models}

\section{Closed string}

Wess-Zumino-Witten models were introduced in 
\cite{Wess:1971yu, Witten:1983ar, Witten:1983tw} as $\s$-models with Lie groups as target spaces. They represent spacial version of CFTs, where a larger symmetry is present, given by the Lie algebra of the target space. The algebra is described by the generators
\eqn{\left[T^a,T^b\right]=i{f_{ab}}^cT_c\ ,}
where $f_{abc}$ are the structure constants of the group. The trace operator used here is normalised as
\eqn{\Tr\left(T^aT^b\right) = 2\delta_{ab}\ .}

The model has a symmetry which is generated by currents $J$ and $\bar J$.
Their algebra is obtained as central extension of the Lie algebra. This is known as Kac-Moody algebra and is given by
\eqn{
	\left[J_n^a,J_m^b\right] = i {f^{ab}}_c\, J^c_{n+m} + k\,n\,\d^{ab}\d_{n+m,0} \ .
}
Again, there is also an independent anti-holomorphic sector present generated by $\bar J$ with the same commutation relations. The constants ${f^{ab}}_c$ appearing in the algebra are the structure constants of the original Lie group. $k$ determines the central extension and is called the level of the model. From unitarity constraints it follows that $k$ must be a positive integer.

Once this structure is given, the Virasoro algebra can be obtained from it by the so-called Sugawara-construction. The Virasoro generators are then given by
\eqn{\label{wzwvir}
	L_n = \frac{1}{2}\frac{1}{k+h^\vee} \sum_a\sum_m :J_m^aJ_{n-m}^a:
}
involving normal ordered products of the Kac-Moody generators. $h^\vee$ is the dual Coxeter number, which is a constant associated to the Lie group. This Virasoro algebra comes with a central charge
\eqn{
	c=\frac{k\, d}{k+h^\vee}\ ,
}
where $d$ is the dimension of the Lie group. 

The remaining commutation relations are given by
\eqn{
	\left[L_n,L_m\right] &= (n-m)L_{n+m} + \frac{c}{12}(n^3-n)\d_{n+m}\\
	\left[L_n,J_m^a\right] &= -mJ_{n+m}^a\ .
}
These can be used to checked that 
\eqn{
	J(z)=\frac{1}{k}J^a(z)T_a
}
is primary with weight 1, where $T_a$ are the generators of the KM-algebra.

The connection to the worldsheet description can be provided by introducing the map $g$ from the worldsheet into the group. This way the KM-generators can be represented as
\eqn{
	J = -\P_z gg^\1\qquad\bar J=g^\1\P_\bz g\ .
}
The $\s$-model action is given by the two-dimensional integral 
\cite{Witten:1983tw, Gawedzki:1999bq}
\eqn{
	\frac{k}{4\pi i}\Tr\int_\Sigma d^2z\; g^\1\P g g^\1\bP g\ .
}
It turns out that this action, once quantised, is not conformally invariant, although it is classically scale-invariant. It can but be restored by addition of a topological term, the so-called Wess-Zumino term. The full action is given by
\eqn{
	S^{WZW}(g) &= \frac{k}{4\pi i} L(g) 
			+ \frac{k}{4\pi i}\Gamma(g)\ .
}
The first term is given, as states before, by
\eqn{
	L(g) = \Tr\int_\Sigma d^2z\; g^\1\P gg^\1\bP g\ .
}
The second term is more intricate to define. It is assumed that the target group $G$ is connected and simply connected and that the worldsheet $\Sigma$ is boundary-free. In this case the field $g:\Sigma\to G$ map be extended to a map $\tilde g: B\to G$, where $B$ is 3-dimensional whose boundary is the closed string worldsheet, $\P B=\Sigma$. In this case one may define the WZ-form\footnote{for notational convenience there will be no further distinction between $g$ and $\tilde g$.}
\eqn{
	\chi = \frac{1}{3} \Tr\left(dgg^\1\right)^{\wedge 3}
}
and set
\eqn{
	\Gamma(g) = \int_B g^*\chi\ .
}
Note that the pullback has been explicitly included because $\Gamma(g)$ is defined through an integral in the target space.  

Whenever $\chi$ is exact, $\Gamma$ can be reduced to a worldsheet integral, but in the general case $\Gamma(g)$ is multi-valued. This is because there are ambiguities of the form $\int_B \Delta g^*\chi$, where $\Delta g$ stands for the difference of the different extensions. As the holonomy group is $\mathbb Z$ for the groups under consideration, the ambiguities in the action are integer multiples of $\frac{1}{4\pi i}\Gamma(g)$. The only way to make path-integral independent of the choice of the extension is to arrange the coefficient in a way so that the ambiguous contributions are multiples of $2\pi$.

The relative coefficient between $L(g)$ and $\Gamma(g)$ is fixed by the requirement of conformal invariance. Luckily it is possible to adjust the parameter of the model, $k$, in a way that yields well-defined correlation functions, namely by demanding that $k$ is positive integer.

One should note that, although the WZ-term is given by an integral over a 3-dimensional extension of the worldsheet, its variation is still a worldsheet integral. 
In the presence of a $B$-field (a two-form $B$ living in the target space (group manifold) $G$),
the variation of the $B$-field part of the action is
\begin{align}\begin{split}
\d\int_\Sigma g^* B &= \int_\Sigma (g+\d g)^*B - \int_\Sigma \d g^* B \\
	&= \int_{\P^\1 (g_*\Sigma \cup (g+\d g)_*\Sigma)} dB = \int_{\P^\1(g_*\Sigma \cup (g+\d g)_*\Sigma))} \chi,
\end{split}\end{align}
where the first two integrals are over the worldsheet $\Sigma$, and the 
last two integrals are over a submanifold of the target space (defined by the map
$g: \Sigma \to G, g(\Sigma) =: g_*\Sigma$). 
The expression $\P^\1 O$ denotes the volume enclosed by the boundary-less surface $O$.
Clearly, $g_*\Sigma \cup (g+\d g)_*\Sigma)$ has no boundaries, because closed strings are considered,
and the variation at initial and final times vanishes.

$\chi$ is a three-form living in the target space. As soon as it is no longer exact 
(i.e. $\chi \neq dB$) it cannot easily be expressed from
a worldsheet point of view (indeed, the invariance of the action under $G\times G$
determines $\chi$). One way to proceed is to extend the worldsheet $\Sigma$ 
by a third coordinate $y$, so that the integration domain can be expressed by means
of an extended map $\tilde g (z, \bar z, y)$ and a three-dimensional extension $B$ of the worldsheet:
\begin{align}\begin{split}
	\tilde g_* B &= \P^\1(g_*\Sigma \cup (g+\d g)_*\Sigma)) \\
	\Sigma &\subset B \\
	\tilde g(z, \bar z, 0) &= g(z, \bar z) \\
	\tilde g_*|_{y=0} B &= g_*\Sigma
\end{split}\end{align}

The equations of motion are obtained through the two contributions to the variation of the action. In addition to the topological term, the local worldsheet integral provides another term, so that both combined give the equations of motion
\eqn{
	\bP J = 0 = \P\bar J\ .
}
These state that the classical WZW-currents are holomorphic or anti-ho\-lo\-mor\-phic, respectively.

Finally we list the OPEs in several relevant cases, as these have been used in later calculations. They are given by
\eqn{
	J^a(z)J^b(w)&\sim \frac{k\d^{ab}}{(z-w)^2} + i{f^{ab}}_c\frac{J^c(w)}{z-w}\\
	J^a(z)g(w,\bar w) &\sim-T^a\frac{g(w,\bar w)}{z-w} \\
	\bar J^a(z)g(w,\bar w) &\sim\frac{g(w,\bar w)}{z-w}T^a\ .
}

\subsection{Polyakov-Wiegmann identity}

The WZW-action also satisfies a very useful identity, which is due to Pol\-ya\-kov and Wiegmann \cite{Polyakov:1983tt}. When the field $g$ is composed of two field as $g=g_1g_2$, the terms in the action split into
\eqn{
	L(g_1g_2) &= L(g_1) + L(g_2) + \Tr\int d^2z \left(g^\1\P g_1\bP g_2 g_2^\1 + g_1^\1\bP g_1 \P g_2g_2^\1\right) \\
	\Gamma(g_1g_2) &= \Gamma(g_1) + \Gamma(g_2) -\Tr\int d^2z\left(g_1^\1\P g_1\bP g_2g_2^\1 - g_1^\1\bP g_1\P g_2g_2^\1\right)\ ,
}
which combine to
\eqn{
	S(g_1g_2) &= S(g_1) + S(g_2) + W(g_1,g_2)
}
with
\eqn{
	W(g_1,g_2) &= 2\Tr\int d^2z\; g_1^\1\bP g_1\P g_2g_2^\1\ .
}	
This is valid for the closed string case. In the open string case, where $\Sigma$ has a boundary, more complications appear.

\subsection{Chiral symmetry}

The holomorphicity of the currents points towards a chiral symmetry of the model.
It can easily be verified that
the action has a  symmetry group $G_L \times G_R$, which acts as
\begin{align}\begin{split}
	g \to h_L(z) g h_R(\bar z)\ .
\end{split}\end{align}
Most easily this can be seen when 
which can be seen when one applies the Polyakov-Wiegmann formula:
\begin{align}\begin{split}
	L[h_Lgh_R] + \Gamma[h_Lgh_R] &=
		L[h_L] + \Gamma[h_L] + L[gh_R] + \Gamma[gh_R] + W[h_L, gh_R] \\
		&= L[h_L] + \Gamma[h_L] + L[g] + \Gamma[g] + L[h_R]  \\ 
		&\qquad + \Gamma[h_R] + W[h_L, gh_R] + W[g, h_R] \\
		&= L[g] + \Gamma[g]
\end{split}\end{align}
The fresult in the final line is obtained by noting that $L[h_L]$ vanishes because it contains derivatives $\bar\P$; the same is true for $L[h_R]$, $W[h_L, gh_R]$ and 
$W[g, h_R]$. $\Gamma[h_L]$ is zero because it is given by $(h_L^\1 \P h_L dz)^{\wedge 3}$ and holomorphicity of $h_L$ causes any three-form constructed out of an extension into the interior of the worldsheet to vanish. This argument is also true for $\Gamma[h_R]$. Therefore the action possesses the symmetry claimed.

\section{Open string}

As expected, the situation changes drastically when a worldsheet boundary is introduced into the model \cite{Klimcik:1996hp, Alekseev:1998mc, Stanciu:1999id}. The most obvious effect it has is, that the symmetry group cannot be fully preserved any more. We will see below that conformal boundary conditions imply that the worldsheet boundary lies in a (twisted)
conjugacy class, so that only part of the full symmetry can be realised.

In the case of open strings variation of the $B$-field parts yields
\begin{align}\begin{split}
	\d\int_\Sigma g^* B &= \int_\Sigma (g+\d g)^*B - \int_\Sigma \d g^* B
\end{split}\end{align}
as well, but the surface $g_*\Sigma \cup (g+\d g)_*\Sigma$ has a boundary, so that the operation
$\P^\1$ is not well defined. However, this problem can be overcome in the presence of D-branes,
because they provide a unique way to close the holes by using them as "caps". This is possible,
because $(g_*\Sigma \cup (g+\d g)_*\Sigma) \cap (\text{D-branes})$ are 1-cycles.
When $D_1$ and $D_2$ denote
those parts of the D-brane hyperplanes, which are bounded by these 1-cycles, the domain of
integration can be defined as
$ M:=\P^\1 (g_*\Sigma \cup (g+\d g)_*\Sigma \cup D_1 \cup D_2)$. The $D_i$ are not uniquely determined,
but that this does not matter as long as the variation is correctly defined as
\begin{align}\begin{split}
	\int_M \chi - \int_{D_1} \a_1 - \int_{D_2} \a_2.
\end{split}\end{align}
Independence of the choice of the interpolating hyperplanes $D_i$ is achieved by demanding that
\begin{align}\begin{split}
	d\a_i = \chi|_{D_i}
\end{split}\end{align}
By this definition $\a$ is not determined uniquely, but only up to an exact one-form $\b$:
\begin{align}\begin{split}
	\a \to \a + d\b
\end{split}\end{align}
Thus the action should read
\begin{align}\begin{split}
	S^{WZW}[g] &:= \frac{\kappa}{4\pi i} L[g] + \frac{\kappa}{4\pi i} \Gamma[g] -\frac{\kappa}{4\pi i} \int_D (\a+d\b) \\
		&= \frac{\kappa}{4\pi i} L[g] + \frac{\kappa}{4\pi i} \Gamma[g] 
			-\frac{\kappa}{4\pi i} \int_D \a - \frac{\kappa}{4\pi i} \oint_{\P D} \b \\
		&= \frac{\kappa}{4\pi i} L[g] + \frac{\kappa}{4\pi i} \Gamma[g] 
			-\frac{\kappa}{4\pi i} \int_D \a + \frac{\kappa}{4\pi i} \oint_{\P \Sigma} \b \\
\end{split}\end{align}
This means that an action
\begin{align}\begin{split}
	S \sim L[g] + \int_{g_*\Sigma} B + \int_{g_*\Sigma \cap \text{D-brane}} A
\end{split}\end{align}
is only well defined up to
\begin{align}\begin{split}
	(\chi, \a_1, \a_2, \b_1, \b_2) := 
		\frac{\kappa}{4\pi i} \Bigl[\int_M \chi - \int_{D_1} \a_1 - \int_{D_2} \a_2 
		+ \oint_{\P\Sigma}\b_1 + \oint_{\P\Sigma}\b_2 \Bigr]
\end{split}\end{align}

\subsection{D-branes}

Having identified conformal boundary conditions for the WZW model (although, as remarked before, there do often exist boundary conditions which are not of the maximally symmetric type described by conjugacy classes; maybe not even twisted conjugacy classes are general enough to capture all possible boundary conditions) we are now interested in their spacetime interpretation.

Following \cite{Alekseev:1998mc} we use the currents\footnote{The definition of $J$ differs by a minus sign
compared to \cite{Alekseev:1998mc}.}
\begin{align}\begin{split}
	J = \P g g^\1 \qquad \bar J = g^\1 \bar\P g
\end{split}\end{align}
and re-write them in term of tangential and normal vectors on the disk. In the open string picture these are given by
\begin{align}\begin{split}
	&\P_t = \2(\P + \bar\P) \qquad \P_n = \2(\P - \bar\P) \\
	&\P = \P_t+\P_n \qquad \bar\P = \P_t - \P_n.
\end{split}\end{align}
With
\begin{align}\begin{split}
	\Ad_f X := f X f^\1
\end{split}\end{align}
the currents can be re-written as
\begin{align}\begin{split}
	J &= \P_t g g^\1 + \P_n g g^\1 = \Ad_g (g^\1 \P_t g~) + \Ad_g (g^\1 \P_n g)\\
	\bar J &= g^\1 \P_t g - g^\1 \P_n g
\end{split}\end{align}
$J \pm \bar J =0$ are conformal boundary conditions.

\paragraph*{The Dirichlet case}

The sum of both is given by
\begin{align}\begin{split}
\label{JJ1}
	J + \bar J = (1+\Ad_g) g^\1 \P_t g + (1-\Ad_g) g^\1 \P_n g\ .
\end{split}\end{align}
In the commutative limit, which may be defined as the limit in which all those structure constants in the current-current OPE vanish, which are not contracted with the identity operator (i.e.\ only $C^0_{ij}$ survived), 
(\ref{JJ1}) reduces to $\P_t X = 0$. This is a well-known condition from open bosonic string theory. It describes Dirichlet boundary conditions. On the other hand we will find, that the combination $J - \bar J = 0$ corresponds to Neumann conditions.
This is verified as follows.

Acting on a fixed group element $y$, $\Ad_g$ describes an orbit in the group. We can split the tangent space
of the group into a part $T^n_y$ which is normal to the orbit, and a part $T^t_y$ tangential to it.
Note that therefore the action of $\Ad_g$ on $T^n_y$ is trivial, i.e.
\begin{align}\begin{split}
	\Ad_g \Bigr|_{T^n} = \text{id},
\end{split}\end{align}
whereas we cannot say anything about its action on the tangential part.
Therefore, on $T^n$ the Dirichlet conditions becomes
\begin{align}\begin{split}
	(J+\bar J)\Bigr|_{T^n} = (1+\Ad_g) g^\1 \P_t g \Bigr|_{T^n} = 0
\end{split}\end{align}
Thus indeed the Dirichlet condition $g=\text{const}$ is satisfied as long as
the boundary of the worldsheet maps into a conjugacy class.

On the conjugacy class, i.e. evaluated on $T^t$, the operater $1-\Ad_g$ is invertible (because it
vanishes only on $T^n$, except for degenerate cases). It is possible to define a 2-form
\eqn{
\w = g^\1 dg \frac{1+\Ad_g}{1-\Ad_g} g^\1 dg\ .
}
It can be shown that $d\w \propto (g^\1 dg)^3$.
This provides a geometric interpretation of the boundary condition. With this point of view we see that conjugacy classes are in fact D-branes of WZW models.

Note that in the normal directions the Neumann condition $\P_n g$ is {\em not} satisfied.

\paragraph*{The Neumann case}

\begin{align}\begin{split}
	J - \bar J = (1-\Ad_g) g^\1 \P_t g - (1+\Ad_g) g^\1 \P_n g,
\end{split}\end{align}
Again, we can split the tangent space like before. But now
the operator $1-\Ad_g$ vanishes and we end up with Neumann boundary conditions
$\P_n g=0$ on the conjugacy class.

Note that in a WZW setting there is less freedom to choose D-branes of different dimensions than in ordinary tensor products of flat bosonic CFTs. In particular it is not always possible to impose Dirichlet conditions in all direction, i\.e\. to construct a D0-brane,
because this would be in conflict with the commutation relations between the currents.

\subsection{Conjugacy classes}

As has been already indicated above a way to construct D-branes in WZW model is provided by the use of conjugacy classes \cite{Fredenhagen:2000ei}.
Generally, the only requirement for conformal invariance is the vanishing of the energy momentum tensor on the boundary,
\eqn{
	\Tr T = T(z)-\bar T(\bar z) \equiv 0\ .
}
One way satisfy this condition is by employing the Sugawara construction of $T$ by taking $T(z)\sim :J(z)J(z):$ and $\bar T(\bar z)\sim :\bar J(\bar z)\bar J(\bar z):$. For the currents $J$ and $\bar J$ this means that they must be glued together by fulfilling
\eqn{
	J(z) = \Lambda \bar J(\bar z)
}
at the boundary. $\Lambda$ denotes an automorphism of the algebra, and its choice is restricted by conformal invariance. It is possible to associate a geometry to the choice of $\Lambda$, as demonstrated in the previous paragraph, where a rather explicit description of the D-brane through the vanishing of the normal bundle was given. Generally the automorphisms are of the form
\eqn{
	\Lambda = \Omega \Ad_g\ ,
}
where $\Omega$ is an outer automorphism, i.e.\ one that does not depend on location $g$ in the group. It is clear then,
that in the simplest case, where $\Omega=$id the D-branes are given by conjugacy classes

\begin{align}\begin{split}
	g |_{\P\Sigma} \in C(f)
\end{split}\end{align}
(if $f \in g(\P\Sigma)$ then $hfh^\1 \in g(\P\Sigma)$), where $C(f)$ is a conjugacy class
\begin{align}\begin{split}
C(f) := \{gfg^\1, g \in G\}\ . 
\end{split}\end{align}

This construction coincides with the example in the previous paragraph, so that generalised Dirichlet boundary conditions can be associated to conjugacy classes with trivial outer automorphism, whereas the Neumann boundary conditions correspond to conjugacy classes with $\Omega = -$id. In general cases other outer automorphism may be of relevance, leading then to twisted conjugacy classes, which are but of not further importance for what follow.

The action is then constructed as
\begin{align}\begin{split}
	L_{g_*\Sigma} + \Gamma_M - \int_D \a
\end{split}\end{align}
where
\begin{align}\begin{split}
	d\a(g) = \chi(g) \qquad \forall g \in C(f) 
\end{split}\end{align}
It is possible to evaluate the topological term $\Gamma$ on the conjugacy class. In fact, it reduces to a local contribution then, as can be seen from the following calculation:
\begin{align}\begin{split}
\label{GAMMA-CONJUGACY}
\Gamma[gfg^\1] &= \Gamma[g] + \Gamma[fg^\1] 
	-\int_M d (g^\1 dg d(f g^\1)(f g^\1)^\1) \\
	&= \Gamma[g] + \Gamma[g^\1] + \int_M d(f^\1 g^\1 dg f g^\1 dg) \\
	&= \int_{\P M} f^\1 g^\1 dg f g^\1 dg 
\end{split}\end{align}
Then $\a$ is defined as
\begin{align}\begin{split}
	\a = f^\1 g^\1 dg f g^\1 dg 
\end{split}\end{align}
In addition the ambiguities $\d S$ must be integer multiples of $2\pi$, which
imposes a constraint on valid conjugacy classes. In particular the consequence is that the WZW level $k$ will be quantised.

For illustrational purposes,
and as preparation of chapter \ref{ch-ren}, we briefly explain the maximally symmetric D-branes of a WZW model with target group SU(2).

\subsection{Example: D-branes in $SU(2)$}

As was noted above, branes in WZW models are, at least in the maximally symmetric case, described by conjugacy classes of the target group. The geometry of such subgroups of $SU(2)$, which is isomorphic to $S^3$ is well-known and given by a $S^2$-bundle which degenerates at the poles of $S^3$.

For these D-branes a conjugacy class $C$, which is a closed manifold, can be contracted in two ways. The difference between the possible contractions leads to the ambiguity
\eqn{
	\Delta S = \int_C \omega + \frac{k}{4}\int_B \chi
}
in the action.
In the second integral appears the integration domain $B$, which is the ball in $S^3$, which is bounded by $C$.

In order for the path-integral to be well defined the ambiguity must be an integer multiple of $2\pi$. An explicit evaluation shows that these integers run from 1 to $k-1$. Thus there are $k-2$ D2-branes in the $SU(2)$-WZW model, which are characterised by the fact that they pass through the points diag$(\exp\pi i\frac{n}{k}, \exp-\pi i\frac{n}{k})$. At the points $\pm 1$ the 2-branes degenerate to points, i.e.\ they are D0-branes.

%
%
%

\chapter{Spacetime interpretation}

In previous sections the $\s$-model approach has been introduced as a way to compute scattering amplitudes starting from a conformal two-dimensional worldsheet theory. It provides a way to compute S-matrix elements in the closed string case and in the open string case. For a full description of the theory, one would like to have a full effective action, though. 

According to the philosophy behind the $\s$-model description, the couplings of the worldsheet operators gain the interpretation of spacetime fields in the effective theory. An effective theory containing all possible fields which can be generated by the string, should encode complete information about the vacuum structure of string theory, also non-perturbatively. 

We will review the basic arguments for the bosonic case, leading to the conjecture that the effective action equals the generating functional ${\cal S}={\cal Z}$. Later we will see that BSFT modifies this conjecture.

\section{The string path-integral}

In the attempt to formulate string theory in a way which resembles the description of quantum field theories, a path-integral formalism has been established \cite{Green:1987sp, Green:1987mn}, which resides on an action principle minimising the area of $1+1$-dimensional string worldsheets. A suitable formulation has been given by the Polyakov-action
\eqn{\label{smaction}
	S = \frac{T}{2}\int d^2\s
		\left(
		\sqrt{h}h^{ab}G_{\m\n} + i\varepsilon^{ab}B_{\m\n}\right)\P_a X^\m\P_b X^\n
		+\frac{1}{4\pi} \int d^2\s \sqrt{h}R\Phi\ ,
}
where $\s^{1,2}$ and $h$ are the worldsheet coordinates and worldsheet metric, $G, B, R$ and $\Phi$ are spacetime metric, Kalb-Ramond field, Ricci scalar and the dilaton field. The string tension $T$ is given by $T=\frac{1}{2\pi\a'}$.

A configuration which is supposed to reproduce usual quantum field theories is given by flat spacetime $G=\eta$, vanishing Kalb-Ramond field $B=0$ and constant dilaton. In this case the theory is quantised by choosing the standard measure in the path-integral.

Still, there are fields present which do not have an obvious interpretation from the spacetime point of view, in particular the worldsheet metric $h$. In fact, an investigation of the partition function shows that there is an anomaly present, which inhibits the theory from being scale invariant.

It had been a great success of string theory to find a way how to avoid that so-called Weyl-anomaly. It turned out that the vanishing conditions for it require that the worldsheet theory is a conformal field theory with a central charge of $c=26$ in the bosonic case and $c=15$ in the supersymmetric case. This implicitly fixes the number of spacetime dimensions in case the (tensor products of) CFTs describe free particles.
For bosonic string theory one obtains $D=26$ spacetime dimensions and for the supersymmetric case $D=10$.

The obtained action is scale invariant and therefore describes string theory at a RG fixed point. Therefore one has obtained a classical vacuum of the theory. Setting stability considerations of the vacuum aside, it is possible to consider small perturbations in the spacetime fields and expand the path-integral in the couplings. In this way one can calculate S-matrix elements. 

The problem with this approach is that a string can be excited in infinitely many ways. In other words, there will be infinitely many couplings present which must be taken into account. The spacetime fields will be functions of the string map $X$, so that the space of couplings is determined by a power expansion as well as by a derivative expansion of the target space fields in the string map. Such an expansion is ambiguous, since partial integration can change the structure of the derivative expansion.

Let ${\cal Z}$ be the partition of the $\s$-model. It is also a generating functional for expectation values of vertex operators $\<V_1\dots V_n\>$
\cite{Fradkin:1984pq, Fradkin:1985ys}. 
It has been argued that during renormalisation of the action, massless poles are subtracted, thus the partition function obtained from the renormalised action is conjectured to be related to the effective action for the massless modes 
\cite{Tseytlin:1986ti, Tseytlin:1986zz}.

Usually, for the computation of correlation functions a certain Weyl-gauge of the worldsheet metric is chosen, in particular the conformal gauge when the computations are done in a CFT framework. Nevertheless the $\s$-model is naturally defined `off-shell', i.e. away from conformal gauge and the $D=26$ or $D=10$ condition. Thus one might suspect, that the $\s$-model description is indeed a good starting point for writing down an off-shell extension of the S-matrix.

Indeed we shall see in later chapters, that such an extension is possible and is provided in form of boundary string field theory (BSFT) for the bosonic open string. In the case of the closed string, a similar off-shell action cannot be written down, at least not while attempting to use similar methods as in the open string sector. However, it will be explained in some detail that BSFT indeed is able to capture information about the closed string background. The reason for that is a non-trivial correspondence between the open and closed string sector.

\section{Effective action and renormalisation group flow}

For background fields, which do not represent a flat background spacetime, the $\s$-model action (\ref{smaction}) is subject to restrictions which ensure Weyl invariance. This is necessary to set up a consistent theory, which is invariant under scale transformation. To first order the Weyl variation can be measured by the trace of the worldsheet energy-momentum tensor (in a particular renormalisation scheme) \cite{Green:1987sp, Polchinski:1998rq}
\eqn{
	T^a_a = - \frac{1}{2\a'}\b^G_{\m\n}h^{ab}\P_a X^\m\P_b X^\n 
			- \frac{i}{2\a'}\b^B_{\m\n}\varepsilon^{ab}\P_a X^\m\P_b X^\n - \frac{1}{2}\b^\Phi R\ .
}
The $\b$-functions which appear in this expression have been obtained in lowest order in the string coupling as
\eqn{
	\b^G_{\m\n} 
	&= \a' R_{\m\n} + 2\a'\nabla_\m\nabla_\n\Phi - \frac{\a'}{4}H_{\m\l\omega} {H_\n}^{\l\w} + {\cal O}(\a'^2) \\
	\b^B_{\m\n}
	&= -\frac{\a'}{2}\nabla^\w H_{\w\m\n} + \a'\nabla^\w\Phi H_{\w\m\n} + {\cal O}(\a'^2) \\
	\b^\Phi
	&= \frac{D-26}{6} - \frac{\a'}{2}\nabla^2\Phi + \a'\nabla_\w\Phi\nabla^\w\Phi
		-\frac{\a'}{24}H_{\m\n\l}H^{\m\n\l} + {\cal O}(\a'^2)\ .
}q
In this expression, $\nabla$ is the covariant derivative and $H=dB$.
Scale invariance is obtained through the conditions
\eqn{
	\b^G=\b^B=\b^\Phi=0\ .
}
For the flat background spacetime these $\b$-functions vanish, except for the first term in $\b^\Phi$, which gives again the condition on the number of spacetime dimensions $D$. 

The most remarkable property of these equations is that they can be integrated. In fact it is possible to obtain them as variations of a spacetime action. This action is given by
\eqn{
	{\cal S} &= \frac{1}{2\k_0^2}\int d^Dx\sqrt{-G}e^{-2\Phi}
		\left[ -\frac{2(D-26)}{3\a'} + R -\frac{1}{12}H^2+4|\P_\m\Phi|^2 + {\cal O}(\a')\right]\ .
}
We will encounter a technically similar situation, though for open strings, in the investigation of the topological theory. The string map is then restricted to constant maps, so that the integration of the $\b$-functions yields an effective static spacetime potential.

\section{Open string effective action}

In the open string case similar methods can be applied to the $\s$-model action. The difference here is that now additional boundary fields are present. The boundary action which can be added has the form
\eqn{
	\int dx^0\left[ T(X) + A_\mu(X)\dot X^\mu + \cdots\right]\ ,
}
where $T$ is the open string tachyon, $A$ denotes the photon field and the dots stand for massive fields.
While this $\s$-model partition function approach was successful for the massless 
string modes leading to covariant expressions to all orders in powers of gravitons and 
dilatons in the closed string case and the vector field strength in the open string case, 
it produced unfamiliar expressions when applied to the tachyon field $T$. The expression for the partition function 
${\cal Z}[T]$ computed by expanding in 
derivatives of $T$ has the following structure in the critical bosonic string theory (both in 
the closed string case on 2-sphere and open string case on the disk):
\eqn{
	{\cal Z} &= a_0 \int d^DX e^{-T}\left[ 1+ a_1\a'\P^2T +{\cal O}(\a'^2)\right]\ .
}
The constants $a_0$ and $a_1$ are renormalised constants, where $a_1$ is scheme dependent.
Again, as in the closed string case, some extra input or guiding principle is necessary to fix an off-shell extension of scattering amplitudes. 

Reverting to the original boundary $\s$-model with tachyonic and massless modes only, one notices that the model is renormalisable within the standard derivative expansion, i.e.\ the space of boundary couplings is closed under renormalisation group operations.
As has been argued in
\cite{Tseytlin:1986ti, Tseytlin:1987ww, Tseytlin:1989md},
the effective action for the massless fields should be given by the renormalised partition function. This conjecture holds up to the first few orders 
\cite{Andreev:1988cb}.
However, when in addition to the photon field also a tachyon field is admitted, then one finds that the tachyon generates a potential. This requires but a modification of the conjecture ${\cal S}[T,A] = {\cal Z}[T,A]_{ren}$, and we will see in chapter \ref{ch-ren} how this can be implemented.

\part{Closed string deformations in open string field theory}

%
%
%

\chapter{Boundary string field theory}

\section{Generalities}

The formulation of string theory as it has been presented in the previous chapter is a perturbative formulation. Implicitly it is assumed that the theory can be formulated as expansion around a fixed configuration point, a consistent vacuum. This is most obvious when employing the language of conformal field theories as they, by definition, describe a renormalisation group fixed point. This immediately raises the question, whether it is possible to formulate, in the spirit of quantum field theory, a {\em string field theory}. Despite the success of the conformal field theory description of string theory it is clear, that in the end the theory can only be finalised by setting up a `second quantised' version.

Such a string field theory would incorporate all possible vacua as classical configurations, would contain an understanding of non-perturbative phenomena like solitonic connections between distant points in moduli space and would possibly also contain M-theory as limit.
In any case it is reasonable to assume that such a second quantised version should be evidently built upon fundamental principles of string theory -- principles, which are certainly not (easily) accessible in perturbative formulations. Maybe the lacking of such basic insight is the greatest flaw in modern string research; or the greatest challenge. A big problem is, for example, to identify the overall dynamical degrees of freedom which are not only valid in a certain region of moduli space. In connection with this it seems that a subtle relation between open and closed strings play an important role.

Most excitingly a certain correspondence between them is obvious already in the earliest calculations of scattering amplitudes. Although initially open and closed strings are treated as different objects, it becomes clear that open string amplitudes contain closed strings as intermediate states as soon as one goes beyond tree-level. The heuristic picture of one-loop open strings which look like tree-level closed strings support this observation.
Over the years a number of evidences has been found from various areas in string theory pointing towards a certain duality of open and closed strings \cite{Gross:1970eg, Lovelace:1971fa, Strominger:1986zd, Bonora:2006tm, Katsumata:2004cc, Garousi:2004pi, Sen:2003iv, DiVecchia:2003ae, Shatashvili:2001ux, Mayr:2001xk, Kapustin:2004df}. The reason for progress in this direction was a increasingly better understanding of non-perturbative open strings. At last the availability of {\em open string field theories}, which are in the centre of interest for such issues, has given various new insights, among them the famous Sen conjecture (see \cite{Sen:2004nf} and references therein).

An open string field theory comprises the consistent truncation to the open string sector concerning the relevant degrees of freedom, at least non-perturbatively. In modern language it can be considered as the worldvolume theory that includes all open string modes living on a D-brane. There are basically two formulations available, cubic OSFT \cite{Witten:1985cc} and boundary string field theory (BSFT) \cite{Witten:1992qy,Witten:1992cr,Shatashvili:1993kk,Shatashvili:1993ps}, the relation between both being not entirely clear. Both come with advantages and disadvantages, predestinating them for application in different realms. 

In works of Sen it has been shown that BSFT provides an answer to the vacuum selection problem for open strings. Namely, in the limit of small string coupling constant the closed string background can be fixed and it then is possible to ask, what are the possible D-brane configurations for such a background. This question is answered by {\em classical open string field theory}, whose equations of motion directly give the desired vacua. In this picture D-branes are viewed as solitons  of the open string tachyon. Since it is possible to calculate the potential for the tachyonic degrees of freedom and to consistently truncate the theory to massless (and tachyonic) fields, the minima of the potential correspond to static vacua. 

The next step would be the incorporation of closed string degrees of freedom in this framework. The approach followed by Zwiebach \cite{Zwiebach:1997fe} rests on the idea to include them by and be very careful with counting them. This is because at loop-level closed strings are expected to appear alone from the open string sector already, so that an overcounting of the closed string degrees of freedom must be avoided. From a conceptual point of view open and closed strings are treated on equal footing and appear both as fundamental degrees of freedom. In Witten's cubic OSFT on the other hand it only necessary to explicitly work with open string modes in order to generate closed string poles in the S-matrix. This has been shown in \cite{Giddings:1986wp, Zwiebach:1990az}.
Therefore, closed string may be viewed as derived objects in this approach.

It is not clear how similar processes work in BSFT. It has been shown that BSFT is capable of describing the decay of D-branes into the vacuum (which is most presumably the closed string vacuum). Also, it was argued in \cite{Gerasimov:2000ga} that open string degrees of freedom are removed during this decay, and using some intuition obtained from \cite{Sen:2004zm} one is led to think that information about closed strings at the endpoint of the decay must be contained in the BSFT action. The problem with this approach is the close connection to the worldsheet formulation of string theory (which is an advantage from the CFT point of view, since it is easy then to establish a connection between the worldsheet and the spacetime description). But the worldsheet formulation is supposed to be `local' in moduli-space, i.e.\ it is not expected to be a good formulation of the physics of the tachyon condensate. Nevertheless the main result of this thesis is to show that indeed BSFT can in principle be used to answer questions about the degrees of freedom at the endpoint, although they appear in a rather involved way.

Before tackling this issue an introduction to BSFT will be provided. Its relation with effective actions will be discussed and some well-known though important results for the bosonic tachyon potential are presented.

\section{Introduction to BSFT}

Following \cite{Witten:1992qy, Witten:1992cr} and \cite{Shatashvili:1993kk, Shatashvili:1993ps} the basic idea behind the construction of BSFT is the application of the BV-formalism \cite{Henneaux:1992ig} on a field theory with infinitely many degrees of freedom, which exactly constitute the open string degrees of freedom under consideration. Supposed now such a theory does exists, it must be formulated over a supermanifold containing fields, anti-fields, ghost and anti-ghosts. Moreover a closed non-degenerate odd symplectic $\omega$ must exist together with a well-defined ghost number operator $U$. The supermanifold has Darboux coordinates $q^a$ and $\theta^a$ of grading 0 and 1. Locally, $\omega=d\theta_adq^a$, so that the BV-antibracket is given by
\be
	\left\{A,B\right\} = \frac{\PR A}{\partial u^k}\omega^{kj}\frac{\PL B}{\partial u^j} \ ,
\ee
where $u^a$ are local coordinates on the supermanifold. The BV-action is then determined through the master equation
\be
\label{mastereq}
	\left\{ S,S\right\} = 0.
\ee
In addition the existence of a vector field $V$ is postulated which generates the symmetries of $\omega$. It is then easy to show that the action must satisfy\footnote{\ref{dSlocal} implies \ref{mastereq}.}
\be
\label{dSlocal}
	dS = i_V\omega\ .
\ee
The important step is now to make contact between the abstract formulation and string theory by identifying the relevant objects on both sides. First of all, the coordinates are taken to be the boundary degrees of freedom of a two-dimensional field theory on a disk with conformal bulk. This makes sure that the resulting theory is formulated in terms of open strings, and that the closed string background is completely fixed. Neglecting issues of well-definedness it is intuitive to think of this as the space of all 2d field theories fibered over all possible bulk configurations, so that its tangent space is constituted by the coupling constant $u$ of local {\em open string} vertex operators ${\cal V}$ only. BRST invariance is taken into account by demanding that ${\cal V}$ is closed up to exact forms, i.e.\
\be
	\left\{Q_{BRST},{\cal V}\right\} = d{\cal O}
\ee
for some operator ${\cal O}$.

In the next step the symmetry generated by $V$ is identified with the BRST symmetry, and $V$ therefore with its current. Finally, the odd symplectic form can be defined as worldsheet correlation function
\be
	\omega\left({\cal V}_1,{\cal V}_2\right) = \int_{\P\Sigma}dt_1dt_2\langle{\cal O}_1(t_1){\cal O}_2(t_2)\rangle\ ,
\ee
where the integration goes over the boundary of the worldsheet with coordinate $t$. The quantity on the right hand side can be computed by means of conformal field theory.
Putting the parts together one arrives at the following differential expression for the action
\be
\label{dSexpl}
	dS = \frac{1}{2}\int_{\P\Sigma}dt_1dt_2\langle d{\cal O}(t_1) \left\{ Q_{BRST}, {\cal O}(t_2)\right\}\rangle\ .
\ee
Note that as this formula possibly includes operators of conformal weight $<1$ it makes only sense when equipped with a cutoff.

Further simplification is achieved by assuming that the field content can be split in a ghost sector and a matter sector, i.e.\ that local operators are always given as product of a ghost part and a matter part. Also, this form must be kept up so that matter-ghost mixing operators are not allowed. This is reasonable for most sensible theories. Concretely this means that instead of considering operators ${\cal O}$ of ghost number one as basic building blocks of the state space, the operators ${\cal V}$ are used alternatively. They can be incorporated in the worldsheet formulation of correlators by adding a term
\be
	\Delta I = \int_{\P\Sigma}dt{\cal V}
\ee
to the path-integral. With a suitable definition of the $b$- and $c$-ghosts which is consistent with the usual usage in CFT  the relation between the states can be rephrased as
${\cal V}=b_{-1}{\cal O}$. As matter and ghosts are not allowed to mix, this relation is invertible, so that ${\cal O} = c{\cal V}$. This enables one to compute the BRST-action on ${\cal O}$ as follows:
\be
\label{noncov}
	\left\{Q_{BRST},{\cal O}(t)\right\} = -\sum_i\left(h_i-1\right) u_i{\cal V}_i(t) (c\P c)(t)\ ,
\ee
where $h_i$ denotes the conformal weight of ${\cal V}_i$ in the expansion ${\cal V}=u^i{\cal V}_i$.

The term $h_i-1$ in (\ref{noncov}) looks like the classical part of a $\beta$-function, and indeed this is its origin. It was shown in \cite{Shatashvili:1993kk, Shatashvili:1993ps} that the energy-momentum tensor contained in $Q_{BRST}$ receives corrections in higher order in the boundary couplings. This is due to contact terms arising from the boundary of the open string worldsheet. This means that $\frac{\P}{\P t^i}Q_{BRST}=0$ is not true anymore. 
Most remarkably it is possible to integrate the differential expression for the action (\ref{dSexpl}.) Skipping the most tedious calculations, one arrives at the final result
\be
\label{SBSFT00}
	S=\left(1-\beta^i\frac{\P}{\P u^i}\right){\cal Z}(u)\ .
\ee
Here ${\cal Z}$ is the generating function as obtained by determining the vacuum expectation value of the unit operator in the theory perturbed by arbitrary operators ${\cal V} = u^i{\cal V}_i$. As announced, in this expression appear $\beta$-functions, which are associated to the coupling $u$. 
This form suggestively points out the relation to renormalisation group flow, and indeed the classical configurations of this action are characterised by BRST-invariance including conformal invariance.

It is clear that (\ref{noncov}) does not transform correctly under coordinate repara\-me\-tri\-sa\-tions, because $u^i$ does not transform like a vector. This problem is resolved in the formulation of (\ref{SBSFT00}), as all objects appearing there like $\beta^i$ and the Zamolodchikov metric $G_{ij}$ do have the correct transformation behaviour \cite{Shatashvili:1993kk}. This renders the expression for the action invariant under local changes of coordinates in the space of coupling constants.

In order to demonstrate the usage of this formalism it is instructive to look at a simple example.
Heuristically, in the most basic (and most investigated) case of bosonic flat string theory, one can imagine the following proceeding: First choose some open string boundary condition, e.g.\ a spacefilling D25-brane. This theory is conformal, therefore the correlation functions can be evaluated in the associated BCFT. The boundary interaction which can be added to the standard action $(2\pi\alpha')^{-1}\int d^2z\P X\bar\P X$ in this case has the form
\eqn{
	\Delta I &= \int dt {\cal V}\\
	{\cal V} &= T(X) + A_\mu(X)\P_t X^\mu + B_{\mu\nu}(X)\P_tX^\mu\P_tX^\nu + C_\mu(X)\P_t^2 X^\mu + \dots
}
The coupling constants $u$ appearing in \ref{SBSFT00} are given by the modes $\phi_i=(A_\mu,$$B_{\mu\nu},$$C_\mu,$ $\dots)$ of the fields $A(X), B(X), C(X), \dots$. Thus the classical equations of motion obtained from $S$ by varying with respect to $\phi_i$ exactly correspond to configurations where $\beta_i(\phi) = 0$.

One should note that these expressions are rather formal. In practice, as mentioned before, a cutoff has to be introduced in order to deal with non-renormalisation interactions. This is not a problem, since as long as {\em all possible} interactions are taken into account, it is feasable to find a renormalisation scheme that leads to a theory without divergences and therefore to a well-defined fixed point. Perturbative calculations on the other hand suffer from all the usual problems and ambiguities one encounters in RG theory. Therefore it has been of great value exact solutions for the BSFT action have been constructed by Sen. Some of them will be reviewed in the next section.

\section{BSFT and the renormalisation group}

The appearance of (worldsheet) $\beta$-functions suggests that there is a close relation between processes in BSFT and renormalisation groups flows. It is immediately clear that a vacuum configuration, describe by certain values $u^*$ for the couplings represents a fixed point of worldsheet renormalisation group flow, since there the $\beta$-functions vanish, and
\be
	S(u^*) = {\cal Z}(u^*) \ .
\ee
Supposed there exist two vacuum solutions, which are located at the fixed points $u^*_1$ and $u^*_2$. At these points ${\cal Z}(u^*_{1,2})$ has the interpretation of a boundary entropy \cite{Affleck:1991tk, Kutasov:2000qp}. In the special case where these two vacua are related through flat directions in moduli space, the BSFT action evaluated on the interpolating line is a g-function. On the other hand, tachyon condensation provides an example for a complementary situation, where the two vacua are connected through a line of off-shell configurations. By construction, the BSFT action still is a meaningful quantity then. For this reason it can be viewed as an off-generalisation of the boundary entropy.

With these preparations it is possible to re-derive the g-theorem in a most simple way. Indeed, scale transformations of $S$ are given by the Callan-Symenzik equation \cite{Kutasov:2000qp}
\be
	\mu\P_\mu S = -\beta^i\P_i S = - \beta^i\beta^jG_{ij}\ .
\ee 
Here it has been used that $S$ does not contain any explicit scale dependence. Therefore $S$ decreases along RG trajectories, as long as $G$ is positive definite. The positivity of $G$ is guaranteed for unitary theories, for which also the expression (\ref{SBSFT00}) has been derived\footnote{However, when the restriction of matter-ghost separation is abandoned, then there could be propagation ghost degrees of freedom, which destroy unitarity.}.

In order to understand the role of renormalisation group flow in BSFT better it is instructive to consider the most simple example of tachyon condensation in this framework \cite{Shatashvili:1993kk,Andreev:2000yn}.
According to the general construction outlined above the open string tachyon is included in the action by a boundary term
\be
	\Delta I = \oint T(X)\ .
\ee
The tachyon field $T(X)$ can be Taylor-expanded in the string map $X$ as follows:
\be
	T(X) = a + b_\mu X^\mu + u_{\mu\nu}X^{\mu\nu} + \cdots\ ,
\ee
where $a,b,u,\dots$ are coupling constants. The inclusion of these boundary interaction terms breaks conformal invariance, since the tachyon operator is relevant. But this does by construction not affect the conformal properties of the theory in the bulk, therefore the renormalisation group flow is only expected to appear in the open string sector as a boundary flow. Rephrased differently, the bulk is kept on-shell, while the boundary theory is taken off-shell. Therefore it makes sense to consider the boundary renormalisation group flow alone without further reference to the bulk.

\section{BSFT and boundary states}
\label{ch-bs}

As has been shown in section \ref{sec-bs} it is possible to represent conformal boundary conditions in BCFT by so-called boundary states, which are certain linear combinations of Ishibashi-states. This is generally true for conformal field theories with boundaries. The purpose of this section is to review the construction of boundary states corresponding to flat D-branes in 
bosonic string theory. In a next step this concept will be generalised, and what emerges is a rather obvious connection at BSFT. This can be used to calculate D-brane tensions and discuss aspects of open string tachyon condensation in the closed string sector (employing open/closed duality at the conformal points).

\subsection{Boundary states for D-branes}

First the construction of D-brane states will be reviewed. This section follows \cite{DiVecchia:1999rh, DA}, where also the supersymmetric constructions are presented, but here only the bosonic case is of interest. The construction works as follows.

Let $X$ be the mapping of the string worldsheet into the target space. In the case of open strings the worldsheet has a boundary ${\P\Sigma}$ on which conditions must be specified in order to obtain a conformal theory. For flat space all possible D-branes are characterised by imposing either Dirichlet or Neumann condition in each direction. The Dirichlet condition becomes
\be
	X^i|_{\P\Sigma} = {\rm const} ,
\ee
while for Neumann condition
\be
	\P_t X^a|_{\P\Sigma} = 0 
\ee
must be imposed.
At the conformal point open and closed string partition functions are connected via modular transformations, which means that one-loop open string diagrams can be represented by closed string tree-level diagrams. After performing the transformation to the closed string channel one obtains analogous conditions on the string map, as the D-brane enters as condition on the initial and final closed string state.

More explicitely, in the case of closed strings it is convenient to use holomorphic coordinates on the worldsheet given by
\be
	z=e^{2i(\tau-\sigma)}\qquad \bar z=e^{2i(\tau+\sigma)}\ ,
\ee
where $\tau$ denotes the temporal direction and $\sigma$ the spatial. In a Euclidean setting, where $\tau\to-i\tau$, the coordinates are related by complex conjugation, so that $\bar z=z^*$.
The bosonic string map $X(z,\bar z)$ can be written as
\be
	X^\mu(z,\bar z) = \frac{1}{2} X^\mu(z) + \frac{1}{2} \tilde X^\mu(\bar z)\ ,
\ee
where
\be
	X^\mu(z) = x_0^\mu -i\sqrt{2\alpha'}\ln z \; \alpha_0^\mu + i\sqrt{2\alpha'}\sum_{n\ne 0}\frac{\alpha_n^\mu}{n}z^{-n}
\ee
and
\be
	\tilde X^\mu(z) 
		= x_0^\mu -i\sqrt{2\alpha'}\ln \bar z \; \tilde\alpha_0^\mu 
			+ i\sqrt{2\alpha'}\sum_{n\ne 0}\frac{\tilde\alpha_n^\mu}{n}\bar z^{-n}\ .
\ee
After performing the conformal transformation which allows to switch from open string variables to closed string variables, the conditions which must be imposed on the initial are
\be
\label{bc01}
	\P_\tau X^a|_{\tau=0}|B\rangle = 0
\ee
in Neumann directions $a$ along the D-brane, and
\be
\label{bc02}
	X^i|_{\tau=0}|B\rangle={\rm const}
\ee
in Dirichlet directions perpendicular to the D-brane. The same construction must be applies to the final state. As these conditions preserve conformality on the boundary, $|B\rangle$ is called a boundary state.

In a next step it is possible to look for more explicit realisations of a boundary states. In the case of string theory in a flat background such a construction is indeed possible. Just by re-writing (\ref{bc01}) and (\ref{bc02}) in terms of the closed string oscillators one obtains
\ba
	\left( \alpha_n^a+\tilde\alpha_{-n}^a\right)|B\rangle &=& 0 \\
	\left( \alpha_n^i-\tilde\alpha_{-n}^i\right)|B\rangle &=& 0 \qquad \forall n\ne 0
\ea
and for the zero modes
\ba
\label{BSC00}
	p^a|B\rangle&=&0\\
	\left(x_0^i-y^i\right)|B\rangle&=& 0 \ ,
\ea
where $y^i$ is constant.
It can be easily checked that the boundary state has the representation
\be
	|B\rangle = N_p\delta^{d-p-1}(x_0^i-y^i)e^{-\sum_{n=1}\frac{1}{n}\alpha_{-n}^\mu S_{\mu\nu}\tilde\alpha_{-n}^\nu} |p=0\rangle
\ee
for a D$p$-brane embedded in $d$-dimensional space. Here $|p=0\rangle$ denotes the closed string vacuum with zero momentum. The matrix 
\eqn{
S_{\mu\nu}={\rm diag}(-1, \one_\|, -\one_\perp)
}
has been introduced for a condensed notation, where the directions denoted by $\|$ are parallel to the brane, while $\perp$ denotes directions perpendicular to the brane. $N_p$ is a normalisation constant which has to be determined independently.

This state is compatible with conformal invariance, but in order to obtain BRST invariances the state needs to be supplemented with an appropriate ghost part. In a completely analogue way boundary conditions for the ghost fields $(b,c)$ can be found from which the boundary state can be determined. Details of the construction can be found in \cite{DiVecchia:1999rh, DiVecchia:1999fx}. The complete BRST invariant boundary state is then given by a product of the matter and ghost part.

\subsection{Path-integral representation and generalised boundary states}

In the situation presented above the construction of the boundary states turns out to be very simple, the reason for this being the flat closed string background and the absence of any open string background fields. In more general situations it can be useful to work with a path-integral representation of the boundary state. 

D-branes in flat space are rigid flat hypersurfaces, characterised by setting certain cartesian variables to zero. In general one would expect that the D-brane on which the string endpoints are fixed are as submanifolds in a curved embedding space. Therefore the string map at the boundary of the worldsheet should not be constant, but an arbitrary function.

This can be accomplished by changing the open string boundary conditions slightly,
\be
	X^i|_{\P\Sigma} = q^i(\tau)\ ,
\ee
where $q(\tau)$ is an arbitrary function with support on the boundary. 
Translation into closed string variables yields a modified condition on the boundary state
\ba
	p^a|B\rangle&=&0\\
	\left(x_0^i-y^i(t)\right)|B\rangle&=&0\ .
\ea
Contrary to (\ref{BSC00}), $y(t)=\sum y_ne^{int}$ is now a function regular on the boundary. In terms of oscillator modes this becomes
\be
	\left(\alpha_n-\tilde\alpha_{-n}-y_n\right)|y\rangle \ ,
\ee
which implicitely defines the state $|y\rangle$. Technically this is a coherent state is given by
\be
\label{genBS}
	|y\rangle = N_p \delta^p(x_0-y_0) e^{-\sum_{n>0} \frac{1}{n}\left( y_ny_{-n} - \alpha_{-n}\tilde\alpha_{-n} 
		- \alpha_{-n}y_n + \tilde\alpha_{-n}y_{-n}\right)}|0\rangle \ .
\ee
It is possible to impose a reality condition $y^*=y$, so that $y_{-n}=y_n^*\equiv\bar y_n$.

The normalisation $N_p$ can be fixed by demanding, that unity has a representation through coherent states, 
\be
	\one = N_p^*N_p \int \left[\prod_n\frac{dy_n}{\sqrt{2\pi}}\right]|y\rangle\langle y| \ .
\ee

Given the construction of these coherent states, the idea is to express arbitrary boundary states with conformal boundary condition $B$ by
\be
	|B\rangle = \int Dy{\cal W}_B(y) |y\rangle \ ,
\ee
where ${\cal W}_B(y)$ is an appropriate weighting functional of $y$.

\subsection{Neumann and Dirichlet states}

The Neumann and Dirichlet states, which have been constructed before, can be associated to certain choices of ${\cal W}_B$. An explicit and short calculation shows that
\be
	{\cal W}_N(y) = 1
\ee
gives the Neumann boundary state after performing the path-integral.
Also the Dirichlet state can be constructed this way, by setting
\be
	{\cal W}_D(y) = \delta(y_0-\hat y_0)\prod_{n\ne 0}\delta(y_n)
\ee
for a D-brane located at $\hat y_0$.

\subsection{Transitions between boundary states}

The coherent state formalism allows to make explicit contact with the worldsheet formulation of CFT. This is achieved by noting, that the pure $y$-dependent part in (\ref{genBS}) is given by the integral over the modes of the sigma model field on the boundary of a disc of the classical action. Therefore
\be
	\langle0|y\rangle = e^{-S_\P(y)}\ ,
\ee
where $S_\P(y)=\frac{1}{2}\oint y\bar\P\bar y$
is the boundary term obtained by evaluating the Polyakov action on is classical solutions $y(\tau,\sigma)$.

In particular it is possible to include arbitrary boundary interactions in this formalism. This provides the definition of a boundary state in the presence of arbitrary open string fields, whose impact is collected in an boundary interaction term $I[T,A,\dots]$,
\be
	|B[T,A,\dots]\rangle=\int Dy \;e^{- I[T,A,\dots](y)}|y\rangle\ .
\ee
In particular this state provides the correct coupling of closed string state to D-branes in the presence of open string background fields. Therefore the conformal configurations are given by open string couplings $u^*=(T^*,A^*, \dots)$, which are determined as RG endpoints or as solutions of the BSFT action, as has been discussed above.

It is possible to verify in the level boundary states, that solutions of BSFT correspond to different conformal boundary conditions. To see this in the simplest case, consider a spacefilling brane together with a tachyon depending only on one coordinate. I.e.\ take the interaction term
\be
	\Delta I = \frac{1}{2\pi\alpha'}\oint T(y) 
\ee
with a quadratic tachyon profile in the 25-direction
\be
	T(y) = u |y_{25}|^2\ .
\ee
The boundary state obtained in dependence of the coupling $u$ is
\be
	|B[u]\rangle = N_u\prod_{n>0}e^{-\frac{1}{n}\alpha_{-n}S_n\tilde \alpha_{-n}}|0\rangle
\ee
with
\ba
\label{TCBSS}
	\left(S_n\right)_{0,0} &=& -1 \\
	\left(S_n\right)_{a,b} &=& \delta_{ab} \qquad {\rm for\;\;} 1\le a,b\le 24 \\ 
	\left(S_n\right)_{25,25} &=& \frac{1-\frac{2\pi\alpha' u}{n}}{1+\frac{2\pi\alpha' u}{n}}
\ea
and a normalisation
\be
\label{TCBSN}
	N_{u} = \frac{1}{2} |u|^{-\frac{1}{2}}\prod_{n>0}\left|1+\frac{2\pi\alpha' u}{n}\right|^{-1}
		= \frac{1}{2}\left|2\pi\alpha'\sqrt{u}\Gamma(2\pi\alpha' u)\right|\ .
\ee
Here the tension of the 25-brane has been normalised to 1. In the last step of (\ref{TCBSN}) the infinite product has been conducted using $\zeta$-function regularisation. By consecutively performing the condensation in different directions it is possible to compute the correct relative D-brane tensions $N_{p}/N_{p-1}$. 

More interestingly one sees that the coupling $u$ is able to interpolate between a D25- and a D24-brane. For this consider first the limit $u=0$. In this case, $(S_n)_{25,25}=1$ so that the associated boundary state is a spacefilling brane. But as soon as $u$ is switched on, RG flow will increase its value and drive it towards its fixed point at infinity. This should also be a conformal point, and indeed one finds that $(S_n)_{25,25}=-1$, so that the resulting state is a D24-brane.
This shows very intuitively, how tachyon condensation processes can be understood from the boundary states' point of view.

One might wonder at this point, if it is also possible to describe a condensation into a pure closed string vacuum in this formulation, where no (perturbative) open string excitations are present. Indeed this is possible, but some more preparations are necessary. This issue will be taken up again in \ref{sec-csv}.

%
%
%

\chapter{Open string tachyon condensation}

The presence of tachyons, which generically appear in many models of string theory, indicate that the vacuum, in which these models are formulated, is not a stable vacuum. Similar statements are known from quantum field theory, from where it is known that the minimum of the potential determines the expectation values of the background fields and therefore determines the correct vacuum around which the theory should be expanded. This however makes it necessary to work with a second quantised theory, so that the tachyon itself bears a physical interpretation. In string theory the situation is a little bit different, as one mostly works in a first quantised formalism. The analogy to quantum field theory suggests but to rather attack such issues with a version of string field theory.  

String field theory indeed has made significant progress when, in the context of BSFT, condensation processes have been discovered, which are triggered by open string tachyons. This phenomenon has been affirmed by calculations in cubic SFT, while due to the nature of this theory, the calculations are laborious and can only be conducted numerically. Tachyon condensation processes add substantial insight into open string field theory, as they open the path to understanding dynamic formation and decay of D-branes.

Informally speaking, tachyon condensation appears, viewed from the CFT point of view, in the boundary sector. There the RG flow is under slightly better control since, at least at tree level, the central charge of the CFT cannot change. Closed string tachyon condensation on the other hand, almost inevitably causes an RG flow in the closed string sector, so that it is usually impossible to keep the condensation under control. Also, there are no as general predictions for the end-point of such condensations as for the open string sector, except in very controlled settings \cite{Witten:1992cr}. %

From this point of view it is remarkable that the open string condensation can be handled rather well and allows access of several interesting non-perturbative effects. It becomes possible to derive effective actions for tachyonic configurations interpolating between different D-brane configurations, even time-dependent ones. To some extent it is also possible to investigate the closed strings coupling to condensing D-branes, which means some further step in understanding open-closed string correspondences. In supersymmetric settings, where BPS conditions reflect the stability of D-branes, transition rules may be formulated which culminate in an K-theoretic formulation and generalisations thereof.

Integrating out heavy non-tachyonic fields in the open string path-integral yields an effective theory for the massless and tachyonic string modes. Its static part provides a potential whose minima determine the possible vacua of the theory as renormalisation group fixed points. This will be explained in the next section, where also some properties of the potential 
are discussed and summarised as Sen's conjectures. A formalism based on boundary states is introduced. 

Also condensation processes with the closed string vacuum as end-point will be discussed. This is done from several points of view, and it is shown how open string degrees of freedom are completely removed from the spectrum after condensation.

The final comments in this chapter deal with the open string completeness conjecture, which touches some aspects of the coupling to closed strings.

The condensation processes discusses here take place in a flat background geometry. Although it is believed that similar phenomena appear in curved backgrounds, almost no concrete examples nor proofs for this concept are available.
 In chapter \ref{ch-ren} however, an example is presented which gives the desired hints that condensation on curved branes in curved target spaces do take place.

\section{Tachyon effective action and lower dimensional branes}

In order to study the properties of target space fields it is usually advised to investigate their low energy effective action. This is obtained by integrating out heavy fields, and to lowest order at tree-level these can be eliminated by inserting their classical equations of motion. The restriction to light degrees of freedom as the only dynamical ones makes sure the action gives a consistent theory. In the case of field theories with tachyons the situation is more involved. As they contribute a negative mass squared, the mass is no longer a good expansion parameter, as combinations of heavy and tachyonic particles might show up in the energy region of light fields. Therefore, in a strict manner, one cannot approach the problem by simply integrating out heavy fields. But one can try to derive {\it some} kind of effective action and do a formal analysis. This is the usual approach for tachyon effective actions, and this is also the most intricate part of the analysis. The reason why this approach works lies in the properties of the RG flow induced by tachyons. In fact it can be shown that the flow of boundary tachyons as the most relevant fields and the massless fields decouples from the rest \cite{Affleck:1991tk, Tseytlin:1987ww}. In this sense it is feasible to work with tachyon effective actions.

Although the most important models of string theory do not contain tachyons in their perturbative spectrum (which is the reason of their importance) tachyon generically appear in the spectrum when D-branes enter the game. In fact, strings on non-BPS branes or stretching between D-branes have tachyons in their spectrum, and even in a supersymmetric setting those may survive the GSO projection. The condensation process triggered by such tachyons may result in a decay of D-branes (like in the case of annihilating brane-anti-brane systems) or formation of lower dimensional configurations. It is the aim of effective tachyon action to capture and describe this behaviour. Much of this information is contained in the tachyon potential $V(T)$, which is obtained as the static part of the tachyon effective action\footnote{In this context time dependent solutions are also of immense interest, but for the purposes of the later chapters these aspects are of inferior relevance.}

Employing techniques which originated in the study of the $\sigma$-model in the path-integral approach \cite{Tseytlin:1987ww, Andreev:1988cb, Tseytlin:1988tv, Tseytlin:1989md, Tseytlin:2000mt} and BSFT it has been possible to observe some basic properties of tachyon effective actions, which are conveniently summarised in Sen's three conjectures. Although these have not been proven in the strict sense there is little doubt on their correctness. In particular it can be regarded as significant fortification that, up to now, almost every aspect of these conjectures has been verified also in cubic string field theory \cite{Schnabl:2005gv, Ellwood:2006ba}.

\subsection{Sen's conjectures}

Before presenting illustrative examples we will formulate the conjectures. The first one addresses the issue of (local) minima of the static potential. Their existence is essential for providing the mere possibility to find a stable vacuum. But in addition to that the first conjecture makes a statement on the interpretation of the energy differences in the tachyon potential (see \cite{Sen:2004nf}):

\begin{enumerate}
\item The tachyon potential $V(T)$ does have a pair of global minima at $T=\pm T_0$ for non-BPS D-branes, and a one-parameter $(\a)$ family of global minima at $T=T_0e^{i\a}$ for the brane-anti-brane system. At this minimum the tension of the original D-brane configuration is exactly cancelled by the negative contribution of the potential $V(T)$. Thus
\eqn{V(T_0)+E_p = 0\ ,}
where $E_p$ is the energy associated with the  D-brane. In the supersymmetric case
\eqn{E_p=
\begin{cases}\tilde T_p\qquad\;\,\text{for non-BPS D$p$-branes}\\
						2 T_p\qquad\text{for D}p\text{-}\bar{\rm D}p\text{ brane pair}
\end{cases}\ .
}
Thus, as general rule, the total energy vanishes at the minimum of the potential.
\end{enumerate}

The second conjecture makes a statement about the endpoint of the condensation process, suggesting a closed string vacuum as the result of a full decay:

\begin{enumerate}
\setcounter{enumi}{1}
\item Since the total energy density vanishes at $T = T_0$, and furthermore, neither the non-BPS D-brane nor the brane-antibrane system carries any RR charge, it is natural 
to conjecture that the configuration $T = T_0$ describes the vacuum without any 
D-brane.
\end{enumerate}

A pure closed string vacuum as it is conjectured is expected to contain no perturbative open string excitations at all. This leaves to obvious questions, namely: what exactly happens to the open string degrees of freedom, and: is open string field theory able to describe the closed string vacuum? Both question will be addressed later.

The third conjecture reads as follows:
\begin{enumerate}
\setcounter{enumi}{2}
\item Although there are no perturbative physical states around the minimum of the potential, the equations of motion derived from the tachyon effective action $S_{\rm eff}(T, \dots)$ 
does have non-trivial time independent classical solutions. It is conjectured that 
these solutions represent lower dimensional D-branes. 
\end{enumerate}

These conjectures are supported by many examples \cite{Sen:1998ex,Horava:1998jy,Sen:1998ii,Sen:1998tt,Sen:1999mh,Sen:1999xm, Maldacena:1997re,Schnabl:2005gv,Ellwood:2006ba}.

\subsection{Example: the lump solution}

Here we will provide an example, which illustrates a method to obtain an exact solution for a tachyon field, which describes a lower dimensional D-brane. This solution is time-independent and uses methods of CFT in order to explicitly construct it. Each D-brane configuration corresponds to a certain CFT furnished with appropriate boundary conditions. Adding tachyon interactions to the action modifies the BCFT, which will in general result in RG flow. However, in some cases, as in the presented example in this section, it is possible to control this RG-flow and determine the value of the coupling constant at the end point of the flow.

To be specific, a boundary term
\eqn{\int_{\P\Sigma}d\t\;T(\t)}
is added to the action, where $T=T\left(X(\t)\right)$ has an expansion in the string map $X(\t)$ at the boundary, but contains no worldsheet derivatives. From the point of view of the BCFT, $T(\t)$ is a vertex operator, by which in general conformal invariance is broken. But when $T(\t)$ does not break conformal invariance, the modified action represents a classical solution of string field theory and thereby a vacuum solution. Thus there is a correspondence between such classical solutions and conformally invariant vertex operators.

The conformally invariant field theory describing a spacefilling brane in bosonic string theory is well known. Starting with such a solution in a flat uncompactified target space means choosing a boundary interaction $T\equiv 0$. This describes string theory on a space-filling D25-brane. It is possible to construct a D24-brane which can be regarded as a solution of the theory on the D25-brane with tachyon interaction, the so-called 'lump solution'.  

Sen's approach to this problem was, to find the exact BCFT associated to this lump solution by conducting a series of marginal deformations. For this the theory is in one direction compactified on a circle with radius $R$. Radius deformations are exactly marginal, therefore it is possible to transport the theory to the point $R=1$. This is the self-dual radius, which is also distinguished by the presence of a extended symmetry of the bulk CFT.
The currents of this $SU(2)_L\times SU(2)_R$ symmetry are given by
\eqn{
	J_L^3=i\bP X_L\qquad J_R^3=i\P X_R\qquad J_{L,R}^1=\cos 2X_{L,R}\qquad J_{L,R}^2=\sin 2X_{L,R}\ ,
}
where $X$ has been split up into a left-moving and a right-moving part.

The Neumann boundary conditions for the upper half plane as worldsheet imply that $X_L=X_R$ at the boundary, therefore the boundary operator $\cos X = \cos (X_L+X_R) = \cos 2X_L$ is exactly marginal. Therefore it is possible to include a interaction term
\eqn{
	-\a\int d\t\cos X
}    
with $\a$ an arbitrary constant. This can be clearly interpreted as switching on a background tachyon field.

The value of $\a$ cannot be taken arbitrarily once the radius $R$ is taken back to infinity. As was shown in \cite{Sen:1999mh}, the one-point function of the operator already ceases to vanish for generic $\a$,
\eqn{\label{oneptsentc}
	\vev{\cos X(0)/R}_{R;\a} \propto (R-1)\sin 2\pi\a
}
in the vicinity of $R=1$. However, for those values of $\a$, where the sine in (\ref{oneptsentc}) vanishes, the one-point function stays zero for all values of $R$ \cite{Sen:1999mh}. While $\a=0$ describes the original theory with a space-filling brane, the soltuion with $\a=\frac{1}{2}$ has a different interpretation at $R=\infty$. A direct way to argue is to remind oneself of the effect the inclusion of the operator $-\a\int \cos X = -\a\int J_L^1$ has. It merely is a Wilson line which causes a rotation in the enhanced $SU(2)_L$ symmetry, implemented by a phase factor $\exp 2\pi i\a$. In order to ensure that the operators in the theory keep their form during the rotation, a field-redefinition in form of a $U(1)$-rotation of $X_L$ is necessary, so that $X_L\to-X_L$. This does not affect the bulk, but it has effect on the boundary conditions. They change from $X_L=X_R$ to $X_L=-X_R$, therefore from Neumann to Dirichlet. The resulting D-brane is thus a D24-brane.

This construction can be repeated independently for all other directions, too, so that the tachyon obviously can be used to construct D-branes with arbitrary codimension.

This solution has been obtained in a somewhat indirect way by a series of exactly marginal deformations. It should in principle be possible to derive the same result by directly perturbing the theory at $R=\infty$. This will however initiate a renormalisation group flow and is therefore rather hard to deal with exactly. In the next paragraph a perturbative approach to the problem is presented.

\section{Closed string vacuum}
\label{sec-csv}

One of the main insights gained from the study of tachyon condensation in Sen's approach is, that there is indeed a connection between open string vacua and closed string vacua on  the level of string field theory. Further evidence for this is provided by recent calculations \cite{Zwiebach:1997fe,Ellwood:2006ba}, in the framework of cubic string field theory. 

In this section we want to investigate, how this connection can in principle be understood in BSFT. The understanding, which has been established there, is that the perturbation theories of open 
and closed strings are expansions in some background independent universal theory around 
the different vacua. 
The approach followed here goes back to \cite{Gerasimov:2000ga}, where it has been suggested that these expansions are connectied by a Higgs-like mechanism. 

Subsequently we will describe a boundary state approach to the same problem.

\subsection{Fate of open string degrees of freedom}

For this example we will only focus on the massless and tachyonic modes, in a bosonic string theory.
The closed string background is given by the metric $G$ and the $B$-field, in the open string sector we wish to include the photon field $A$ besides the tachyon $T$. In general, the full spectrum must be included of course, but as has been explained in earlier sections, the truncation of the theory to light modes is a valid approximation.

The open string sector is as usual included by a Wilson line taking the form $e^{-\int d\t (T+AdA)}$ of a boundary integral. There are two obvious symmetries of this action, namely a gauge transformation $A\to A+d\Lambda$ and a shift $A\to A+$const. It is important to understand that $d\Lambda$ enters the action in the same way as a closed $B$-field contributions. This is obvious when considering the effective spacetime action obtained from the BSFT construction in form of
\eqn{\label{TC-STA}
\S(G,B&,A,T)= \S_{\rm closed}(G,B)\\&+\int d^{26}X\sqrt{G}\left(e^{-T}(1+T)+e^{-T}|T'|^2+\frac{1}{4}e^{-T}|B-dA|^2+\dots\right)\ .
} 
We have no knowledge over the closed string part, except for its mere existence. This follows from applying the ideas of unitarity of a consistent second quantised theory to BSFT in its expansion around the supposed closed string vacuum, as will be explained below. Therefore this Ansatz  is justified a posteriori.

The spacetime action (\ref{TC-STA}) posses the gauge invariance
\eqn{
	B\to B+da\qquad A\to A+a\qquad a\in\Omega^1\ .
}

From previous considerations we know already two fixed points of the theory. The one for vanishing tachyon field is unstable, since the tachyon itself is not massless. The other, stable, point lies at $T=\infty$. It becomes obvious at once, that the open string part of the action simply vanishes at that point in moduli space. This is not a statement about any modes to become massive/undynamical, so that they can be integrated out; rather one sees that the kinetic term for the open string gauge field, which is the only dynamic field at the endpoint, is multiplied with $e^{-T}\to 0$. Hence the open string modes are simply removed from the spectrum.

The action (\ref{TC-STA}) is to be compared to the standard Higgs Lagrangian
\eqn{\label{TC-Higgs}
S(H, \phi,{\cal A}) = \int d^n X\left(
	\frac{1}{g^2}F({\cal A})^2 + H^2|d\phi-{\cal A}|^2 + |H'|^2 + \lambda(H^2-H_0^2)^2
\right)\ ,
} 
where the Higgs field $\Phi$ has been split in a radial and an angular part
\eqn{
	\Phi=He^{i\phi}\ .
}
The gauge transformations associated to this action
\eqn{
	{\cal A}\to{\cal A}+d\chi
	\qquad
	\phi\to\phi+\chi
}
suggest a formal identification of (\ref{TC-Higgs}) with (\ref{TC-STA})) by means of
\eqn{
	e^{-\frac{T}{2}} \to H
	\qquad
	A\to\phi
	\qquad
	B\to{\cal A}\ .
}

The analogy with between the two theories is as follows. In the Higgs model there is a phase where the symmetry is broken. This is indicated by the fact the $H$ takes a non-vanishing expectation value and the angular field $\phi$ is fixed. This corresponds on the string theory side to a situation where the tachyon $T$ vanishes and the gauge field $A$ takes a specific value. Unlike in the Higgs model, this is not a stable configuration for the string theory.

On the other hand there is a phase in the Higgs-model that corresponds to the configuration where the tachyon becomes infinite, or $\Phi=0$. This vacuum is characterised by a vanishing radial field variable $H$, while the value of the angular field variable $\phi$ is not determined. The fact that $\phi$ is ill-defined at this point has no physical reasons, but is an artefact of the choice of the coordinate system. In fact, such an apparent singularity disappears with the choice of a well-behaved coordinate system on the field space.

From this we can gain some physical insight. The singular field $\phi$ is mapped to the open string field $A$ by the above correspondence. At the endpoint of tachyon condensation the open strings are removed from the spectrum, which means that $A$ is certainly no good variable more for the formulation of the model.

Although the Lagrangian vanishes at $T=\infty$, the model is not ill-defined. The indication for a variable transformation naturally forces one to include more than just the field with which we started. In particular, in order to be able to construct a suitable transformation of the fields all possible interaction terms must be included. This includes also terms which come with arbitrarily high derivatives of the string field $X$. That means, expanded in the old variables, the tachyon can be represented as an infinite series $T=T(X+AX'+CX''+\dots)$. 

All this supports the idea that there is a closed string vacuum at the endpoint of the open string tachyon condensation, which carries no open string degrees of freedom.

\subsection{Boundary states}

If the end point of open string tachyon condensation does truly describe closed string vacuum, one should expect that this process also gives some way of interpolating between closed string correlators on the disk and on the sphere. This question has been asked already in \cite{Ambjorn:2002im}.

For a set of closed string operator insertions one would like to compare
\eqn{
	\<{\mathcal O}_1\cdots{\mathcal O}_n\>_{\rm sphere}
}
with
\eqn{
	\<{\mathcal O}_1\cdots{\mathcal O}_n\>_{\rm disk}
}
and ask if there is way to interpolate between the two expressions. The existence of a smooth interpolation will show that the coupling to the closed string sector is consistent with the idea of an emerging closed string vacuum, and therefore will offer more support of the above proposed interpretation of the end point of candensation.

Most easily this problem is treated in the boundary state formalism introduced above. We have seen already that a boundary state corresponding to a spacefilling brane will condense to a lower dimensional brane under tachyon condenstion with quadratic tachyon profile. This is reflected in the fact that $(S)_{25,25}$ in (\ref{TCBSS}) interpolated between $\pm1$ for $u=0,\infty$. But we know (or otherwise reassure ourselves with a brief calculation) that expectation values of on-shell closed string states do not distinguish between Dirichlet- and Neumann-boundary conditions. Therefore the sole inclusion of a quadratic tachyon interaction cannot be the full answer. This is, on the other hand, also not expected, since in general an infinite number of additional interaction terms must be taken into account. 

The correct answer can be found by an inspection of the boundary state. As shown above, it is given by
\eqn{
	|B[u]\rangle = N_u\prod_{n>0}e^{-\frac{1}{n}\alpha_{-n}S_n\tilde \alpha_{-n}}|0\rangle\ .
}
Let us look in the 25-direction only, which is valid since the state factorises into contributions from each dimension. The matrix element $(S)_{25,25}$ controls the interaction between the holomorphic and anti-holomorphic sector. This in turn encodes the boundary conditions. Now note that
\eqn{
	\left(S_n\right)_{25,25} &= \frac{1-\frac{2\pi\alpha' u}{n}}{1+\frac{2\pi\alpha' u}{n}}\ .
}
An absence of an interaction between the right and left moving sector can only be achieved when $(S)_{25,25}\equiv 0$. The quadratic tachyon alone is not capable of doing this, thus a more general ansatz is in place. This is achieved by formally assuming an $n$-dependence of $u$ in the above formular, hence let us make the ansatz
\eqn{
	\left(S_n\right)_{25,25} &= \frac{1-r}{1+r}\ .
}
One immediately sees that the matrix element vanishes for $r=1$, while $r=0$ gives the Neumann condition and $r=\infty$ the Dirichlet condition.

The important question is now, what boundary interaction must be added to the string action in order to get this function for $(S)_{25,25}$. It is in fact not too hard to find the associated expression. It is given by
\eqn{
\label{nl-bs-i}
	\Delta I_r &= \frac{r}{2} \sum_{n=1}^\infty y_ny_{-n}
}
in the notation of section (\ref{ch-bs}). Similar boundary interactions were also considered in \cite{Li:1993za}\footnote{in order to preserve locality, only finite sums over terms of this for were considered.}.

This is obviously a non-local interaction term. We will encounter it again in the next chapter, where it appears naturally as a closed string variation in BSFT. In fact it originates in a radius change of the compactification torus in target space. Anticipating the arguments to be presented, one must thus accept (\ref{nl-bs-i}) as a legitimate boundary interaction term\footnote{One must understand, that these arguments involve a string field theoretic treatment which goes beyond conformal field theories.}. Therefore this also supports the existence of a closed string vacuum in open string field theory.

%
%
%

\chapter{Factorisation of BSFT action}
\label{ch-fact}

BSFT is defined by the path integral over $\sigma$-model fields for a 
fixed closed string background $\X$ with the dynamical open string 
degrees of freedom $t$ corresponding to
boundary deformations of the CFT on the disk.  
It is, however, important to note that these deformations are not 
required to be local on the boundary of the worldsheet
\cite{Witten:1992qy, Witten:1992cr, Shatashvili:1993kk, Shatashvili:1993ps}
Once non-local boundary perturbations are included the 
distinction to open and closed degrees of freedom on a 
worldsheet with boundary becomes ambiguous.
In fact in the early days of background independent open string
theory it was realised that the notion of locality on the worldsheet was a
major question to be addressed since deformations on the boundary were
described by a limiting procedure of taking the closed string
operator from the bulk and moving it to the boundary. The simplest
way to identify $\X$ is by means of the closed string $\sigma$-model
Lagrangian. $\X$ then defines a conformal $\sigma$-model background in
the absence of boundaries.
In the examples studied in this chapter we will make some
natural choices in this regard and then demonstrate a relation between
them. The presentation follows \cite{Baumgartl:2004iy} closely.

The key ingredient in our approach is the factorization property for the
BSFT space-time action $ \S(\X| t)$,
\be
    \S(\X| t) = \Z_0(\X) \S_0(\X|t)\ .
\ee
Here
$\Z_0(\X)$ is the D-instanton partition function and $\S_0(\X|t)$ is
described purely in terms of the quantum mechanical degrees of
freedom $\Phi_b$ on the boundary. 
Given the relation between 
the worldsheet partition function and the BSFT
action (see equation (2.1) below)
this property is a consequence of the 
following 
conjecture for 2d conformal field theories on a manifold with
boundary:
\be
\label{ant}
	\int_{\Phi|_{\partial \Sigma}=
		\Phi_b(\theta)} D[\Phi]\, e^{{i \over {\alpha'}}I_{\Sigma}(\Phi)}=
			\Z_0(\X)\,e^{{i \over {\alpha'}}I^\bdry(\Phi_b)}\ .
\ee
Here $I_{\Sigma}(\Phi)$ is a
Lagrangian for the worldsheet conformal field theory on a 2d surface with
boundary, and $\Z_0(\X)$ is the D-instanton partition function 
which is given by (\ref{ant}) for $\Phi_b=0$.  We verify this
property the case of when $\Sigma$ is a disk, in the
situation where ghosts and matter decouple and for $\X$ such that
the  closed string worldsheet is conformal and described in terms
of a WZW (or related) model. These technical assumptions are
necessary since not much is known about BSFT when ghosts and matter
do not decouple.

The logic underlying our approach is the following: To each 2d CFT
with  boundary corresponds a boundary action $I^\bdry(\Phi_b)$. Due
to the factorisation property, $I^\bdry(\Phi_b)$ is independent of
$\alpha'$ but certainly depends on the CFT chosen on the left hand
side of (\ref{ant}).
The ambiguity in this process is under control. 
On the other
hand, in the reconstruction of the bulk CFT for a given boundary action
there may be further ambiguities.  
We then claim that there is a distinction
between the class of bulk theories reconstructed from boundary
actions $I^\bdry$, differing by (non-local) functionals of the
boundary field $\Phi_b$. 

Note that 
the boundary action plays a central role
in BSFT since one integrates over
all maps from the worldsheet to the target space
without specifying the boundary conditions.
One starts from a boundary action and considers the
class of its boundary deformations; this class contains all other
boundary actions with the same number of boundary fields $\Phi_b$ (or
less).
The boundary actions corresponding to boundary conformal field theories on
the worldsheet are, by definition of the string field theory action, 
solutions of the classical equations of motion for $\S(\X|t)$. 
These are in turn critical points in $t$
for fixed $\X$ and denoted by $t_*$. The space-time action
$\S(\X|t_*+t_q)$ expanded around $t_*$ to $n$-th order in
$t_q$ is supposed to reproduce the $n$-point open string
amplitudes for the background defined by
$t_*$. This is known to be true on classical
level in the space-time field theory
corresponding to disk amplitudes on the worldsheet.

Concretely we start with a closed string background $\X$ and find $t_*^1$.
Then we look for a second critical
point of $\S_0$: $t^2_*$; since $\X$ is a ``hidden variable''
in the open string field theory action $\S_0$ we need to reconstruct
it for the new critical point $t_*^2$.
This in general is a difficult problem and in principle might be ambiguous. Even so
we can argue that in the set of critical points of
$\S_0(\X| t)$ there are critical points $t=t_*^1$ and $t=t_*^2$  such
that the expansion around $t_*^2$ is identical to the expansion around
$t_*^1$ but for different closed string background $\X'$, i.e.
\be
\label{EQNConjecture}
    \S_0(\X|t_*^1+t_q) = \S_0(\X'|t_*^2+t_q)\ .
\ee
Thus, a deformation from $t_*^1$ to $t_*^2$ can be interpreted as deforming the closed string
background from $\X$ to $\X'$.

A simple realisation of the conjectured property (\ref{EQNConjecture}) leads to
the Seiberg-Witten map \cite{Seiberg:1999vs}: It is well known that a constant magnetic
$B$-field can be seen equivalently as a closed string background $\X$ or a
perturbation on the boundary
of the open string worldsheet, i.e.
$\S_0(\X=(G,B)|t_q) = \S_0(\X'=(G,0)|t_*+t_q)$.
The result of \cite{Seiberg:1999vs} can then be formulated as the 
statement that the expansion around $t_*$ leads to 
non-commutative field theory in Minkowski space.
The generalisation to a non-constant,
closed $B$-field leads to Kontsevich's deformation 
quantisation \cite{Kontsevich:1997vb}. 
At present we allow for arbitrary $B$ compatible with bulk conformal invariance.

Note that the factorisation of the worldsheet partition function
into bulk and boundary contributions is crucial for the closed
string degrees of freedom to be contained in open string 
field theory.
Indeed if bulk $\alpha'$-corrections entered in the definition of
$I^\bdry(\Phi_b)$ one would get different $\alpha'$-expansions for the open
and closed string $\beta$-function. 
The factorisation property, which guarantees that closed string fluctuations 
do not feed back into the definition of the open string field theory,
is instrumental for the open-closed string correspondence to work. 
This appears to be a very subtle distinction
between bulk
conformal field theories in 2d and general 2d QFT where this factorisation does not hold in
general.

In a first the the promised properties will be shown in the context of BSFT on a torus. In this setting the radius deformations are very well under control, and the basic ideas can be applied easily.
As an example for a curved closed string background we
then prove the factorisation property for boundary WZW models with
arbitrary boundary conditions to all orders in perturbation theory
in section 3. This requires a definition of WZW models with boundary
conditions {\em which are not of the class} $J=R\bar J$
\cite{Klimcik:1996hp, Alekseev:1998mc, Stanciu:1999id, Alekseev:1999bs, FigueroaOFarrill:2000kz}
rather only implying
$T-\bar T=\beta^i(t)V_i(t)=0$, where $V_i(t)$ is a boundary
perturbation and $\beta^i(t)$ its $\beta$-function. 
An application will be presented in the next chapter.

\section{BSFT on a torus}
\label{SecBSFTonTorus}

In the case when ghost and matter fields decouple the definition of
the space-time action in flat space is written in terms of the disk
partition function $\Z(t)$ and boundary $\beta$-function as
\cite{Shatashvili:1993ps}\footnote{Note that this expression is written without use of a metric on the
space of boundary interactions, only the vector field $\beta^i$ is
required.}
\be
\label{EBSFT}
	{\cal S}(\X|t) =
	\left(1-\beta^i\frac{\P}{\P t^i}\right)\Z (\X|t)\ .
\ee
Here $t^i$ are
the couplings representing the open string degrees of freedom and
$\beta^i$ denotes the $\beta$-function associated to the coupling
$t^i$.

For our purpose we suggest a different normalisation of the
space-time action by replacing $\Z(\X|t)$ by $\Z^\bdry(\X|t)\equiv
\Z(\X|t)/\Z_0(\X)$, where $\Z_0(\X)$ is the ``D-instanton" partition
function, which is independent of the open string background
$\{t\}$. Of course this normalisation assumes the factorisation of
the CFT on the disk, which we will prove shortly.
Our normalisation
does not alter the dynamics of the open string fields $t^i$,
therefore we can work with $\S_0(\X|t)$ instead of $\S(\X|t)$,
\be
\label{EBSFTnew}
{\cal S}_0(\X|t) =
\left(1-\beta^i\frac{\P}{\P t^i}\right)\Z^\bdry (\X|t)\ .
\ee

To start with we consider the free action for maps $X$ from the disk
into a circle of radius $R$
\be
\label{Efreeaction}
    S(X) = \frac{R^2}{4\pi i\a'}\int_D \P X\bP X\ .
\ee
where $\P \equiv dz\P_z$. The radius $R$ plays the role of a
closed string modulus. According to BSFT we are instructed to
integrate over maps with free boundary conditions, which leads to the
notion of the boundary field $f$ defined through $X(z,\bar z)|_{\P
D} = f(\t)$; boundary deformations are functionals of $f$, in
general non-local. This field $f$ can be unambiguously extended from
the boundary to the interior of the disk via harmonicity condition
(harmonic functions are solutions of the worldsheet equations of
motion). Every field $X(z, \bar z)$ may thus be split into a
harmonic boundary field and a bulk field which obeys Dirichlet
conditions,
\be
\label{expens}
    X(z, \bar z) = X_0(z, \bar z) + X_b(f)\ ,
\ee
such that $X_0|_{\P D} = 0$ and $X_b(z, \bar z)|_{\P D} = f(\t)$
with $\Delta X_b = 0$, so $X_b(f)$ is a harmonic function with value
$f(\theta)$ on the boundary.

Note that the boundary field can always be expanded as $f=\sum_{n}
f_n e^{in\t}$, which suggests a separation into chiral and
anti-chiral modes corresponding to positive and negative
frequencies. Thus, $f = f^+ + f^- + f_0$ can then be extended to $X_b(f) =
f^+(z) + f^-(\bz) + f_0$. Moreover there is a reality condition ${f^+}^* =
f^-$. The zero mode $f_0$ plays the role of the space-time
integration variable in the space-time action.

Plugging this ansatz into the free action (\ref{Efreeaction}) the mixed
terms containing $X_0$ and $X_b$ vanish after partial integration.
The action splits into 
\be
\label{Epolyakovbb}
    S(X) = \frac{R^2}{4\pi i\a'}\int \P X_0\bP X_0 + \frac{R^2}{4\pi i\a'}\int
\P f^+ \bP f^-\ .
\ee 
Given the translation invariance of the measure in this example
the factorisation property is obviously satisfied. The partition function then reads
\be
\label{EQNpartfnI}
    \Z(R) = \Z_0(R)\int D[f]\, e^{-\frac{R^2}{4\pi i\a'}\int \P f^+ \bP f^-}\ ,
\ee
where
\be
\label{EQNpartfnII}
    \Z_0(R) = \int D[X_0]\, e^{-\frac{R^2}{4\pi i\a'}\int \P X_0\bP X_0}\ ,
\ee
supplemented by the $b,c$ ghost system is the ``D-instanton''
partition function\footnote{ Here we take the conventional boundary
conditions for $b$ and $c$, because decoupling of matter and ghost
sector is assumed. }. Since $f^\pm$ is harmonic its contribution
takes the form of a non-local boundary interaction
\be
\label{hil}
    I^\bdry(f) = \frac{R^2}{4\pi}\oint f H(f)
        = \frac{R^2}{4\pi} \oint \oint d\t d\t' f(\t) H(\t,\t') f(\t')=
        \frac{R^2}{2\pi}\oint f^+ \partial_{\theta}f^-\ ,
\ee
where $H$ is a Hilbert transform $H(f)=\P_n f = \partial (f_- -  f_+)$. The kernel is given by
$H(\t,\t')=$ $\frac{1}{4\pi i}$$\sum_{n}$$e^{in(\t-\t')}|n|$. Integration over $f$ with
this boundary interaction then produces the partition function of a
D-brane extended along the $X$-direction.

To be more general we can add local interactions on the boundary, parame\-trized by couplings $\{t_q\}$.
They are given by
functionals of $f$, so that the local and non-local contributions can be collected into
\be
\label{EQNbdryint}
    I^\bdry(t, X) = I^\bdry(t, f)\ ,
\ee
\be
\label{boundarypar}
    \Z^\bdry(R|t)= \int D[f] e^{\frac{i}{\a'}I^\bdry(t,f)}\ .
\ee
From the above it is now clear that a
change in the closed string modulus $R\to R+\d R$ appears as a deformation
of the boundary interaction
\ba
\label{addbndryint}
    I^\bdry &\to& I^\bdry + \d I^\bdry\\
    \d I^\bdry(R) &=& \frac{R\d R}{2\pi}\oint f H(f)\ .
\ea
In the presence of open string degrees of freedom
this is a non-trivial modification of the boundary theory.
For instance  for the Euclidean D1-brane wrapping $S^1$
the condition for marginality of the boundary operator $\exp ikX_b$ with
$k=n\in {\mathbb Z}$ is changed by the shift\footnote{Similarly,
strings attached to the D-instanton can wind around $S^1$. Their contribution to
the boundary partition function
is represented by the insertion of boundary vertex operators
$\exp i\frac{wR}{\a'}X_b$.} $R\to R+\delta R$.
We thus conclude that
the modulus $R$ of the closed string background
$\X=S^1_R$ enters as a non-local boundary interaction.
In particular,
\be
\label{EQNShiftI}
    \S_0(\X=S^1_{R+\d R}| t^1_*+t_q)
    =  \S_0(\X=S^1_R|t^2_* + t_q)\ ,
\ee
in accord with (\ref{EQNConjecture}).
Note that the theory without additional boundary interactions
is conformally invariant for any $R$.
Therefore there is no $\beta$-function associated to the radius deformation.
But the $\beta$-functions for other couplings depend on the non-local part
(and therefore on the bulk moduli) of the boundary interaction.

After this warm-up we will now consider interacting CFTs.
In the next section we show
that the factorization property also holds for boundary conformal
theories on group manifolds.

\section{Boundary WZW model}
\label{SecbWZW}

The prototype example for open strings propagating in curved
space-time is the WZW model which is also an example where the $B$-field is not closed. Here we will discuss this case in
detail.
Other curved target spaces can be treated in a similar fashion.

As is well known 
\cite{Klimcik:1996hp,Alekseev:1998mc, Stanciu:1999id}
in this case
worldsheets $\Sigma$ with boundary $\P\Sigma$  require some care in
the definition of the topological term $\Gamma(g)=
\int_{\Gamma}\tr(dg g^{-1})^3$ with $\partial\Gamma=M$. For a closed 2d surface 
$M$ this term is defined as an integral of a 3-form over a 3-manifold 
$\Gamma$ with the 2d surface $M$ as its boundary.
If the 2d surface has a boundary the unambiguous definition of this term is problematic.

We need the condition $H^3(G)=0$ on the group $G$ in order to define the
topological term in the WZW model in terms of a globally
well-defined 2-form $w_2$ such that $dw_2(g)=w_3(g)=\tr(dg g^{-1})^3$
(since $w_3$ is a closed 3-form, $dw_3=0$, such $w_2$ always exists
locally). We write this formally as $w_2(g)=d^{-1}w_3(g)$. 
If
$H^3(G) =\mathbb Z$ 
there is no such globally well-defined $w_2$, but
$\Gamma(g)=\int_M w_2$ is still globally well-defined modulo
$\mathbb Z$ as long as $M$ has no boundaries. If $M=\Sigma$ has a
boundary, one needs the condition $H^3(G)=0$ in order to define
$\int_{\Sigma} w_2(g)$ for an arbitrary map $g:\Sigma\to G$. This is
the case, for instance, for $SL(2,\mathbb R)$ which we will now
consider. However, even in this case $\Gamma(g)$ is not unique since
any $w_2'$ that differs from $w_2$ by an exact 2-form,
\be
\label{omegatwo}
w'_2=w_2(g)+d\beta(g)\ ,
\ee
leads to the same $w_3$. In general  $d\beta$
is closed but not necessarily exact. Thus, the action
$\Gamma(g)$ is defined by the 3-form $w_3$ up to an ambiguity that
comes from the 1-form $\beta$, which contributes to the action only
through a boundary term
\be
\label{amb}
	\Gamma^{\beta}(g)=\int_{\Sigma} w_2(g) + \int_{\partial \Sigma} \beta(g^b)\ .
\ee
We denote by $g^b$ the restriction of $g$ to the boundary. If $\beta$ is not well-defined globally, $\int_{\Sigma} d\beta$ still makes sense and 
depends only on $g^b$ since for two different continuations of $g^b$ into the bulk the difference is
$\int_{S^2}d\beta=0 \,\; {\rm mod}\,\; \mathbb Z$.  

For  
$SL(2,\mathbb R)$,  (\ref{amb}) can serve as definition of a class
of WZW actions together with the standard kinetic term

\be
\label{stand}
	I_{WZW}=\frac{\kappa}{4\pi i}\int_{\Sigma} \tr\, 
		({\partial}_{\mu}g g^{-1})^2+\frac{\kappa}{4\pi i}\Gamma^{\beta}(g)\ .
\ee
One expects the theory to be exactly conformally invariant for particular choices
of the boundary term $\int_{\partial \Sigma} \beta$. Classifying such 1-forms $\beta$ is
an interesting question, in particular, in view of
solutions to the quantum conformality condition
$T=\bar T$ on the boundary which do not reduce to the
condition that $g^b$ belongs to a fixed conjugacy class, which in turn follows from
the equations for the currents $J=\bar J$ on the boundary. The latter constraint is, in fact, stronger than the conformality condition. 

Let us now see how the procedure described
for free scalar field in the previous section is modified in this case. From 
$dw_2=w_3$  
it follows immediately on the level of differential forms that
\be
\label{difffor}
	\gamma(g_1,g_2) \equiv w_2(g_1g_2)-w_2(g_1)-w_2(g_2)+\tr\, g_1^{-1}dg_1 dg_2 g_2^{-1}\ .
\ee
is a well-defined closed 2-form. We note in passing that (\ref{difffor}) is closed without restriction to $H^3(G)=0$.
Furthermore, $\gamma$ defines a 2-cocycle on the loop group $\hat{LG}$. Indeed, if
we integrate the closed 2-form (\ref{difffor}) over the disk with boundary $S^1$, we get
$\alpha_2(g^b_1,g^b_2)=
\int_{D}\gamma(g_1,g_2)$,
where $g^b$ is the restriction of $g$ to the
boundary and this $\alpha_2$ satisfies the cocycle condition. To see that
$\alpha_2$ only depends on the boundary data of $g_1$ and $g_2$,
we note that for two different extension
$g_i^+$ and $g_i^-$ 
of $g_i^b$
the difference
\be
\label{EQNdiff}
    \int_{D^+} \gamma - \int_{D^-} \gamma=
    \int_{S^2} \gamma = 0 \,\; {\rm mod}\,\; \mathbb Z
\ee
as a consequence of (\ref{difffor}).
Since $g_i^+$ and $g_i^-$ are the same on the boundary and are otherwise independent,
the result follows. The fact that $\alpha_2$ satisfies cocycle condition can be checked
by direct algebraic computation using (\ref{difffor}) (see also 
\cite{Faddeev:1985iz, Mickelsson:1985sb, Mickelsson:1989hp, Losev:1995zf}).

To continue we will use the following decomposition (motivated by the free field example
in the previous section) for a generic map from the disk $\Sigma$ to the group $G$:
\be
\label{decom}
	g(z,\bz)=g_0(z,\bz)k(z,\bz); \quad \quad {g_0}|_{\partial \Sigma}=1; \quad \quad
k|_{{\partial \Sigma}}=f(\theta)\ ,
\ee
so
$g_0$ describes the D-instanton and $k$ is purely defined by the boundary data
$f(\theta): S^1 \rightarrow G$. We will give a concrete definition of $k$ below. 
For $H^3(G)=0$ each 2-form appearing on the rhs of (\ref{difffor})
is separately well-defined,
so that
\be
\label{coc}
	\int_{\Sigma}w_2(g_0k)=\int_{\Sigma}w_2(g_0)+\int_{\Sigma}w_2(k)-
\int_{\Sigma}\tr\, g_0^{-1}dg_0 dk k^{-1} {\rm\; mod\;} \mathbb Z
\ee
since the 2-cocycle 
$\alpha_2(g^b_1\equiv 1,g^b_2\equiv f)=0 $ 
mod $\mathbb Z$ \cite{Pressley:1988qk}. Combined with the kinetic term in (\ref{amb}) this leads to
the expression 
\eqn{
\label{actionsl}
	I_{WZW}(g)&=I_{WZW}(g_0)\\&+\frac{\kappa}{4\pi i}\int_{\Sigma}
\tr\, (\partial_{\mu}k k^{-1})^2+\frac{\kappa}{4\pi i}\Gamma^{\beta}(k)+ \frac{\kappa}{2\pi i}\int_{\Sigma}\tr\,
g_0^{-1}{\bar \partial} g_0 \partial k k^{-1}\ .
}
This action is
well-defined though the theory depends on the 1-form $\beta$ through
the boundary integral $\int_{S^1}\beta(f)$ in $\Gamma^{\beta}(k)$.

In order to proceed we will now specify the extension $k(z, \bz)$
of the boundary data $f(\theta)$ by
solving the Riemann-Hilbert problem
for $f(\theta)$. This means we decompose $f(\theta)$ as
\be
\label{rh}
	f(\theta)=h_+(\theta)h_-(\theta)\ ,
\ee
where $h_+$ can be holomorphically continued to $h(z)$ into the
disk $\Sigma$ and $h_-$ anti-holomorphically to ${\bar h}(\bz)$.
Thus, we have for $k(z,\bz)$
\be
\label{kzz}
	k(z,\bz)=h(z){\bar h}(\bz)\ .
\ee
Here $h$ and $\bar h$ are fields on the complexified group\footnote{The factors $h$ and $\bar h$ can be constructed by solving the equation of motion in Minkowski signature, that is, $k(\sigma^+,\sigma^-)=h(\sigma^+){\bar h}(\sigma^-)$ where $h$ and $\bar h$ are independent functions and then define $h(z)$ and ${\bar h}(\bz)$ by analytic continuation.}.
This $k(z,\bz)$ solves the WZW equations of motion and, together with $g_0$,
gives an unique decomposition of an arbitrary field
$g=g_0k$ on the disk. We will take (\ref{actionsl})
with this decomposition as definition of the WZW model on the
disk for arbitrary boundary fields $f(\theta)$ taking values in the
group manifold.
In background independent open string field theory we are instructed to
integrate over $f(\theta)$.
As we emphasised above this WZW theory on the disk depends on
the 1-form $\beta$ on the boundary, and since this 1-form is completely
arbitrary we include it in the definition
of the boundary perturbation in BSFT.
We do not specify for which $\beta$ this theory is conformal -- this
is a good question and the only comment we will make is that
the string field theory action is one candidate
for the solution -- its critical points correspond to conformal boundary interactions
parametrised by $\beta(f)$.
We conclude that the WZW theory on the disk for the case $H^3(G)=0$
is given by the action (\ref{actionsl}) with the definitions (\ref{rh}), (\ref{kzz}) and (\ref{decom}).


\subsection{Bulk-boundary factorisation}

Unlike for the free field case, in the classical WZW action (\ref{actionsl})
the boundary field $k$ does not decouple from the bulk fields $g_0$ 
on the level of the classical action.
The interaction between these two fields is given by 
\be
\label{EQNW}
    \frac{\kappa}{2\pi i}\int_\Sigma d^2z\bar J_{g_0} \P K\ ,
\ee
where $\bar J_{g_0}$ is the anti-holomorphic $g_0$-current, and
the holomorphic function $K(z)$ is defined via $\P K = \P k k^\1$ using the fact that $\P k k^\1$ is a holomorphic 1-form. 

Nevertheless, we will show below that this cross term in (\ref{actionsl})
between $g_0$ and $f$ (which parametrises $k$) does not
contribute to the path integral over $g_0$. Concretely we will prove
that the $n$-point function
\be
\label{npt}
	\biggl\langle\Bigl(\int_{\Sigma}\tr\,
g_0^{-1}{\bar\partial} g_0 \partial k k^{-1}\Bigr)^n\biggr\rangle_{g_0}=0\ .
\ee
Thus for any choice of $\beta$
\be
\label{antwzw}
	\int_{g|_{\partial \Sigma}=
f} D[g]e^{-I_{WZW}(g)}=
\Z_0 e^{-W(f)}\ ,
\ee
where
\be
\label{wlive}
	W(f)=\frac{\kappa}{4\pi i}\int_{\Sigma} \tr\, (\partial_{\mu}k k^{-1})^2+\frac{\kappa}{4\pi i}\Gamma^{\beta}(k)
\ee
and
\be
\label{znot}
	\Z_0=\int_{{g_0|}_{\partial \Sigma}=1}D[g_0]e^{-I_{WZW}(g_0)}\ ,
\ee
verifying our conjectured factorisation in this class of models. This is the main
technical result of this work.

We will now give a qualitative argument for the vanishing 
of the n-point function (\ref{npt}). The explicit proof of 
this claim is given in
the appendix. Consider the functional integral over $g_0$ 
at fixed
$k$. This is the WZW theory with boundary conditions in the 
identity conjugacy class.  
The correlators of $\bar J$'s are functions with poles in $\bz$,
but no positive powers of $\bz$ occur. These correlators are then
multiplied by functions $\partial K$ which are polynomials of
positive powers of $z$. These products are proportional to a
positive power of $e^{i\t}$ so that the $\t$ integral vanishes as
long as no singularities occur and the $U(1)$-action by $e^{i\t}$ is
unbroken. In \ref{proofvanish} we proof that no such singularities
appear\footnote{It should be noted that this argument works only because
we integrate over the disk. One-dimensional
integrals  of such perturbations over the boundary of the disk would give rise to divergences
\cite{Bachas:2004sy}.}. 
But before that we show an important property of correlation functions involving a chiral current.

\subsection{Decoupling of chiral currents}

The crucial property for these arguments to work is that the
$n$-point functions of the antichiral currents $\bar J$ on the disk
are functions of $\bz$ only. In general one might expect
interactions of the currents with their images. This would generate
terms which behave singular at the boundary. But in this particular
case no such terms appear, and this is due to the following
argument: It is important that in this $n$-point function
only chiral bulk fields are involved in the WZW theory with
$g_0=1$ at the boundary (for the trivial conjugacy class),
i.e. we are interested in the expectation values $\<\bar
J(\bz_1)\cdots\bar J(\bz_p)\>_{D}$ with Dirichlet boundary
conditions. This amplitude has an equivalent representation in terms
of a Dirichlet boundary state $|B_D\>$. The explicit construction of
$|B_D\>$ is not needed. We merely need to assume that such a state
exists. Then the expectation value can be written as an unnormalised
correlation function $\<0|\bar J(\bz_1)\cdots\bar J(\bz_p)|B_D\>$.
Expanding the currents in modes we get 
\be
\label{EQNcurrentI}
        \sum_{n_1\cdots n_p}\bz_1^{n_1}\cdots \bz_p^{n_p}\<0|\bar
j_{n_1}\cdots\bar j_{n_p}
        |B_D\>\ .
\ee
All $\bar j_n$ with $n\le 0$ annihilate on the vacuum, thus the
expression contains only terms with $n>0$. 
The boundary state is
defined by $\bar J d\bz|B_D\> = Jdz|B_D\>$, thus it maps $\bar j_n$
to $j_{-n}$. Since the holomorphic and anti-holo\-mor\-phic currents
commute, the $j_{-n}$ can be moved all the way to the left to act on
the vacuum, which it annihilates. This then implies that bulk normal
ordered monomials of chiral operators have a vanishing expectation
value also for Dirichlet boundary conditions. For 
the ordinary product of antiholomorphic currents we then conclude that 
\be
\label{EQNcurrentII}
    \<0|\bar J(\bz_1)\cdots\bar J(\bz_p)|B_D\> \propto
        \<0|\bar J(\bz_1)\cdots\bar J(\bz_p)|0\>\ ,
\ee
that is, the boundary state enters only in the normalisation.
Thus the only singularities are those of coinciding  
$\bar J$'s,
which can then be treated in the manner described above.

To summarise, this line of argument shows that, although the bulk and boundary fields to not decouple in the classical action, the partition
function is independent of the interaction term 
$\int \bar J_{g_0}\partial K$ 
to any
order in perturbation theory. 
Thus the boundary
degrees of freedom decouple from the bulk and the partition function
factorises. To complete the argument we note that the translation
invariance of the functional Haar measure $D[g]$ implies that no
Jacobian occurs when integrating out the bulk fields $g_0$. Thus
\be\label{EQNconclIII}
    \int_{g|_{\partial\Sigma}=f} D[g] e^{-I_{WZW}(g)} = \Z_0 e^{-W(f)}\ ,
\ee
where
\be\label{EQNconclIV}
    W(f) = \frac{\kappa}{4\pi i}\int_\Sigma\tr\,(\partial_\mu kk^{-1})^2 + \frac{\kappa}{4\pi i}\Gamma^\beta(k)\ .
\ee
An immediate consequence of the above result is that the boundary partition
function on a group manifold is related to the flat space partition
function by a non-local boundary deformation in agreement with the
correspondence stated in the introduction.

\subsection{Vanishing of chiral current $n$-point functions}
\label{proofvanish}
\label{SecAppendix}

Here we give an explicit proof of the claim that that 
(\ref{EQNW}) does not contribute to the path integral over the bulk field $g_0$.
We choose coordinates $z=\r e^{i\t}$ on the disk ($|z|\le 1$). The
operator $\exp -\int \bar J_{g_0}\partial K$ is expanded as $\sum (n!)^\1 (-1)^n I_n$, where
\ba
\label{EQNappI}
    I_n &\equiv& \int d^2 z_1 \P_1 K(z_1) \cdots \int d^2 z_n \P_n K(z_n)
{\cal A}_n(\bar z_1, \dots, \bar z_n) \\
    {\cal A}_n &\equiv& \vev{\bar J_{g_0}(\bz_1)\cdots\bar J_{g_0}(\bz_n)}\ .
\ea
Here $\bar J_{g_0}=g_0^{-1}\bar\partial g_0$ is the anti-holomorphic bulk current. 
The basic ingredient for computing the integral (\ref{EQNappI}) is the OPE of the anti-holomorphic currents
\be\label{EQNappII}
    \bar J^a(\bz_1)\bar J^b(\bz_2) \sim \frac{\k \d^{ab}}{(\bz_1-\bz_2)^2}
+ \frac{if^{abc}}{\bz_1-\bz_2}
    \bar J^c(\bz_2)\ ,
\ee
where $\bar J = \k^\1 \bar J^a T^a$, $T^a$ are the generators of the
algebra, $f^{abc}$ the structure constants and $\d^{ab}$ the Cartan
metric. But we will see that the calculation does not depend on
details like symmetry structures of the group.

As general strategy we evaluate the indefinite integrals in
order to treat the singularities correctly. The result is then shown
to be a regular function of all variables, so that the boundaries
can be inserted and no singularities occur.

It is clear that the one-point function vanishes, $I_1$=0.
The two-point function is more involved since the Wick theorem does
not hold and there are self-interactions of the currents. The
amplitude is
\be\label{EQNappIII}
    {\cal A}_2(\bar z_1, \bar z_2)
    = \VEv{ \frac{k\d^{ab}}{(\bar z_1 - \bar z_2)^2} + \frac{ i
f^{abc}}{\bar z_1-\bar z_2} \bar J^c(\bar z_2)  } T^a T^b
    \propto \frac{1}{(\bar z_1 - \bar z_2)^2}\ .
\ee
We expand the holomorphic field as
\be\label{EQNappIV}
    \P K(z) = \sum_{m>0} m K_m z^{m-1}\ .
\ee
Thus, $I_2$ consists of (a sum of) terms
\be\label{EQNappV}
    m_1m_2\oint d\t_1 e^{i(m_1-1)\t_1}\oint d \t_2
e^{i(m_2-1)\t_2}\int_0^1 d\r_1  \int_0^1 d\r_2
\frac{\r_1^{m_1}\r_2^{m_2}e^{2i\t_1}}{(\r_1-\r_2e^{-i(\t_2-\t_1)})^2}\ .
\ee
The structure becomes more obvious when a relative boundary
coordinate $\t=\t_2-\t_1$ is introduced,
\be\label{EQNappVI}
    m_1m_2\oint d\t_1 e^{i(m_1+m_2)\t_1}\oint d \t e^{i(m_2-1)\t}\int_0^1
d\r_1  \int_0^1 d\r_2
        \frac{\r_1^{m_1}\r_2^{m_2}}{(\r_1-\r_2e^{-i\t})^2}\ .
\ee
As $m_i\ge 1$ the $\t_1$-integral makes the whole term vanish as
long as the remaining integrals are not divergent. The $m_i$ are set
to $1$, because higher powers of $\r_i$ will, at best, smoothen the
singularities. We conduct the $\r_1$-integral and the relevant part
becomes
\be\label{EQNAppCalcI}
    \int d\t \int d\r_2
        \left[
        \r_2\ln(\r_1-\r_2e^{-i\t}) -
\frac{\r_2^2e^{-i\t}}{\r_1-\r_2e^{-i\t}}
        \right]\ .
\ee
The second part of (\ref{EQNAppCalcI}) is
\ba\label{EQNAppCalcII}
            &\int& d\t e^{-i\t}\left[
            \2\r_2^2e^{i\t} +
\r_1\r_2e^{2i\t}+\r_1^2e^{3i\t}\ln(\r_1-\r_2e^{-i\t})
        \right] \\
    &=& \r_1^2\int d\t e^{2i\t}\ln(\r_1-\r_2e^{-i\t}) + {\rm regular\; terms}\ . 
\ea
Conducting the $\t$-integral yields
\be\label{EQNappVII}
	-\frac{i}{2}\left(\r_1^2e^{2i\t}-\r_2^2\right)\ln\left(\r_1e^{i\t}-\r_2\right)
		+ {\rm regular\; terms}\ .
\ee
which is non-singular in all variables. Therefore the whole
expression is non-divergent and vanishes finally under the
$\t_1$-integral.

The first part of (\ref{EQNAppCalcI}) is, after $\r_2$-integration,
\be\label{EQNappVIII}
    \2\int d\t \left[
        \r_2^2\ln(\r_1-\r_2e^{-i\t}) + e^{-i\t}\int
d\r_2\frac{\r_2^2}{\r_1-\r_2e^{-i\t}}
        \right]\ .
\ee
The whole expression becomes, using the result from (\ref{EQNAppCalcII}),
\be\label{EQNappIX}
    -\2\int d\t
        e^{2i\t}(\r_1^2-\r_2^2e^{-2i\t})\ln(\r_1-\r_2e^{-i\t})
            + {\rm regular\; terms}\ .
\ee
This term is regular even without $\t$-integration. Therefore all
terms are finite and finally vanish under the $\t_1$-integral. Thus
$I_2=0$.

The three-point-amplitude is proportional to
\be\label{EQNappX}
    {\cal A}_3(\bar z_1, \bar z_2, \bar z_3)
    \propto \frac{1}{ (\bar z_1- \bar z_2)(\bar z_1-\bar
z_3)(\bar z_2 - \bar z_3)}\ .
\ee
$I_3$ contains terms of the form
\be\label{EQNappXI}
     \int d\r_1 d\t_1 \cdots d\r_3 d\t_3
    \frac{m_1m_2m_3\r_1\r_2\r_3}
{(\r_1e^{-i\t_1}-\r_2e^{-i\t_2})(\r_2e^{-i\t_2}-\r_3e^{-i\t_3})(\r_1e^{-i\t_1}-\r_3e^{-i\t_3})}\ .
\ee
Again we set $m_i=1$ in order to single out the most singular part.
The indefinite integration over $\r_1$ gives 
\ba\label{EQNAppCalcIII}
\nonumber
    &&\oint d\t_1\int d\r_1 \frac{\r_1\r_2\r_3}
{(\r_1e^{-i\t_1}-\r_2e^{-i\t_2})(\r_2e^{-i\t_2}-\r_3e^{-i\t_3})(\r_1e^{-i\t_1}-\r_3e^{-i\t_3})}\\
    \nonumber&&=\frac{\r_2^2e^{-i\t_2}\r_3}{(\r_2e^{-i\t_2}-\r_3e^{-i\t_3})^2} \oint
d\t_1
        e^{2i\t_1} \ln(\r_1e^{-i\t_1}-\r_2e^{-i\t_2})\\
         &&\qquad\qquad- \left[ \r_2e^{-i\t_2} \leftrightarrow \r_3e^{-i\t_3} \right]\ .
\ea
Now we conduct the $\t_1$-integral\footnote{We multiply the integrand
with $e^{-i\t_1}$, which does not change the degree of divergence.
We could also use the integrand without modifications, but the computation is
slightly longer.} 
\ba\label{EQNAppCalcIV}
    \nonumber&&\int d\t_1 e^{i\t_1} \ln(\r_1e^{-i\t_1}-\r_2e^{-i\t_2}) \\
    &&=i\r_1\left[
\frac{\r_1e^{-i\t_1}-\r_2e^{-i\t_2}}{\r_1e^{-i\t_1}\r_2e^{-i\t_2}}\ln(\r_1e^{-i\t_1}-\r_2e^{-i\t_2})
        -\frac{e^{i\t_2}}{\r_2}\ln(\r_1e^{-i\t_1}) \right]\ .
        \qquad
\ea 
Restoring the pre-factors from (\ref{EQNAppCalcIII}) we see that
(\ref{EQNAppCalcIV}) is less singular than
\be\label{EQNappXII}
    \frac{i\r_2\r_3e^{i\t_1}}{{\bar z_{23}}^2}
        \left(
        \bar z_{12}\ln\bar z_{12}
        - \bar z_1\ln\bar z_1
        \right) - \left[ \r_2e^{-i\t_2} \leftrightarrow
\r_3e^{-i\t_3}\right]\ .
\ee
The expression in the bracket is completely regular. As pre-factor
we recognise the contribution from the 2-point function. Thus, we
conclude that $I_3$ must have the same or a less singular behaviour
than $I_2$. Thus, the overall $\t_1$-integration, which is also
present for the three-point function, makes the whole expression
vanish, $I_3=0$.

This argument can now be applied recursively to $n$-point functions.
For the sake of a clear presentation we switch to a rather symbolic
notation. The recursion then works like (modulo some permutations)
\ba\label{EQNappXIII}
\nonumber 
    &&\int d z_n \frac{(\cdots)}{(\bz_1-\bz_2)(\bz_2-\bz_3)\cdots
        (\bz_{n-1}-\bz_n)(\bz_n-\bz_1)} \\
   &&\propto \frac{(\cdots)}{(\bz_1-\bz_2)(\bz_2-\bz_3)\cdots
        (\bz_{n-1}-\bz_1)} + {\rm less\; singular\; terms}
\ea
until one ends up with a three-point amplitude. Thus, all these
indefinite integrals are indeed regular.

Now we argue that in fact all the integrals $I_n$ must vanish. We
extract $e^{i\t_1}$ from each factor $(\bz_i-\bz_j)^\1$ and shift
all the other boundary coordinates $\t_i\to \t_i'=\t_i-\t_1$. This
gives a global factor of $\exp i\sum_{i}m_i\t_1$. The $m_i$ are
always positive, thus the $\t_1$-integration makes the whole
expression vanish. We arrive at the central result of this
calculation:
\be\label{EQNappXIV}
    I_n = 0\ .
\ee
The immediate consequence is that the operator 
$\exp -\int \bar J_{g_0}\partial K$ 
is marginal
and therefore the partition function does not depend on it.


\subsection{Groups with $H^3(G)\neq 0$ }

Let us now turn to the case when $H^3(G)=\mathbb Z$ 
such as $G=SU(N)$.
In this case $w_2$ is not globally defined 
(it is ill-defined on a high codimension
submanifold of the target, which is just a point for the case of the $SU(2) \cong S^3$ group manifold).
Let us follow the arguments for the $H^3(G)=0$ case and see where the problems show up. In order to reduce
the action to the form (\ref{actionsl}) we use the decomposition $g=g_0k$, with $k$ as in
(\ref{rh}). 
Since (\ref{difffor}) is still well-defined we can formally arrive at the equation (\ref{coc}).
In particular, the non-trivial 2-cocycle $\alpha_2(g_0^b,f)$ after formula (\ref{coc}) is 
again zero for $g^b_0=1$. 
There is no
problem to globally define the first
and the last term on the rhs of (\ref{coc}). The difficulty resides in the second term $w_2(k)$.
Thus the problem is with the definition of the WZW action on solutions of the 
classical equations of motion 
${\bar \partial}(\partial k k^{-1})=0$, with $k|_{\partial\Sigma}=f(\theta)$, where $f$ is arbitrary. 
In this case the classical Lagrangian turnes out to be
not a function anymore \cite{Gawedzki:2001ye, Gawedzki:1999bq}. 
However, this might be expected, because
a path integral with boundary conditions defines
a wave-function, which corresponds to a 
section of some bundle.
Note that although the action is ambiguous 
the equations of motion derived from it are well-defined.

Recall that the reason we want to consider boundary conditions which are not in a conjugacy class 
is that according to the philosophy of BSFT one has to integrate over all degrees of freedom 
including the boundary fields with boundary interactions parametrised by the 1-form $\beta$. From  the expression  
(\ref{EBSFT}) for the string field theory action, it follows that 
it is the space-time action (\ref{EBSFT}) that needs to be well-defined
for boundary deformations and not the worldsheet classical action 
$W(f)=\frac{\kappa}{4\pi i}\int\tr\,(\partial_\mu kk^{-1})^2 + \frac{\kappa}{4\pi i}\Gamma^\beta(k)$. 
That is, an integral over boundary maps  
\ba\label{ZBB}
    \Z^\bdry &=& \Z/\Z_0 \\ &=&\int 
	D[f]e^{-W(f)}\ ,
\ea
shall be well-defined, where $D[f]$ is a Haar
measure for $k$ written in terms of $f$ after expressing $k$ via the 
solution of the Riemann-Hilbert problem described above. 
Even if this path integral diverges ultimately, it is the combination entering in (\ref{EBSFT})
that shall lead to a well-defined space-time action.

Since there are infinitely
many choices for $\beta$ one would like to classify them according the conformality condition for
the corresponding quantum theory. As we mentioned for $SL(2,\mathbb R)$, this is exactly
the question that background independent open string field theory studies.

For $H^3(G)\neq 0$, one way to remove the topological obstruction in defining
$\Gamma^\beta(k)$, is by deleting a high codimension submanifold in $G$ and
repeating (\ref{difffor}) for $g_1=g_0$ and $g_2=k$.
Since these
relations are algebraic we still can safely derive the formal relation (\ref{coc}). 
One might then suggest that in this case a $d\beta$ can be found, so that
the integral over boundary fields is still well-defined 
as mentioned above (with appropriate regularisation procedure).
We recall that a similar situation appears for the
analogous 
quantum-mechanical problem for trajectories with
boundaries in a compact phase-space (associated with coadjoint orbits, and related), 
where the classical action on the world-line
is ill-defined due to non-trivial $H^2$ of the phase space though the path integral can be properly defined
in order to get a correct wave-function
\cite{Alekseev:1988vx}.
In short --
although the action is ill-defined on high codimension submanifolds the path integral
on the manifold with boundary still gives a well-defined and 
correct ``wave-function" (matrix element).
According to \cite{Alekseev:1988ce, Alekseev:1990vr} 
our current problem is an infinite-dimensional version of the quantum mechanical problem.
We believe that the same is true for the family of 2d field theories related
to WZW models on the disk for group manifolds with non-trivial $H^3$.

Critical points of the string field
theory action (\ref{EBSFT}) are supposed to lead to well-defined 
conformal boundary conditions and well-defined $\Z^\bdry=\Z/\Z_0$,
which is the value of the space-time action on-shell according to (\ref{EBSFT}) 
(these boundary interactions, in particular, do contain the
restriction to conjugacy classes as a sub-set of the conformal conditions).

So at the moment
we simply assume that (\ref{stand}) be given via (\ref{actionsl})
for all groups including those with $H^3(G) \neq 0$ (as we mentioned for $SL(2,\mathbb R)$,
everything is properly defined in (\ref{actionsl}) and this case is very intersting
on its own right) and define
the string field theory action via standard methods.

At this point a comment about the measure $D[k]$ is in order. If we
start with the Haar measure for $g$, the natural measure for $k$
comes out to be the functional Haar measure for $k(z,\bz) = h(z)\bar
h(\bz)$. Note, however, that $k$ is uniquely determined in terms of
the boundary data. When pulling back $D[k]$ to the
boundary a Jacobian occurs and introduces a further non-locality in
the boundary interaction. So the total non-local boundary
deformation resulting from a shift in the closed string background
is given by $W(f) = \frac{\kappa}{4\pi i}\int_\Sigma\tr\,(\partial_\mu kk^{-1})^2 
+ \frac{\kappa}{4\pi i}\Gamma^\beta(k)$ plus the Jacobian generated.
In the next section
we will give an illustration by considering the large radius limit
of the $SU(2)$ model.

\subsection{The $SU(2)$ boundary action}
\label{SECBOUNDARYACTION}

We now specify the closed string background to be the group manifold of $SU(2)$ (or its complexification) and set the one-form $\b$ to zero temporarily.
For convenience we substitute $\l\equiv\k^{-\frac{1}{2}}$ and obtain the $SU(2)$-boundary action\footnote{Please note that here we denote the level of the WZW model by $\kappa$ rather than by $k$ as in previous chapters.}\footnote{The trace is normalised in a way so that the quadratic part of the action is given by the standard term $\sum_{m>0}mf_m\bar f_m$ in flat space.}
\be
S=\frac{1}{(i\l)^2}\Tr \int \P kk^{-1}\bP kk^{-1}\ .
\ee
Expanding the action up to fourth order in $f$ and $\bar f$ one finds after some tedious algebra
\be\label{EQNAC}
	S = s\sum_{m=1}m f_m^\a\bar f_{m\a}
		+ \a (V_\a - V_{\bar\a}) + \b V_\b + \g (V_\g + V_{\bar\g})\ ,
\ee
with
\ba\label{EQNACII}
	V_\a &=& \sum_{c,b,a=1}(b-a)\d_{c,a+b} \epsilon_{\m\n\l}   f^\m_a f^\n_b  \bar f^\l_c\\
	V_\b &=& - \sum_{a,b,c,d=1}\frac{(c-b)(a-d)}{a+d} \d_{c+b,a+d} f^\m_c  f^\n_b \bar f_{a\m}  \bar f_{d\n} \\
	V_\g &=& - \frac{2}{3}\sum_{a,b,c,d=1}(a-d-b)\d_{c,b+a+d}
			   f^\m_c \bar f_{a\m} \bar f^\n_b \bar f_{d\n}\ ,
\ea
where $V_{\bar\a}$ and $V_{\bar\g}$ is obtained by exchanging $f$ with $\bar f$. The original action (the starting point of the renormalisation group flow) is found for $s=1, \a=\frac{\l}{2}, \b=\g=\frac{\l^2}{2}$.
Although the action can be computed exactly we truncate its expansion at ${\cal O}(\l^3)$ (we will see that this gives $\beta$-functions which are exact up to ${\cal O}(\l^5)$).

There is another contribution to the action coming from a Jacobian due to the change of variables. Starting from the standard Haar measure on the group $[dk]=k^{-1}dk$ we obtain the following Jacobian:
\ba\nonumber
J &&= \frac{\left(\left [\d kk^{-1}\right ]^+, \left [\d kk^{-1}\right ]^-\right)}{\left(\d f, \d\bar f\,\right)} \\
	&&= \begin{pmatrix} \frac{\left [\d hh^{-1}\right ]^+}{\vphantom{\hat f}\d f}\quad \, 
				\frac{\left [h\d \bar h\bar h^{-1}h^{-1}\right ]^+}{\vphantom{\hat f} \d \bar f}\cr
		 \qquad 0\qquad\, \frac{\left [h\d \bar h\bar h^{-1}h^{-1}\right ]^-}{\vphantom{\hat f}\d \bar f}
		 \end{pmatrix}
		= \begin{pmatrix}\;J_{11}\;\,J_{12}\;\cr \;0\,\; \;\;J_{22}\end{pmatrix},
\ea
where $\pm$ indicates restriction to the holomorphic/anti-holomorphic part.
Thus $\Det J=\Det J_{11} \cdot \Det J_{22}$. The rather lengthy calculation can be found in the appendix. Here we note simply the result. The measure contributes
\be
I_\text{measure} = 4\l^2\sum_{n=1}f_{n\mu}\bar f^\mu_n + {\cal O}(\l^4)
\ee
to the action. 
This is  a mass term for the boundary field $X_b$ which, due to its classical dimension, flows to $\infty$ in the infrared thus forcing $X_b$ to zero, i.e.  Dirichlet boundary conditions in all directions. Thus we see that  the tachyonic decay of the space-filling brane in the $SU(2)$ WZW-model is already encoded in the measure\footnote{In fact, the measure for the boundary field $k$ is not uniquely determined by the bulk theory. Here we have taken the Haar measure 
as the starting point. Alternatively, one could consider the flat symplectic measure for $f$ and $\bar f$. In this case the tachyon arises as a one-loop counter term (see next chapter).}.

%
%
%

\chapter{Renormalisation of the boundary action}
\label{ch-ren}

This chapter is devoted to a concrete study of a special case of a factorised boundary action. The action is treated perturbatively and expanded around a flat space background. This results in non-local terms, which are found to contribute to the $\b$-functions of the system. The renormalisation of the local and non-local couplings is conducted explicitly. The results can be interpreted in the framework of tachyon condensation, where indications are found that the end point of the condensation is a spherical 2-brane in $S^3$ target space. The presentation here follows \cite{Baumgartl:2006xb}.

\section{Three-sphere boundary action}

Once the concrete form of the action has been obtained, we can analyse the quantised theory. It is clear that the action as it stands is not scale-invariant due to the presence of the mass term. To account for the mass and the `cosmological constant' we introduce the
tachyon coupling $T(X) = a + uf\bar f$. Note that we do not expect the expansion up to fourth order to lead to a renormalisable theory. The exact action should, however, be renormalisable since the bulk theory from which it has been obtained is renormalisable. In particular we expect the renormalised action to describe field configurations in the same group manifold. Thus the structure of the interaction terms (which respects the group symmetry) should be untouched. Therefore we will assume $\l=\k^{-\frac{1}{2}}$ to be scale dependent, but keep the relative couplings fixed. Accounting for wave-function renormalisation we will allow for $s$ to be scale dependent.

\subsection{The action}

Let $T_\m$ be the generators of $SU(2)$.
We define the operators
\be
{\rm ad}_f = \left [f^\m T_\m, \cdot\;\right ] \qquad {\rm Ad}_h = h\;\cdot\;h^{-1}
\ee
and derive (with $\w(\theta) \equiv 2\sqrt{f^\mu(\theta)f_\mu(\theta)}\,\,\,$)
\ba
	{\rm Ad}_h  &=& {\rm id} + \frac{i\sin \lambda\w}{\w}{\rm ad}_f  
			+ \frac{\cos\lambda\w-1}{\w^2}{\rm ad}_f^2 \\
	\d hh^{-1} &=& \left [i\lambda {\rm id} + \frac{\cos \lambda\w-1}{\w^2}{\rm ad}_f 
		+ i\frac{\sin \lambda\w-\lambda\w}{\w^3}{\rm ad}_f^2\right ]\delta f^\m T_\m\\
	h^{-1}\d h &=& \left [i\lambda {\rm id} - \frac{\cos \lambda\w-1}{\w^2}{\rm ad}_f 
		+ i\frac{\sin \lambda\w-\lambda\w}{\w^3}{\rm ad}_f^2\right ]\delta f^\m T_\m\ .
\ea
With these preparations the action can be obtained exactly in these coordinates. An expansion in the perturbation parameter $\l$ up to order $\l^3$ yields the expressions (\ref{EQNAC}) and (\ref{EQNACII})\footnote{A normalisation of boundary integrals has been used, which absorbs factors of $2\pi$ in a convenient way.}.

\subsection{The Jacobian}

Here we present details about the calculation of the Jacobian as advocated in (\ref{SECBOUNDARYACTION}). Unlike for the action, it is not possible to obtain an explicit expression for arbitrary $\l$, but a perturbative expansion is possible. 
Let us first focus on the determinant of the matrix $J_{11} = \frac{[\d hh^{-1}]^+}{\d f}$. In components it can be expressed as
\ba
(J_{11})_{nm}^{\m\n} &=& 
		i\l \d_{nm}^{\m\n} \\
	\nonumber
		&&- 2i\oint d\t e^{i(n-m)\t}
			\left [{\epsilon_\rho}^{\m\n} f^\rho(\t)\frac{{\rm d}}{{\rm}d\lambda} + 2 \epsilon^{\m \l \r}{\epsilon_\r}^{\k\n}f^\l (\t)f^\k(\t)\right ]\cal{A}(\lambda)\\
			\cal{A}(\lambda)&=&
			\frac{\sin\lambda\omega-\lambda\omega}{\omega^3}\ .
\ea
The functions $f(\t)$ are given by the holomorphic function $f(z)$ with coordinates restricted to the boundary $z=e^{i\t}$. As $f$ has no zero mode, the integral can only be non-zero when $m>n$. In particular only the very first term contributes to the trace of $J_{11}$. Higher powers of $J_{11}$ contain terms $\oint e^{i(n-p_1)\t_1}\oint e^{i(p_1-p_2)\t_2}\cdots\oint e^{i(p_k-m)\t_{k+1}}$ with $n<p_1<\cdots<p_k<m$. Under the trace these terms vanish again. Therefore (we suppress irrelevant factors coming from tracing over space indices)
\be
	\Tr\,J_{11}^n = (i\l)^n \Tr\, 1\ .
\ee
Using the expansion of the determinant in traces, 
\be
	\ln\Det\frac{J_{11}}{i\l} = \Tr \int \frac{ds}{s}e^{-s\frac{J_{11}}{i\l}}\ ,
\ee
we get
\be
	\Det \frac{J_{11}}{i\l} = 1\ .
\ee
For the computation of $J_{22}$ we expand
\ba
	h\d\bar h\bar h^{-1}h^{-1}
	&=& \sum_{i=0}^8 \d b_i\\
	\d b_0 = i\l\d\bar f &&\qquad 
		\d b_1 = \bar Z_1{\rm ad}_{\bar f}\d\bar f\\
	\d b_2 = \bar Z_2{\rm ad}_{\bar f}^2\d\bar f &&\qquad 
		\d b_3 = i\l Z_3{\rm ad}_f\d\bar f\\
	\d b_4 = Z_3\bar Z_1{\rm ad}_f{\rm ad}_{\bar f}\d\bar f &&\qquad 
		\d b_5 = Z_3\bar Z_2{\rm ad}_f{\rm ad}_{\bar f}^2\d\bar f\\
	\d b_6 = i\l Z_1{\rm ad}_f^2\d\bar f &&\qquad
		\d b_7 = Z_1\bar Z_1{\rm ad}_f^2{\rm ad}_{\bar f}\d\bar f\\
	\d b_8 = Z_1\bar Z_2{\rm ad}_f^2{\rm ad}_{\bar f}^2\d\bar f\ .&&
\ea
with the abbreviations
\ba
	Z_1 &=& -\frac{\l^2}{2} + \frac{\l^4 \w^2}{24} - \frac{\l^6\w^4}{720} + {\cal O}(\l^7) \\
	Z_2 &=& -\frac{i\l^3}{6} + \frac{i\l^5\w^2}{120} + {\cal O}(\l^7) \\
	Z_3 &=& i\l + \w^2 Z_2\ .
\ea
The functional matrices, which enter the determinant are
\be
	\left [\frac{\d b_i}{\d \bar f}\right ]^-_{(\m n)(\n m)}\ ,
\ee
where the upper index `$-$' indicates projection on the antiholomorphic modes.
Due to $m\le n$, they are all upper triangular matrices, hence the determinant is just the product of the diagonal entries. We re-write it in the following way:
\be\label{EQNDET}
	\Det \frac{J_{22}}{i\l} \equiv e^{\Tr \ln  B_{nm}^{\m\n}}\ ,
\ee
where
\ba
	B_{nm}^{\m\n} = \d^{\m\n}_{mn} &-& \frac{4i}{\l}\oint Z_3\bar Z_1f^\l\bar f^\k \epsilon^{\k\n\r}\epsilon^{\r\l\m}\\
					&-&\frac{8}{\l}\oint \left [Z_3\bar Z_2 f^\l \bar f^\k\bar f^\r
						+ Z_1\bar Z_1 f^\l f^\k\bar f^\r\right ]
							\epsilon^{\r\n\s}\epsilon^{\k\s\tau}\epsilon^{\l\tau\m}\\
		&+&\frac{16i}{\l}\oint Z_1\bar Z_2f^\l f^\k\bar f^\r\bar f^\s 
		\epsilon^{\s\n\tau}\epsilon^{\tau\r\w}\epsilon^{\w\k\xi}\epsilon^{\xi\l\m}\ .
\ea
Next we expand the logarithm (\ref{EQNDET}) and derive an expression for the contribution to the action. A straight forward calculation reveals
\be
	I_{\rm Jacobian} = 4\l^2 \sum_{n>0}f_n^\m f_{n\m} + {\cal O}(\l^4)\ . 
\ee
The lowest order of the Jacobian, which modifies the action, has therefore the form of a tachyon interaction.

\section{$\beta$-functions}

For the calculation of the $\b$-functions we evaluate $n$-point functions expanded in loops. These correlators are IR finite because the theory is considered on a one-dimensional compact space. For large momentum the amplitudes are typically divergent, thus it is convenient to introduce a momentum cutoff $\Lambda$. This regularisation seems appropriate as we are dealing with discrete sums so that $\Lambda$ simply appears as upper bound. The divergent parts of diagrams can be found by investigating the behaviour for large $\Lambda$. Higher loop diagrams are treated in the following way. All loops naturally appear with sums over positive momenta only. Therefore sums of the type $\sum_{a=1}^\Lambda\sum_{b=1}^\Lambda f(a,b)$ can be transformed into an expression of the form $\sum_{\mu=2}^\Lambda\sum_{\nu=1}^{\mu-1} f(\nu,\mu-\nu)$. With this method only one divergent sum appears, even for higher loops.

Once the divergent part is extracted the renormalisation procedure can be performed. Here we decide to start from the normal ordered theory with respect to the free field vacuum\footnote{Such a normal ordering prescription can in general not be held at higher loops. In the approximation used here, however, it does hold, because all nested singularities of higher-loop diagrams ($\ge 2)$ are already removed through the 1-loop counter-terms. It turns out that this is not due to cancellations between different diagrams, but all diagrams become finite separately. The inclusion of self-contractions would only modify some numeric coefficients in the 1-loop counter-terms, which does not influence the finiteness of the 2-loop diagrams. The 3-loop-diagrams on the other hand vanish identically.}
and add counter-terms, which cancel the divergent part of the amplitudes. The counter-terms for the two- and three-point-functions ($p$ is the external momentum) are given by the following expression, which must be subtracted from the classical action: 
\ba\label{EQNCOUNTER}
	\Sigma^{(2)}(p,\Lambda)
		&=& \Lambda\left\{32\frac{\a^2}{s^2}\right\}
			+ \ln\Lambda\left\{-96 p\frac{\a^2}{s^2}-64\frac{u}{s}\frac{\a^2}{s^2}\right\} \\
	\Sigma^{(3)}(p,p',\Lambda) 
		&=&
			-(p'-p)\frac{4\a}{3s^2}(4\g-3\b) \ln\Lambda\ .
\ea
Here, contributions up to three-loop order must be taken into account (although the 2- and 3-loop-contributions turn out to vanish). Now the $\b$-functions follow from (\ref{EQNCOUNTER}), the canonical dimensions of $a$ and $u$ as well as the vacum energy for free fields :
\be\label{EQNBETA}
	\begin{matrix}
	\b_s &=& -96\frac{\a^2}{s^2} &\qquad \b_a &=& -a-\frac{u}{s} \cr
	\b_u &=& -u-64\frac{u\a^2}{s^3} &\qquad \b_\a &=& -\frac{4}{3}\frac{\a}{s^2}(4\g-3\b)\ .
	\end{matrix}
\ee  
The non-local couplings do not contribute a counter-term for the cosmological constant. Therefore the $\b$-function for $a$ is not modified by $\l$ and takes its usual form.

All these terms are one-loop contributions and therefore scheme independent. The two- and three-loop contribution to these $\beta$-functions vanish.
After absorption of the coupling $s$ into the field normalization and setting $\a=\frac{\l}{2}, \g=\b=\frac{\l^2}{2}$, the $\b$-functions become
\ba
	\b_{a} &=& -a-u \\
	\b_{u} &=& -u (1- 8{\l}^2) \\
	\b_{\l} &=& -\frac{47}{6}\l^3 \ .
\ea
From these equations we can draw the following conclusions:

1. The coupling $\l$ for the non-local interaction increases under the renormalisation group flow. This should not be taken as an indication that the curvature of the bulk background increases since the bulk theory, which is decoupled, is always on-shell. This coupling should rather be interpreted as an `auxiliary' coupling which mimics the effect of the closed string background on the open string dynamics. 

2. Tachyon condensation {\it inevitably} takes place. As we have seen above the tachyon is non-zero from the beginning due to the contribution of the measure. Furthermore, even if it were set to zero by an appropriate choice of the measure, a tachyon would be generated due to the one-loop counter-term. 

3. The running of $u$ is modified in a curved background. The way how $\l$ enters in $\b_u$ indicates that the tachyon flow in this example has a richer structure than in flat space.

At this point one could wonder about the end-point of the condensation. At perturbative level and with a finite set of couplings taken into account it is not possible to make definite predictions. The obtained $\b$-functions suggest that condensation to lower-dimensional branes can take place, in the same way as in flat space tachyon condensation. An infinite $u$ forces $f^\m\bar f_\m$ to zero, so that the resulting model will describe a D0-brane. The existence of a D0-brane is expected because it also arises in the WZW model (and is therefore compatible with the symmetries of the space). However, it is also possible identify a condensation process towards a higher-dimensional brane as endpoint. We present evidence for this in the next section.

\section{Tachyon condensation on the 3-brane}

The $\b$-functions (\ref{EQNBETA}) exhibit a complicated RG pattern, from which information about possible endpoints of the flow can be deduced. The trivial conformal point is $a=0, u=0, \l=0$, which is just the free boson theory without tachyon. Another well-known fixed point is obtained through tachyon condensation at $a=u=\infty$ (the Zamolodchikov metric vanishes at this point). We want to argue that there is another fixed point, which corresponds geometrically to a 2-brane. This must be expected from the study of D-branes in the WZW model \cite{Alekseev:1998mc, Klimcik:1996hp}.

In order to arrive at this conclusion, it is helpful to consider the boundary action with a tachyon insertion given by
\be
	\oint \beta(X_b) =	\oint \r(X_b^2-c^2)^2\ .
\ee 
In the case of finite $c$, condensation of $\r$ will lead to a localisation on a spherical submanifold. Expansion up to third order in the fields yields an interaction term
\be
	- 4\r c^2 f\bar f + \r c^4\ ,
\ee
from which an identification with couplings $u$ and $a$ can be obtained:
\be
	\r = \frac{1}{16}\frac{u^2}{a} \qquad   c^2 = -4\frac{a}{u}\ .
\ee
The corresponding $\b$-functions are then given by
\be\label{EQNBETAC}
	\b_{c^2} = 
		4 - 8 c^2\l^2
\ee
\be\label{EQNBETARHO}
	\b_\r = 
		-\frac{\r}{c^2}\left( c^2- 16c^2{\l}^2 +4 \right)\ .
\ee

A close look at (\ref{EQNBETAC}) shows that the presence of curvature, parametrised by $\l$, has a stabilising effect on the radius $c^2$. 
The corresponding $\b$-function vanishes for
\be\label{EQNVANISH}
	c^2 = \frac{1}{2} \l^{-2} =  \frac{1}{2} \kappa\ .
\ee 
Indeed, a spherical 2-brane would be characterised by a finite radius $c$ proportional to $\sqrt\kappa=\l^{-1}$, which is expected from the WZW model. Thus the deformation of the flat background prevents the spherical 2-brane from collapsing. But the condition for vanishing $\b_{c^2}$ still depends on $\l$, which itself is driven by its RG flow and increases.

\begin{figure}[htbp]
   \centering
   \includegraphics[width=40ex]{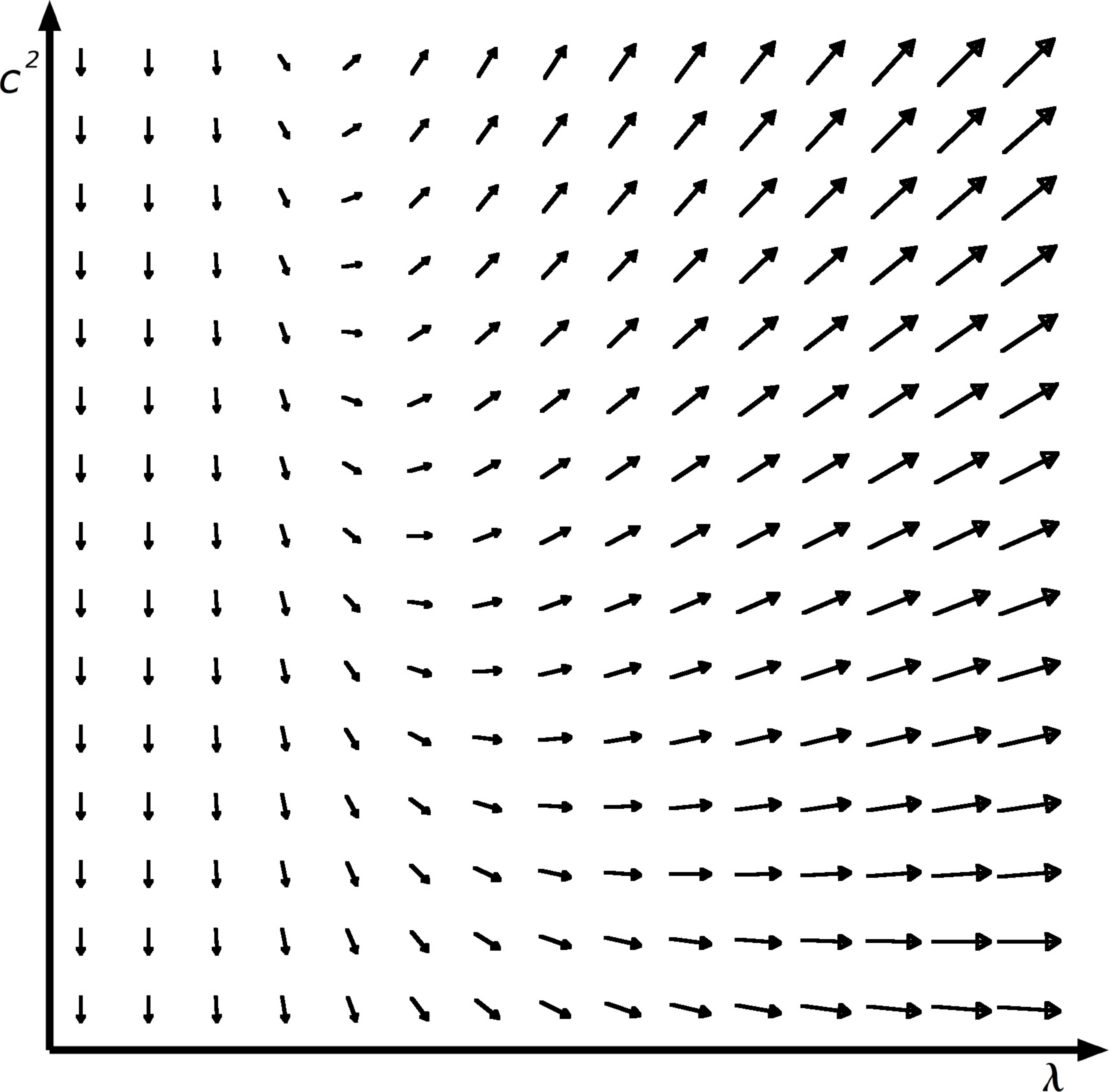} 
   \caption{\small The projection of the flow diagram to the $(\lambda,c^{2})$-plane in the parameter region where condensation to a D2-brane starts. Only the presence of non-vanishing $\l$ enables a flow towards finite $c$. The flow can only be trusted for small $\l$.}
   \label{fig:lcflows}
\end{figure}

For small $\lambda$, where this approximation is valid, $\beta_\rho$ is negative and stays negative after substituting (\ref{EQNVANISH}). Therefore the coupling $\r$ will increase and trigger a tachyon condensation process.

The perturbative $\beta$-functions on the 3-brane suggest that $c^2 = \frac{1}{2} \l^{-2}$ is not the endpoint of the flow. However, while $\l$ evolves along its RG trajectory, $\r$ increases at a much higher rate. Large $\rho$ on the other hand suggests that the perturbative treatment of the flow on the 3-brane is not applicable any more. Rather than trying to follow the flow all along the RG trajectory it is more reasonable to investigate the conjectured endpoint, a spherical 2-brane.

Note that the original couplings $a$ and $u$ both become infinite, so that the condensation process looks quite like usual tachyon condensation with a configuration with vanishing Zamolodchikov metric as endpoint. But the transformation into $(\r,c^2)$-coordinates reveals, that this condensation is not quite as simple, due to the presence of the non-local coupling $\l$. The submanifold given by this equation is lower-dimensional but curved, a phenomenon which is impossible to observe when tachyon condensation in the presence of only local couplings is studied.

The major drawback of the derivation presented here is the treatment within perturbation theory while only retaining a finite set of couplings. Exact results are out of reach with this method, and including higher perturbative corrections in the calculations significantly increases the necessary efforts. In order to substantiate the above results we will present another check for the conjectured end point of the renormalisation group flow and the existence of a spherical 2-brane with the same qualitative properties in the next section.

\section{Stability of the two-brane}

Motivated by the results of the previous section we want to check, if a two-brane is perturbatively stable. We start with a boundary path-integral localised on $X_b^2=c^2$. This constraint is compatible with the group symmetries of $SU(2)$ and describes spherical 2-branes, which are known to be stable. In order to insert this constraint into the action we expand it in the fields $f$ and $\bar f$ to lowest order in $\l$. Moreover we assume that $f^3$ is of the order of the radius $c$ of the 2-sphere, and the other coordinates are small compared to $c$. Explicitly,
\be\label{EQNFIII}
	f^3_n = - \frac{1}{2c} \left[f^\a f_\a + 2 f^\a\bar f_\a\right]_n,  
\ee
where the rhs is projected on the (positive) momentum $n$, and the index $\a$ runs over the directions $\left\{1,2\right\}$. 

Substituting (\ref{EQNFIII}) into the action generates several new vertices. In particular the 3-vertex is removed and the 4-vertex shows a much more complicated structure. The interaction consists of several terms, proportional to different combinations of $\l$ and $c$. It is of the form
\be\label{EQNIIBRANE}
	\oint\beta(X_b) = \frac{\l}{c}A + \frac{s}{c^2}B + \l^2 C\ .
\ee
The three types of interactions, distinguished through their dependence of combinations of $c$ and $\l$, are given by
\ba
	A &=& \sum_i A_i - \sum_i \bar A_i\\
	B &=& B_1+B_2+B_3+\bar B_3\\
	C &=& C_1+C_2+\bar C_2
\ea
with
\eqn{
	A_1 &= \frac{1}{2}\sum_{a=1}\sum_{d=2}\sum_{e=1}^{d-1}(d-a) \epsilon_{\a\b}f^\a_a\bar f^\b_{a+d}{f_e}_\g f_{d-e}^\g 
		\\
	\bar A_1 &= A_1^*\\
	A_2 &= \sum_{a=1}\sum_{c=1}\sum_{d=2}(d-a)\epsilon_{\a\b}f^\a_a\bar f^\b_{a+d}{f_{c+d}}_\g\bar f^\g_c
		\\
	\bar A_2 &= A_2^* \\
	A_3 &= \frac{1}{4}\sum_{a=1}\sum_{b=1}\sum_{g=1}^{a+b-1}(a-b)\epsilon_{\a\b}f_a^\a f_b^\b \bar f_{g\g}\bar f^\g_{a+b-g}
		\\
	\bar A_3 &= A_3^* \\
	A_4 &= \frac{1}{2}\sum_{a=1}\sum_{b=1}\sum_{c=1}(a-b)\epsilon_{\a\b} f^\a_a f^\b_b f_{c\g}\bar f_{a+b+c}^\g
		\\
	\bar A_4 &= A_4^* \\
	B_1 &= \frac{1}{4}\sum_{m=2}\sum_{a=1}^{m-1}\sum_{b=1}^{m-1} m f_{a\a}f^\a_{m-a}\bar f_{b\b}\bar f^\b_{m-b}
		\\
	B_2 &= \sum_{m=2}\sum_{a=1}^{m-1}\sum_{b=1}^{m-1} m \bar f_{a\a} f^\a_{m-a} f_{b\b}\bar f^\b_{m-b}
		\\
	B_3 &= \frac{1}{2}\sum_{m=2}\sum_{a=1}^{m-1}\sum_{b=1}^{m-1} m f_{a\a}f^\a_{m-a} f_{b\b}\bar f^\b_{m-b}
		\\
	C_1 &= -\sum_{a,b,c,d=1}\frac{(a-b)(c-d)}{c+d} f^\a_a\bar f_{c\a} f^\b_b\bar f_{d\b} \d_{a+b,c+d} \\
	C_2 &= \frac{1}{3}\sum_{a,b,c,d=1}(c-a-b)\bar f^\a_a\bar f_{b\a}\bar f_{c\b} f^\b_{a+b+c}\\
	\bar C_2 &= C_2^*\ .
}
A contribution from the 3d measure has been not included, since after imposing the constraint it would appear as a cosmological constant. For the $\b$-functions this is taken into account anyway as part of the tachyon couplings.

We want to check, if this action leads to a conformal fixed point describing a  2-brane. Due to the complexity of the generated non-local interaction terms, it is hard to show conformal invariance directly. Therefore we restrict ourselves to the investigation of tachyonic instabilities. For this we need to show that no tachyon is present, neither due to the measure nor due to counter-terms arising at the quantum level. A tachyon would destabilise the 2-brane and initiate a further condensation.

Such information is contained in the various counter-terms appearing in the renormalisation procedure. Furthermore the various $\b$-functions should vanish for the theory to be scale-invariant. For this one needs to know the logarithmically divergent counter-terms. More-than-logarithmic divergences tell us, if certain couplings can be set identically to zero in a consistent way. 

For example, we might set a certain coupling $g$ to zero (in an adequate theory). Renormalisation then might make it necessary to add a counter-term which excites the coupling $g$. Still, it could be possible to set the renormalised coupling $g_{\rm ren}$ to zero as a renormalisation condition. This is then an arbitrary choice and cannot have much physical meaning; it should rather be viewed as a kind of fine tuning of the theory. However, if all counter-terms  vanish, $g=0$ is a solution of the string field theory action.

This is exactly the situation we encounter in our theory (\ref{EQNIIBRANE}). Due to the complexity of the 4-vertex, the calculation could be done only for vanishing tachyon. However, this is enough to see if tachyonic modes destabilise the 2-brane. According to general scaling arguments, the $\b$-function for $u$ is always proportional to $u$.\footnote{As the calculations are done in the limit $u\to 0$ it is impossible to obtain an expression for $\b_u$. This limit involves some care in the regularisation of the theory. In particular, the appearance of the correct combinations of $u$ and $R$ (the radius in the disk, which has been set to 1) must be restored in order see the behaviour of $\b_u$. The scaling then forces the logarithmic divergences to be proportional to $u$.} 
Hence setting $u=0$ makes its $\b$-function vanish. In order to decide if this condition is 
just fine tuning or has physical relevance,
we need to know the counter-terms. For the 2-point function at vanishing external momentum, they are\footnote{The counter-terms have been calculated in the same way as in the previous section. Again we find, that the free field normal ordering prescription can be consistently implemented and is therefore justified a posteriori.}
\ba\label{EQNIIDDIV}
\nonumber
	&&\Sigma^{(2)}(p=0,\Lambda)	
		=\Bigl [\Lambda\ln\Lambda+(\g-1)\Lambda\Bigr ]
			\left\{\frac{4\l^2}{c^2s^3} -\frac{8}{3}\frac{\l^2}{s^2c^2} - \frac{38}{sc^4}\right\}\\
		&&\;\qquad\qquad\qquad+\ln\Lambda\left\{{\rm terms\;proportional\; to\;}u\right\} \ .
\ea

Most remarkably the more than logarithmically divergent counter-terms are not independent of each other. They arrange themselves in a way so that they all appear with the same factor. Therefore it is possible to remove them altogether by imposing one single condition, adjusting the value of the radius $c$. Setting for example $s=1$ gives
\be\label{EQNLOC}
	c^2 = \frac{57}{2}\l^{-2} = \frac{57}{2}\kappa\ .
\ee
Of course, the numerical factor is still modified by wave function renormalisation, which has not been taken into account here.

The logarithmic part of (\ref{EQNIIDDIV}) contains the 2-loop contribution for $\beta_u$. To prove conformal invariance at the 2-loop level one ought to establish the absence of counter-terms for the other couplings as well, which we have not obtained here. Rather we want to stress the absence of higher-than-logarithmic divergences in the counter-terms after imposition of localisation to (\ref{EQNLOC}) as a check for the claim on the end point of RG-flow of the decaying $3$-brane.

It is tempting to view the RG-behaviour of our model as realisation of 't Hooft's naturalness principle, albeit in a different context than confining gauge theories, for which it was originally formulated \cite{tHooft:1980xb}. Natural theories do not need fine-tuning of the couplings in order to cancel counter-terms; therefore, small parameters stay small under a change of scale, which is a property shared by our model. The physical picture behind is, that small couplings are preferred, when their vanishing increases a symmetry. One could speculate about symmetry enhancement in the above model. Reversing the argument would imply that some symmetry exists which fixes $c^2$ to a certain value. This is reminiscent of the quantisation of radii of D-branes in $SU(2)$ and nourishes hope that higher order perturbation theory could reveal a D-brane potential capable of describing localisation on quantised D-branes.

\part{Closed string deformations in open topological string theory}

%
%
%

\chapter{Strings on Calabi-Yau spaces}

\section{${\cal N}=(2,2)$ SCFT}

It is generally believed that classical solutions of string theory are described by two-dimensional conformal field theories. A special case are superconformal theories, which provide a way to build theories which are supersymmetric in the target space. For the case of string theories compactified on Calabi-Yau spaces, conformal theories with two conserved supercharges in the left- and right-running sector are of special interest \cite{Aspinwall:1994ay, Greene:1996cy, Schwimmer:1986mf}.

Such a ${\cal N}=(2,2)$ theory contains a supermultiplet which is formed of the energy momentum tensor $T$ of weight 2 as top component, the supercurrects $G^\pm$ of weight $\frac{3}{2}$ and the $U(1)$-current $J$ with weight 1 (see e.g.\ \cite{DiFrancesco:1997nk}). The OPE is given by
\eqn{
	T(z)T(w) &\sim \frac{c/2}{(z-w)^4} + \frac{2}{(z-w)^2}T(w) + \frac{1}{z-w}\P_wT(w)\\
}
\eqn{
	T(z)G^\pm(w) &\sim \frac{\frac{3}{2}}{(z-w)^2}G^\pm(w) + \frac{1}{z-w}\P_wG^\pm(w)\\
}
\eqn{
	T(z)J(w) &\sim \frac{1}{(z-w)^2}J(w) + \frac{1}{z-w}\P_wJ(w)\\
}
\eqn{
	G^+(z)G^-(w)&\sim\frac{\frac{2}{3}c}{(z-w)^3} + \frac{2}{(z-w)^2}J(w) + \frac{2T(w)+\P_w J(w)}{z-w}\\
}
\eqn{
	J(z)G^\pm(w) &\sim \pm\frac{1}{z-w}G^\pm(w)\\
}
\eqn{
	J(z)J(w)&\sim \frac{c/3}{(z-w)^2}
}
with central charge
\eqn{
	c=\frac{3k}{k+2}
}
expressed in dependence of the level $k$.
Analogous expressions for the superconformal algebra can be written down for the anti-holomorphic sector.

The primary fields in the NS sector are determined as eigenstates of the operators $L_0$ and $J_0$ by
\eqn{
	L_0|\phi\> &= h|\phi\>\\
	J_0|\phi\> &= q|\phi\>\ .
}
The positive modes of all other operators annihilate these states. Thus a primary state carries a definite weight $h$ as well as a $U(1)$-charge $q$. 

In the Ramond sector $G^\pm$ has a zero mode. This makes an additional condition necessary,
\eqn{
	G^\pm_0|\phi\>_R = 0\ .
}

\paragraph*{Spectral flow}

On the level of the superconformal algebra it is well known that the Ramond sector and the Neuve-Schwarz sector can be connected by spectral flow. To see this one notes that the mode expansions of the fields in the current multiplet can be modified with a continuous parameter $\l$ in the following way:
\eqn{
	L_n &\to L_n + \l J_n + \frac{c}{6}\l^2\d_{n}\\
	G_r^\pm &\to G_{r\pm\l}^\pm\\
	J_n&\to J_n+\frac{c}{3}\l\d_n\ .
}
The NS sector is obtained for $\l=0$, whereas the R sector is given by $\l=\frac{1}{2}$.

The important point is now, that there is a spectral flow operator $U_\l$ which connects both sectors.
Given a field $|\phi\>$ with weight $h$ and charge $\phi$, then
\eqn{
	|\phi'\>=U_\l|\phi\>
}
has
\eqn{
	h'&=h-\l\phi+\frac{c}{6}\l^2\\
	q'&=q-\frac{c}{3}\l\ .
}
Since, after GSO projections, the R sector describes fermionic fields and the NS sector describes bosonic field with respect to target space supersymmetry, the spectral flow operator acquires an interpretation as target space supersymmetry operator.

\paragraph*{Chirality}

The space of states of the ${\cal N}=(2,2)$ theory contains an important subspace, which is formed of chiral and anti-chiral fields. Chirality is defined here by the action of the operators $G^\pm_{-\frac{1}{2}}$, so that they satisfy
\eqn{
	G^+_{-\frac{1}{2}} |\text{chiral}\> = 0
	\qquad
	G^-_{-\frac{1}{2}} |\text{anti-chiral}\> = 0\ .
}
The same construction applies to the anti-holomorphic sector.

An important property of chiral fields is that there is a connection between their weight and their charge,
\eqn{
	h=\frac{1}{2}q\ .
}
Moreover it is consistent to truncate to a subspace consisting only of chiral fields, i.e. these fields form a ring, the so-called chiral ring. Usually this ring has only finitely many fields, which is a property that becomes relevant for example in minimal models and in topologically twisted models.

\section{Gepner model}

Representations of the superconformal algebra with central charge $c\le 3$ are called minimal models, because the unitarity constraints
\eqn{
	L_n^\dagger=L_{-n}\qquad J_n^\dagger=J_{-n}\qquad (G^\pm_r)^\dagger=G^\mp_{-r}
}
select a series of discrete values for the triples $(c,h,q)$, which specify the highest weight representations. They are labelled by integers $(l,m,s)$ and are given by
\eqn{
	h &= \frac{l(l+2)-m^2}{4(k+2)} + \frac{s^2}{8}\;\text{ mod } 1\\
	q &= \frac{m}{k+2}-\frac{s}{2}\;\text{ mod } 1\ ,
}
where $l=0\dots k$, $m=-k\dots k+2$ and $s=-1\dots 2$, subject to $l+m+s\in 2{\mathbb Z}$.
In this notation, the fields with even $s$ are NS fields, while those with odd $s$ live in the R sector. Also the fields are identified under the equivalence relation
\eqn{
	(l,m,s)\sim(k-l,m+k+2,s+2)\ .
}
These representations can be organised according to the ADE classification \cite{Eguchi:1996ds}
\eqn{
	A_k &:\; k\ge 1\\
	D_{2j+2}&:\; k=4j, j\ge 1\\
	D_{2j+1}&:\; k=4j-2, j\ge 2\\
	E_6&:\;k=10\\
	E_7&:\;k=18\\
	E_8&:\;k=28\ .
}
These models are equipped with symmetries $g$ and $h$, acting as
\eqn{
	g\phi^l_{m,s} &= e^{2\pi\frac{m}{n}}\phi^l_{m,s}\\
	h\phi^l_{m,s} &= (-1)^s\phi^l_{m,s}\ ,
}
where $n=k+2$ for $A_k, D_{2j+1}$ and $E_6$, and $n=\frac{k+2}{2}$ for $D_{2j+2}, E_7$ and $E_8$.

The Gepner construction \cite{Gepner:1986wi, Gepner:1987qi, Greene:1988ut, Witten:1993yc} now relies on the idea to split the conformal field theory underlying a string theory into a 4-dimensional part and internal CFT, which is realised as an orbifold of suitable minimal models. Orbifolding does not affect the central charge, so the contributions to $c$ of the minimal model simply add up. The required central charge for the internal CFT is $c=9$, so that Gepner's construction chooses a set of levels $k_i$ so that
\eqn{
	\sum_{i}\frac{3k_i}{k_i+2} = 9\ .
}
The internal CFT is thus described by a suitable tensor product
\eqn{
	\text{CFT}_{k_1}\otimes\dots\otimes\text{CFT}_{k_r}\ .
}
In order to ensure spacetime supersymmetry it is still necessary to employ an orbifold-like construction, due to Gepner, so that in addition modular invariance of the partition function is satisfied.
Following his approach, we introduce the vectors
\eqn{
	\l &= (l_1\dots l_r)\\
	\m &= (s_0,m_1\dots m_r,s_1\dots s_r)\ ,
}
where $s_0=-1,0,1,2$ and $(l_i,m_i,s_i)$ are labels of the CFT at level $k_i$.
Let $\b_0$ be the $2r+1$-dimensional vector which has 1 in each entry, and let $\b_i$ be the $2r+1$-dimensional vector which has 2 at its first and $(r+1+i)$th position and zero otherwise. Then a convenient scalar product can be defined as
\eqn{
	\m \bullet \m' &= -\frac{d}{8}s_0s_0' + \frac{1}{2}\sum_{i=1}^r\left(
		\frac{m_im_i'}{k+2}-\frac{s_is_i'}{2}\right) \\
	2\b_0\bullet\m &= -\frac{d}{s}\frac{s_0}{2}-\sum_{i=1}^r\frac{s_i}{2}+\sum_{i=1}^r\frac{m_i}{k_i+2}\\
	\b_i\bullet\m &= -\frac{d}{2}\frac{s_0}{2}-\frac{s_i}{2}\ .
}
Here $d=D-2$, where $D$ is the number of curled up `internal' dimensions. For the case of Calabi-Yau threefolds $d=2$.

The correct GSO projection is then implemented by projecting on states with
\eqn{\label{GEPNERCOND}
	2\b_0\bullet\m &\in 2{\mathbb Z}+1 \\
	\b_i\bullet\m &\in {\mathbb Z}\ .
}
In order to ensure modular invariance of the partition function twisted sectors must be introduced. For this, we need additional indices
\eqn{
	b_0 &\in\{0,1\dots K-1\}\\
	b_i&\in\{0,1\}\ ,
}
where
\eqn{
	K &= \text{lcm}(4,2k_i+4)\ .
}
The partition function with the sought-after properties is then given by
\eqn{
	{\cal Z}_G^{(r)}(\tau,\bar\tau) &=
	\frac{1}{2}\frac{(\text{Im}\tau)^{-\frac{d}{2}}}{|\eta(q)|^{2d}}
	\sum_{b_0,b_j}\sum_{\l,\m} (-1)^{s_0}
	\chi^\l_\m(q)\chi^\l_{\m+b_0\b_0+b_1\b_1+\dots+b_r\b_r}(\bar q)\ .
}
The sum of $\l$ and $\m$ is subject to the constraint (\ref{GEPNERCOND}).

\section{Landau-Ginzburg description}

It has been pointed out by Gepner \cite{Gepner:1987vz, Gepner:1987qi},
that the massless spectrum of his model is the same as that of a non-linear $\s$-model on a CalabiÐYau manifold given as a hypersurface in a weighted projective space. The connection is most easily established over an intermediate step involving a mapping to a Landau-Ginzberg model furnished with a superpotential in projective space. As next step, the Landau-Ginzburg model can be identified with an according non-linear $\s$-model on a Calabi-Yau manifold.

We will first review the Landau-Ginzburg model.

\subsection{The LG action}

In order to describe $(2,2)$-superspace on the worldsheet, in two dimensions, we need two bosonic
coordinates $(x^0,x^1)$ and four fermionic coordinates
$\theta^\pm,\bar\theta^\pm$ (with $(\theta^\pm)^\dagger =
\bar\theta^\pm$). 
In the standard notation
supercharges and derivatives are given by
\eqn{
  \label{eq:Scharges}
  Q_\pm = \frac{\partial}{\partial\theta^\pm} 
        + i \bar\theta^\pm \frac{\partial}{\partial x^\pm} \: ,
\qquad
  \bar Q_\pm = - \frac{\partial}{\partial\bar\theta^\pm} 
        - i \theta^\pm \frac{\partial}{\partial x^\pm} \: ,
}
and
\eqn{
  \label{eq:Sderiv}
  D_\pm = \frac{\partial}{\partial\theta^\pm} 
        - i \bar\theta^\pm \frac{\partial}{\partial x^\pm} \: ,
\qquad
  \bar D_\pm = - \frac{\partial}{\partial\bar\theta^\pm} 
        + i \theta^\pm \frac{\partial}{\partial x^\pm} \: ,
}
where $x^\pm = x^0\!\pm\! x^1$. They satisfy the
supersymmetry algebra
\eqn{
  \label{eq:Salg}
  \{Q_\pm, \bar Q_\pm\} = - 2i\partial_\pm \: ,
\quad
  \{D_\pm, \bar D_\pm\} =   2i\partial_\pm \: .
}

The Landau-Ginzburg model is described by a chiral and an antichiral
superfield $\Phi$ and $\bar\Phi$, which satisfy $\bar D_\pm \Phi = 0$ and
$D_\pm \bar\Phi = 0$. Their component expansion is given by
\eqn{
  \Phi(y^\pm,\theta^\pm) = 
  \phi(y^\pm) + 
  \theta^+ \psi_+(y^\pm) + 
  \theta^- \psi_-(y^\pm) + 
  \theta^+\theta^- F(y^\pm) \;,
}
where $y^\pm = x^\pm\!\!-\!i\theta^\pm \bar\theta^\pm$. Let us write down the supersymmetry variation by introducing the Grassmann parameters $\epsilon_\pm$ and $\bar\epsilon_\pm$.
A general variation is given by the action of
$\delta = \epsilon_+ Q_- -\epsilon_- Q_+ -
\bar\epsilon_+ \bar Q_- +\bar\epsilon_- \bar Q_+$. On the field they take the form
\begin{eqnarray}
  \label{eq:susyAux}
  \begin{array}{cc}
    \begin{array}{l@{\;=\;}l}
      \delta \phi   &   
      + \epsilon_+ \psi_- - \epsilon_- \psi_+ \;,\\[1mm]
      \delta \psi_+ & 
      + 2 i \bar\epsilon_- \partial_+ \phi + \epsilon_+ F \;,\\[1mm]
      \delta \psi_- & 
      - 2 i \bar\epsilon_+ \partial_- \phi + \epsilon_- F \;,
    \end{array} 
  \hspace{1cm} &
    \begin{array}{l@{\;=\;}l}
      \delta \bar\phi   & 
      - \bar\epsilon_+ \bar\psi_- + \bar\epsilon_- \bar\psi_+ \;, \\[1mm]
      \delta \bar\psi_+ & 
      - 2 i \epsilon_- \partial_+ \bar\phi + \bar\epsilon_+ \bar F \;, \\[1mm]
      \delta \bar\psi_- & 
      + 2 i \epsilon_+ \partial_- \bar\phi + \bar\epsilon_- \bar F \;.
    \end{array}
  \end{array}
\end{eqnarray}
The supersymmetric action is constructed in the following way. It consists of a $D$-term part, which is an integral over a function $K(\Phi,\bar\Phi)$, where all fermionic worldsheet coordinates are integrated out. This term contains the usual kinetic terms in the action as well as information about the spacetime metric. The simplest non-trivial choice for $K$ is 
$K(\Phi,\bar\Phi)=\bar\Phi \Phi$. It is possible to work with this ansatz, as the properties of the model we are interesting in do not depend on the details of the $D$-term.

The second contribution, the $F$-term, is given by an integral over a superpotential $W(\Phi)$. This function is holomorphic, and integration goes only over half of the fermionic worldsheet coordinates:
\eqn{
  \int_{\Sigma} d^2x d^2\theta\, W(\Phi)\bigr|_{\bar\theta^\pm=0} +
  \textrm{c.c.} \;.
}

The worldsheet superpotential fully determines the
topological sector of the bulk theory. Up to total derivatives, the
bulk action can be written as 
\begin{equation}
\label{eq:Sbulk}
\begin{array}{ccc}
  S_\Sigma & = & 
    \displaystyle{ \int_{\Sigma} d^2x \left\{
     - \partial^\mu \bar\phi \partial_\mu \phi + 
     \frac{i}{2} \bar\psi_- ( \stackrel{\leftrightarrow}{\partial_0} +
     \stackrel{\leftrightarrow}{\partial_1} ) \psi_- +
     \frac{i}{2} \bar{\psi}_+ ( \stackrel{\leftrightarrow}{\partial_0} -
     \stackrel{\leftrightarrow}{\partial_1} ) \psi_+ \right. } \\
     & &  \left. \displaystyle{
     - \frac{1}{4} |W'|^2 - 
     \frac{1}{2} W'' \psi_+ \psi_- - 
     \frac{1}{2} \bar W'' \bar\psi_- \bar\psi_+
      } \right\} \: ,
\end{array}
\end{equation}
where the algebraic equation of motion $F= -1/2\,\bar W'(\bar\phi)$
was used.

\subsection{Renormalisation invariants}

In general such a theory is not scale-invariant. However, if we let the theory flow under the renormal\-isation group to a non-trivial fixed point in the infrared, this will yield conformally invariant theory. This has been shown in
\cite{Cecotti:1989jc, Cecotti:1989gv}.

The interesting point is that already at the starting point of the flow, all of the characteristic features of the chiral ring can be read off from the action, since they are completely determined 
by the superpotential W.
The reason for that is, for (quasi-) homogenous superpotentials there there are powerful renormalisation theorems at work, which allow only for a wave function remormalisation of the action, but not a change of couplings that would modify the form of the superpotential \cite{Howe:1989qr, Seiberg:1993vc}.
If we assume that $W(\Phi_i)$ is a 
quasi-homogeneous function, i.e. there exist integers $k_i$ and $d$ with 
$W(\l k_i \Phi_i ) = \l^d W(\Phi_i)$,
then this renormalisation is absorbed by an overall rescaling that in effect leaves the superpotential un- 
changed. This implies that the charge of $\Phi_i$ is $\frac{k_i}{d}$. 

From this it is clear that the effect of renormalisation group flow is solely a change of the $D$-term, i.e.\ only the function $K(\Phi,\bar\Phi)$ changes under renormalisation. The determination of $K$ at the fixed point is a hard and a generally unsolved problem.

A subset of the fields in the spectrum organise themselves in a chiral ring, which is determined only through the superpotential. It can be written as
\eqn{
	H &= \frac{  {\mathbb C}[\Phi_1,\dots,\Phi_n]}{\P_1 W,\dots,\P_n W}\ .
}
The dimension of this ring coincides with the $(c,c)$-ring of the Gepner model. The full correspondence, including the construction of the $(a,c)$-ring by using spectral flow, is explained in detail in
\cite{Martinec:1988zu, Vafa:1988uu, Witten:1997wk, Lerche:1989uy}.
It relies on the following identification of the primaries with the fields
\eqn{
	\Phi^l \equiv \phi^l_{l,0}
}
between the Landau-Ginzburg fields and the fields in the minimal model.

The different representations of the superconformal algebra in the minimal models correspond on the Landau-Ginzburg side to the choice of the superpotential. According to the ADE-classification of singularities in catastrophe theory, the following superpotentials can be identified:
\eqn{
	W_{A_{k+1}} &= \Phi^{k+2}\\
	W_{D_k} &= \Phi_1^{k-1} + \Phi_1\Phi_2^2\\
	W_{E_6} &= \Phi_1^3+\Phi_2^4\\
	W_{E_7} &= \Phi_1^3+\Phi_1\Phi_2^3\\
	W_{E_8} &= \Phi_1^3+\Phi_2^5\ .
}
In particular the central charges on both sides match.

\subsection{Calabi-Yau geometry}

In this section we give an idea of how to establish the correspondence between the Landau-Ginzburg model and the non-linear $\s$-model on a Calabi-Yau space. Following
\cite{Greene:1988ut, Greene:1996cy}
we can consider the path-integral over the Landau-Ginzburg action and, as a first approximation, ignore the $D$-term. Thus the following arguments will be exact for anything which is independent of the K\"ahler-term, which are exactly the elements of the chiral ring, which we are interested in.
  
The path integral representing the partition function of the theory now becomes 
\eqn{
	\int D^n\Phi e^{-\int d^2xd^2\t W(\Phi_1,\dots\Phi_n) + \text{ cc}}\ .
}
To be definite, we make an ansatz for the superpotential
\eqn{
	W &= \sum_{i=1}^n\Phi^{k_i+2}\ ,
}
which is quasi-homogenous in the space ${\mathbb C}P^4(\{w_i\})$, where
\eqn{
	w_i &= \frac{1}{k_i+2}\ .
}
Let us consider a path of field space, in which $\Phi_1\ne0$. Then it is possible to introduce new variables
\eqn{
	\xi^{w1}_1 &= \Phi_1\\
	\xi_i &= \Phi_i\xi_1^{-w_i}\ .
}
Using the quasi-homogeneity of $W$ we find
\eqn{
	W(\Phi_1,\dots,\Phi_n) &= \xi_1\hat W(1,\xi_2,\dots,\xi_n)\ .
}
The path-integral is re-written as
\eqn{
	\int D^n\xi_i\; J\; e^{-\int d^2xd^2\t \xi_1\hat W(1,\xi_2,\dots,\xi_n)}\ .
}
Here a Jacobian $J$ has been included, which comes from the field transformation. It is explicitely given by
\eqn{
	J &= \xi_1^{1-\sum w_i}\ .
}
One notices that the Jacobian is 1 when
\eqn{\label{CYweights}
	\sum w_i = 1\ .
}
In this case the integrand becomes -- formally -- a $\d$-function, which localises on
$\hat W(1,\xi_2,\dots,\xi_n)=0$.
Re-written in original coordinates, this becomes
\eqn{
	W(\Phi_1,\dots,\Phi_n) = 0\, 
}
which defines a variety in weighted projective space ${\mathbb C}P^{n-1}[w_1,\dots,w_n]$. For the case of $n=5$ this describes a Calabi-Yau manifold. One can show that the requirement (\ref{CYweights}) translates into the condition that the first Chern class of $W=0$ vanishes. But this is just the definition of a Calabi-Yau manifold.

We note, though, that the change of variables we have used to simplify 
the path integral is not one-to-one. In fact, upon inspection we see that $\xi_i$ are invariant under the 
transformation 
\eqn{\label{phi-trafo}
	\Phi_i \to e^{2\pi i w_i} \Phi_i\ .
}
Because of this invariance, the model we have constructed lives on $W$ divided by (\ref{phi-trafo}) rather.
Since the charge of $\Phi_i$ is $w_i$, this is precisely the quotient by 
$g_0 = e^{2\pi iJ_0}$, which was required in the Gepner model to obtain a consistent (space-time supersymmetric) string vacuum.

All this supports the idea that the Landau-Ginzburg model is equivalent to a non-linear $\s$-model on a Calabi-Yau. The arguments presented here are rather heuristic. See
\cite{Witten:1993yc, Aspinwall:1994cf} for a further treatment.

\section{Topological string theory}

The chiral ring of the $N=(2,2)$ theory can be regarded as a topological subsector in the sense that the operator product expansions of its finitely many elements do not contain singularities. Concretely this means that the correlators do not depend on the actual position on the worldsheet, which is one of the basic qualifications of a topological theory.
\cite{Witten:1988xj, Witten:1989ig} (see \cite{Neitzke:2004ni} for a review).

In order to arrive at a topological theory a procedure called topological twisting must be conducted. For this one observes that the currents $(T,G^+,G^-,J)$ have operator products of the form
\eqn{
	(G^+)^2 &\sim 0\\
	(G^-)^2 &\sim 0\\
	G^+G^- &\sim T+J\ .
}
The same applies for the multiplet $(\bar T,\bar G^+,\bar G^-,\bar J)$.

For the construction of the topological version of the theory one is tempted e.g. by $(G^+)^2\sim 0$ to declare $G^+$ to a BRST operator and take its cohomology as spectrum of the topological theory. This does not quite work, because $G^+$ has spin $\frac{3}{2}$ instead of 1, which is required for a BRST operator.

As explained in \cite{Witten:1988xj} 
this can be overcome by shifting
\eqn{
	T\to T'=T-\frac{1}{2}\P J\ .
}
The result of this is that the spins of all operators are shifted by their $U(1)$-charge. After this twist, $G^+$ and $J$ have spin 1 and $G^-$ and $T$ have spin 2. Now it does make sense to introduce
\eqn{
	\bar Q_+ &= G^+_0
}
as a BRST operator.

The analogous construction can be done of course with the second odd current. In this case, the BRST operator is
\eqn{
	\bar Q_- &= G^-_0, \
}
while the sign in the shift of the charges is flipped
\eqn{
	T\to T'=T+\frac{1}{2}\P J\ .
}

The choices for the BRST operators can be made independently in the holomorphic and anti-holomorphic sector. This results is two (non-iso\-mor\-phic) versions of the twist, known as A- and B-twist. The A-twist corresponds to the choice of $(G^+,\bar G^+)$, and the B-twist to $(G^+,\bar G^-)$ as BRST operators. Therefore the twisting procedure leaves us with the following possible combinations for the BRST operator:
\eqn{
	\text{A}:&\qquad Q = \bar Q_++ Q_-\\
	\text{B}:&\qquad Q = \bar Q_++\bar Q_-\ .
}

In the transition from the original model to the topologically twisted model the point of most importance is that the chiral ring is preserved and indeed becomes the space of physical fields in the twisted theory. This space is indeed the cohomology of the chosen BRST operator. In all our following considerations only the B-twisted model will be investigated, therefore we will ignore the A-model to a large extent.

How is the topological character of the theory reflected in the properties of the action? The shift in the dimensions of the operators of the superconformal algebra is carried over to the dimensions of the fields. This has the prime effect that the fermions change their scaling behaviour. In particular $\bar\psi_+$ and $\bar \psi_-$ acquire scaling dimension 0, while $\psi_+$ and $\psi_-$ scale with dimension 1. This makes it necessary to adjust the appearance of the worldsheet metric in the action.

Under a re-swcaling of the worldsheet metric $h\to \l^2 h$ the Lagrangian (\ref{eq:Sbulk}) changes according to
\begin{equation}
\begin{array}{ccc}
  S_\Sigma & = & 
    \displaystyle{ \int_{\Sigma} d^2x \left\{
     - \partial^\mu \bar\phi \partial_\mu \phi + 
     \frac{i}{2} \bar\psi_- ( \stackrel{\leftrightarrow}{\partial_0} +
     \stackrel{\leftrightarrow}{\partial_1} ) \psi_- +
     \frac{i}{2} \bar{\psi}_+ ( \stackrel{\leftrightarrow}{\partial_0} -
     \stackrel{\leftrightarrow}{\partial_1} ) \psi_+ \right. } \\
     & &  \left. \displaystyle{
     - \frac{\l^2}{4} |W'|^2 - 
     \frac{1}{2} W'' \psi_+ \psi_- - 
     \frac{\l^2}{2} \bar W'' \bar\psi_- \bar\psi_+
      } \right\} \: .
\end{array}
\end{equation}
By variation with respect to $\l$ one can obtain the worldsheet energy-momen\-tum tensor $T$ directly. It is possible to verify using the superconformal algebra, that $T$ is $Q$-exact. Physically this means that the model is not affected by re-scalings. In this sense it describes a fixed point of renormalization group flow, and the model can be called conformal.

Using this scale-invariance, one can show easily that the path-integral localises on constant fields $x^i$ and on critical points of the superpotential, satisfying $\P_i W=0$ 
\cite{Vafa:1990mu}
In particular one arrives at a very simple formula for the computation of correlation functions. If $F(x^i)$ is any polynomial and $H(x^i)$ is the Hessian det$\P_i\P_j W$, then
\eqn{
\label{eq:bulk-corr}
	\<F(x^i)\>_g &= \int dx^n \frac{ F(x^i)H^{g-1}(x^i)}{\P_1W\P_2W\dots\P_nW}\ ,
}
where $g$ is the genus of the worldsheet. The integration path goes around the critical points of $W$, thus (\ref{eq:bulk-corr}) computes the residues of the integrand. We will later see that a similar formula is valid for bulk-boundary correlators.

%
%
%

\chapter{Matrix Factorisations}

In the previous chapter the correspondence between Gepner models and Landau-Ginzburg models has been explained. For all these considerations it has been assumed that only the closed string sector is described, i.e. we are talking about closed worldsheets. As soon as open strings are taken into account, an additional boundary sector of the theory appears. From the CFT point of view it is in principle clear what happens, because the general discussion of chapter \ref{ch-cft} applies here, too. The challenge is thus to understand the description of the boundary sector on the Landau-Ginzburg side.
From the CFT point of view this has been achieved in \cite{Recknagel:1997sb}, where boundary states in Gepner models have been constructed (see also \cite{Recknagel:1998ih, Recknagel:2000ri, Recknagel:2002qq})

An introduction of a boundary break translation invariance at the boundary, thus only half of the supersymmetries can be preserved \cite{Warner:1995ay}. In this case one has two possible choices for the supercharges (see eg \cite{Brunner:2003dc}):
\eqn{
	\text{A}:&\qquad Q = \bar Q_++ Q_-\\
	\text{B}:&\qquad Q = \bar Q_++\bar Q_-\ .
}
Depending on which of these charges one wants to preserve, the boundary spectrum changes. The first choice describes the so-called A-sector with A-branes as boundary states, while the second choice describes the B-sector.

Our main interest will lay in the investigation of the B-sector, thus we restrict ourselves to the choice $Q = \bar Q_++\bar Q_-$. For the supersymmetry variations this means that
\eqn{
	\epsilon \equiv \epsilon_++\epsilon_-\ .
}
It makes sense to introduce adequate combinations of the fermions
\eqn{
	\eta &= \psi_-+\psi_+
}
and
\eqn{
	\t &= \psi_--\psi_+\ .
}
Then the B-supersymmetry variation
\eqn{
	\d = \epsilon\bar Q-\bar\epsilon Q
}
acts like
\begin{eqnarray}
  \begin{array}{cc}
    \begin{array}{l@{\;=\;}l}
      \d \phi  &   \epsilon\eta\;,\\[1mm]
      \d\eta & -2i\bar\epsilon\P_0\phi \;,\\[1mm]
      \d\t & 2i\bar\epsilon\P_1\phi + \e\bar W' \;,
    \end{array} 
  \hspace{1cm} &
    \begin{array}{l@{\;=\;}l}
      \d \bar\phi   &  -\bar\epsilon\bar\eta\;, \\[1mm]
      \d\bar\eta & 2i\epsilon\P_0\bar\phi \;, \\[1mm]
     \d\bar\t & -2i\epsilon\P_1\bar\phi+\bar\epsilon W' \;.
    \end{array}
  \end{array}
\end{eqnarray}
where the auxiliary fields have already been replaced by their equation of motion. For 0-direction on the worldsheet is the tangential coordinate on the strip, the 1-direction is the normal coordinate. In this notation, and with setting $\t^{0,1}=\frac{1}{2}(\t^-\pm\t^+)$, the B-supercharge has the explicit boundary contribution
\eqn{
	\bar q &= \P_{\t^0}+i\bar\t^0\P_0 \qquad q 
	= -\P_{{\bar\t}^0}-i\t^0\P_0\ ,
}
so that
\eqn{
	Q &= Q^{\rm bulk} + q\ .
}
The superfields at the boundary can be constructed as
\eqn{
	\Phi(y^0,\t^0) &= \phi(y^0) + \t^0\eta(y^0) \\
	\Theta(y^0,\t^0,\bar\t^0) &= \t(y^0) 2-\t^0F(y^0) + 2i\bar\t^0\P_1\phi(y^0) - 2i\t^0\bar\t^0\P_1\eta(y^0)\ ,
}
where $y^0=x^0-i\t^0\bar\t^0$.
These fields are not chiral, but they satisfy the equation $D\Theta = -2i\P_1\Phi$ at the boundary.

\section{Warner problem}

The Warner problem deals with the issue of how to correctly implement B-type boundary conditions in $N=(2,2)$ models. An explicit conduction of the variation shows that there will be a surface contribution from the $D$-term of the Landau-Ginzburg Lagrangian. This surface term can be cancelled by addition of a local boundary actions, which has for $K=\Phi\bar\Phi$ the form
\eqn{\label{WARNERBDRY}
	\frac{i}{4}\int dx^0 \left(\bar\t\eta - \bar\eta\t\right)\ .
}
Variation of the $F$-term vanishes only for constant $W$. In case $W$ is not constant, it contributes
\eqn{\label{WARNERBDRY1}
	\frac{i}{2}\int dx^0\left(
		\epsilon\bar\eta\bar W' + \bar\epsilon\eta W'\right)\ .
}
This term cannot be made vanish by adding a boundary interaction in the fashion of (\ref{WARNERBDRY}), unless one imposes D0-boundary conditions. But from the study of the spectrum of boundary states in the Gepner model one knows that there are much more states present. In \cite{Warner:1995ay} a way for the description of the correct boundary conditions has been pointed out. It relies on the idea that additional boundary degrees of freedom much be incorporated. Concretely it turned out to be necessary to add boundary fermions and enlarge the space of boundary fields that way (see also \cite{Witten:1998cd}).

In order to achieve that we add a fermionic superfield $\Pi$ at the boundary, which does not satisfy a chirality condition, but rather
\eqn{
	D\Pi=E(\Phi)\ ,
}
where $E(\Phi)$ is a polynomial in $\Phi$. The components of the superfield are given by
\eqn{
	\Pi(y^0,\t^0,\bar\t^0) &= \pi(y^0)+\t^0l(y^0)-\bar\t^0E(\phi)+\t^0\bar\t^0\eta(y^0)E'(\phi)\ .
}
The component fields are subject to the supersymmetry variations
\eqn{
	\d\pi &= \epsilon l - \bar\epsilon E(\phi) \\
	\d\bar\pi &= \bar\epsilon \bar l-\epsilon \bar E(\bar\phi)\\
	\d l &= -2i\bar\epsilon \P_0\pi + \bar\epsilon \eta E'(\phi)\\
	\d\bar l &= -2i\epsilon \P_0\bar\pi-\epsilon\bar\eta\bar E'(\bar\phi)\ .
}

With these preparations, the simplest way to remove the surface term (\ref{WARNERBDRY1}) is to add
\eqn{\label{EQNLAGI}
	\int d^t \{
		&i\bar\pi\P_\t\pi + \frac{i}{2} \pi\eta^\a\P_\a J + \frac{i}{2}\bar\pi\bar\eta^\a\bP_\a\bar J \\
				&- \frac{1}{2}\bar \pi\eta^\a\P_\a E + \frac{1}{2}\pi\bar\eta^\a\bP_\a\bar E 
		-\frac{1}{2}|J|^2 - \frac{1}{2}|E|^2
		\}
}
to the action. Once the auxiliary fields $l$ and $\bar l$ have been integrated out, $\pi$ and $\bar\pi$ have variations 
\eqn{
	\delta\pi = -i\epsilon \bar J - \bar\epsilon E \qquad \delta\bar\pi = i\bar\epsilon J - \epsilon\bar E
}
$J$ and $E$ are polynomials in $x$. The requirement of $B$-type supersymmetry for the full Lagrangian places a constraint on the boundary potentials $E$ and $J$. The condition is
\eqn{
	E(x) J(x) = -i W(x).
}
This constraint is rather subtle because it allows to establish a connection between Landau-Ginzburg models and the category of matrix factorizations, which in itself is known to be equivalent to the category of D-branes in the B-model 
\cite{KontsevitchUP, Orlov:2003yp}.
The correspondence can be found by quantizing the fermions via
\eqn{
	\left\{\pi,\bar\pi\right\}=1
}
and finding a representation of the action of $Q$. For this it is necessary to split the full B-charge $Q = Q_\text{bulk} + q$ into a bulk part $Q_\text{bulk}$ and a 'boundary part' $q$, which acts on $\pi$ and $\bar\pi$ only. By Noether procedure it can be obtained as \cite{Hori:2000ic} 
\eqn{
	q &= -i\Bigl [\pi J + i\bar\pi E\Bigr ]_0^\pi.
}
By using the anticommutation relations it can be verified that
\eqn{
	\{q,\pi\} &= E \qquad \{q,\bar\pi\} = -iJ.
}
The quantised fermions $\pi$ and $\bar\pi$ satisfy a Clifford algebra and have a representation through matrices. When we fix a basis by requiring that the fermion grading is measured by $\s_3$, we can identify
\eqn{
	\pi \sim \begin{pmatrix}0 & 1\\ 0 & 0\end{pmatrix} \qquad \bar \pi \sim \begin{pmatrix}0 & 0\\ 1 & 0\end{pmatrix}
}
and represent $q$ as
\eqn{
	q \sim \begin{pmatrix}0 & -iJ\\ E & 0\end{pmatrix}.
}
One verifies immediately that $q^2 = -iW1\!\!\;\!\!1_2$, as required above. In this approach finding an admissible $q$ corresponds to finding a matrix factorisation of the superpotential.

Once the matrix representation of $q$ is established the boundary action can be re-written as a super-Wilson-line. In this form generalisations of $q$ can be found by allowing higher-dimensional matrices, still obeying $q^2 = -iW1\!\!\;\!\!1_2$. 
Alternatively it is possible to introduce more boundary fermions and treat the problem on the Lagrangian level. For example, an obvious generalisation is obtained by just blowing up the boundary Lagrangian (\ref{EQNLAGI}) by adding indices to $\pi$ and $\bar\pi$. This can only account for those matrix factorisations, which can be written as graded tensor products\footnote{for more than two pairs of fermions, this procedure can be applied recursively.}
\eqn{
	q = q_1 \odot q_2 = \begin{pmatrix}
			& & J_2 & J_1 \\
			& & -E_1 & E_2 \\
			E_2 & -J_1 & & \\
			E_1 & J_2 & & 
			\end{pmatrix}.
}
For arbitrary matrix factorisation we expect a more complicated Lagrangian, containing interaction terms between the boundary fermions. In particular we will show that non-linear terms in the fermions show up in $q$ (and therefore also in the variations of $\pi$ and $\bar\pi$) as well as in the Lagrangian. In the next section a method is presented for the reconstruction of the action out of a given boundary B-charge. In other words, we construct B-supersymmetric boundary action with non-linear supersymmetry transformations in the boundary fermions.

\subsection{Reconstruction of the boundary action}

Going over to more general boundary interaction terms requires some more notation. 
As we are dealing with an arbitrary number of boundary fermions,
the quantisation condition becomes
\eqn{
	\left\{\pi^i,\bar\pi_{j}\right\} &= \d^i_{j}.
}
Indices are raised and lowered by the constant metric $G^{i\bar j}$. Representing a complex structure, $G$ is given by $G^{ij} = G^{\bar i \bar j} = 0$ and $1$ otherwise.
Normal ordering is defined by placing $\bar\pi$ to the right. Sometimes it is convenient to represent operators as fermionic derivatives via
\eqn{
	\pi^i \to \pi^i \qquad\qquad \bar \pi^{\bar i} = G^{\bar i j}\bar \pi_j \to G^{\bar i j} \P_{\pi^j} = \P_{\pi_{\bar i}}
}

The fermions can be used to build up an exterior calculus, as long as holomorphic and antiholomorphic components are separated. Otherwise contact terms are present. A general normal ordered $(n,m)$-form can be written as
\eqn{
	T_{(n,m)}&=T^{\n_1\dots\n_m}_{\m_1\dots\m_n}\pi^{\m_1}\cdots\pi^{\m_n}\bar \pi_{\n_1}\cdots\bar \pi_{\n_m}
}
A fermionic operator $q$ can be expanded as
\eqn{\label{EQNexp}
	q &= q_{(1,0)} + q_{(0,1)} + q_{(3,0)} + q_{(2,1)} + q_{(1,2)} + q_{(0,3)} + \dots
}
Its basic anticommutators are
\eqn{\label{EQNacomm}
	\left\{q, \pi^\m\right\} &= q^\m + q_{ab}^\m\pi^a\pi^b - 2q_a^{\m c}\pi^a\bar \pi_c  + \dots\cr
	\left\{q, \bar \pi_\m\right\} &= q_\m + q^{bc}_\m\bar \pi_b\bar \pi_c - 2q^c_{a\m}\pi^a\bar \pi_c + \dots\cr
}
where antisymmetry in the (anti-)holomorphic indices has been used. (\ref{EQNacomm}) is exact when only two pairs of boundary fermions are present.

In the following we assume that $q$ is a representation of the B-SUSY operator acting on the boundary fermions. 
The SUSY-operator from the bulk $Q_\text{bulk}$ does by convention not act on the boundary fermions. 
Therefore the full B-SUSY charge is given by
\eqn{
	Q \equiv Q_{\rm bulk} + q.
}

The wanted Lagrangian, must satisfy the following basic properties:
\begin{itemize}
\item It must be $Q$-closed, up to a term $\frac{i}{2}\eta^\m\P_\m W$, which is cancelled by the bulk contribution.
\item It must be real. This implies that it is $Q^\dagger$-closed, up to $-\frac{i}{2}\bar\eta^\m\bP_\m\bar W$.
\item It must contain a term $i\bar\pi^i\P_t\pi_i$. This term is responsible for the quantisation of the fermions, which accomplishes the actual connection to matrix representations.
\end{itemize}

These requirements are enough to re-construct the Lagrangian from the data provided by a matrix factorisation $q$.

For the reconstruction of the boundary action we start with the canonical kinetic term and calculate its $Q$-variation.
Without placing {\it any} restriction on the boundary part of $Q$ we get 
\eqn{
	:\left [Q, \bar \pi_\m\P_t\pi^\m\right ]:
		= :\left\{q,\bar \pi_\m\right\}\P_t\pi^\m: - :\bar \pi_\m\P_t\left\{q,\pi^\m\right\}:,
}
where $:\cdot:$ denotes normal ordering, placing $\bar\pi$ to the right.
Replacing the fermions in the anticommutator by derivatives yields
\eqn{
	:\left [Q, \bar \pi_\m\P_t\pi^\m\right ]:= \left(\P_t\pi^\m \P_{\pi^\m} + \P_t\bar \pi_\m\P_{\bar \pi_\m} \right)q,
}
where a partial integration has been conducted. The operator in the bracket is just a time derivative acting only on the boundary fermions in $q$. After another partial integration we arrive at
\eqn{\label{EQNgen}
	:\left [Q, \bar \pi_\m\P_t\pi^\m\right ]:= -\dot q,
}
where we define
\eqn{
	\dot T_{(n,m)}&\equiv\left(\P_t T^{\n_1\dots\n_m}_{\m_1\dots\m_n}\right)\pi^{\m_1}\cdots\pi^{\m_n}\bar \pi_{\n_1}\cdots\bar \pi_{\n_m}
}
This formula is valid for arbitrary $q$.

If such terms are supposed to appear in $Q$-exact expressions, they must originate in a variation of $\eta$. Moreover one must note that coefficients of $q$ can also depend on the holomorphic spacetime coordinate.
The relevant $Q$-variations are
\eqn{
	\begin{matrix}
		\left[Q, x^\a\right] &=& 0 &\qquad&\left [Q,\bar x^\a\right ] &=&\bar\eta^\a \\
		\left\{Q, \eta^\a\right\} &=& 2i\dot x^\a &\qquad& \left\{Q,\bar \eta^a\right\} &=& 0 \\
		\left[Q^\dagger, x^\a\right] &=& \eta^\a &\qquad&\left [Q^\dagger,\bar x^\a\right ] &=&0 \\
		\left\{Q^\dagger, \eta^\a\right\} &=& 0 &\qquad& \left\{Q^\dagger,\bar \eta^a\right\} &=& 2i\dot{\bar x}^\a. \\
		\end{matrix}
}
For $T$ a general $(n,m)$-form with coefficients in ${\mathbb C}[x]$ 
we can consider $Q$-variations of the form 
\eqn{
	\left [Q, \frac{1}{2}\eta^\a\frac{\P}{\P x^\a} T_{(n,m)}\right ]_\pm = 
		i \dot T_{(n,m)} - \frac{1}{2}\eta^\a \left [q,\frac{\P}{\P x^\a} T_{(n,m)}\right ]_\pm
}
and
\eqn{
	\left [Q, \frac{1}{2}\bar\eta^\a\frac{\P}{\P \bar x^\a} T_{(n,m)}^\dagger\right ]_\pm =
		-\frac{1}{2}\bar\eta^\a \frac{\P}{\P \bar x^\a} \left [q, T_{(n,m)}^\dagger\right ]_\pm.
}
From this we can read off that we must add
\eqn{\label{EQNc}
	\frac{1}{2}\eta^\a\frac{\P}{\P x^\a} q
	-\frac{1}{2}\bar\eta^\a\frac{\P}{\P \bar x^\a} q^\dagger
}
in order to cancel (\ref{EQNgen}).
Due to the action of $Q$ the following terms will appear:
\eqn{\label{EQNc}
	&\left\{Q, \frac{1}{2}\eta^\a\frac{\P}{\P x^\a} q\right\}
		= i\dot q - \frac{1}{2}\eta^\a \left\{q,\frac{\P}{\P x^\a} q\right\}
}
and
\eqn{\label{EQNd}
	\left\{Q, -\frac{1}{2}\bar\eta^\a\frac{\P}{\P \bar x^\a} q^\dagger\right\}
		= \frac{1}{2}\bar\eta^\a \frac{\P}{\P\bar x^\a}\left\{q, q^\dagger\right\}.
}

The second term on the rhs of (\ref{EQNc}) is a wanted contribution, because this term can be used to cancel the surface term of the SUSY-variation in the bulk. The corresponding condition is
\eqn{
	-i\frac{\P}{\P x^\a}W = \left\{q,\frac{\P}{\P x^\a} q\right\} = \frac{1}{2}\frac{\P}{\P x^\a}\left\{q,q\right\},
}
which can be re-written as\footnote{The same condition appears also in the antiholomorphic coordinates $\bar x$, so that the ambiguity is really $\mathbb C$ and not ${\mathbb C}[\bar x]$.}
\eqn{\label{EQNMF}
	-iW = \frac{1}{2}\left\{q,q\right\} + \text{const}.
}
Hence the condition for $q$ being a matrix factorisation appears completely naturally here.

The rhs of (\ref{EQNd}) introduces a coupling between the holomorphic and antiholomorphic fields. In order to cancel it an appropriate bosonic form must be added.
Such a form is generally given by
\eqn{\label{EQNVEXP}
	V \equiv V_{(0)} + V_{(1,1)} + V_{(2,2)} + \dots,
}
where we demand $V^\dagger=V$.
The condition on $V$ is
\eqn{
	\left [Q,V\right ] &= -\frac{1}{2}\bar\eta^\a \frac{\P}{\P\bar x^\a}\left\{q, q^\dagger\right\},
}
which determines
\eqn{
	V = -\frac{1}{2}\left\{q,q^\dagger\right\}.
}
Collecting all terms yields a
Lagrangian
\eqn{\label{EQNACTI}
	L &= i\bar \pi_\m\P_t\pi^\m
			+\frac{1}{2}\eta^\a\frac{\P}{\P x^\a} q
			-\frac{1}{2}\bar\eta^\a\frac{\P}{\P \bar x^\a} q^\dagger
			-\frac{1}{2}\left\{q,q^\dagger\right\},
}
which is together with the bulk Lagrangian $Q$- and $Q^\dagger$-closed by construction.

\section{B-branes}

For the discussion of the boundary spectrum let us focus on the simplest case again, where only one pair of boundary fermions is present.
For only one spacetime direction, an ansatz for the superpotential is
\eqn{
	W(X)=X^{k+2}\ .
}
For the polynomials $J$ and $E$ there are $\left[\frac{k+2}{2}\right]$ choices, corresponding to the factorisations
\eqn{
	W &= X^n \cdot X^{k-n+2}\qquad\qquad 0\le n\le \left[\textstyle\frac{k+2}{2}\right]\ ,
}
because for large $n$ one can just exchange $E$ and $J$. The boundary contribution to the supercharge is then explicitly given by
\eqn{
	q_n &= \begin{pmatrix} 0 & x^n\\x^{k-n+2}&0\end{pmatrix}\ .
}
For the chiral primaries we know that they are annihilated by the supercharge. As they cannot be obtained as variations of other fields, they must lay in the cohomology of the supersymmetry operator. As long as we restrict ourselves to boundary fields, we can work with $q$ alone. Therefore the task is to fimd the cohomology of $q$ in order to determine the spectrum of boundary states.

\subsection{The spectrum between identical branes}

The operator $q$ has the important property that $q^2=W$. That means it is a differential only on the bulk chiral ring, where $W\sim 0$. This justified the introduction of a (twisted) differential operator \cite{Kapustin:2003rc}
\eqn{
	D \Psi &= q\Psi-(-1)^{|\Psi|}\Psi q\ ,
}
where $\Psi$ is any boundary field composed of $x,\bar x$ and $\pi,\bar\pi$, and $|\Psi|$ is its fermion number. The cohomology with respect to this differential is easily obtained. In the fermionic sector we find
\eqn{
	\Psi_l &= \begin{pmatrix} 0 & x^l \\-x^{k+2-2n+l}\end{pmatrix}\qquad l<n
}
and in the bosonic sector
\eqn{
	\Phi_l &= \begin{pmatrix} x^l & 0\\0 & x^l\end{pmatrix}\qquad l<n\ .
}

\subsection{The spectrum between different branes} 

We can associate each matrix factorisation to a particular brane. In case we want to describe a system of two branes, the two factorisations $q$ and $\tilde q$ can be combined into a differential
\eqn{
	D \Psi &= q\Psi-(-1)^{|\Psi|}\Psi \tilde q\ .
}
The calculation of the cohomology works in the same way as above. However, typically not all states from before propagate between the branes, but only a subset of them. To be explicit, we will consider the example
\eqn{
	W = x^nx^{k+2-n} = x^{\tilde n}x^{k+2-\tilde n}\ .
}
The spectrum then turns out to be \cite{Kapustin:2003ga}.
\eqn{
	\Psi_{l} &= \begin{pmatrix} 0 & x^{l} \\-x^{k+2-n-\tilde n+l}\end{pmatrix}
}
and in the bosonic sector
\eqn{
	\Phi_l &= \begin{pmatrix} x^{\tilde n-n+l} & 0\\0 & x^l\end{pmatrix}\ .
}

\section{Connection to CFT}

The space of boundary field is, from the CFT point of view, given by
\eqn{
	{\mathcal H} &= \bigoplus_{[l,m,s]}\left({\mathcal H}_{[l,m,s]}\otimes {\bar {\mathcal H}}_{[l,m,s]}\right)\ ,
}
where the direct sum goes over equivalence classes of $[l,m,s]$. The B-type boundary conditions must satisfy
\eqn{
	\left(L_n-\bar L_{-n}\right)||B\>\!\> &= 0\\
	\left(J_n-\bar J_{-n}\right)||B\>\!\> &= 0\\
	\left(G^\pm_r+i\eta\bar G^\pm_{-r}\right)||B\>\!\> &= 0\ ,
}
where $\eta=\pm 1$ describes the two spin-structures. The Ishibashi states lay in the sectors
$[l,m,s]\otimes[l,-m,-s]$. The B-type boundary states have been constructed in \cite{Maldacena:2001ky}
as
\eqn{
	||L,S\>\!\> &= \sqrt{k+2}\sum_{l+s\in 2{\mathbb Z}}
		\frac{S_{L0S,l0s}}{\sqrt{S_{l0s,000}}}|[l,0,s]\>\!\>\ .
}
Here $L=0,1,\dots,k$ and $S=0,1,2,3$. For even states, $\eta=1$, for odd states $\eta=-1$. Moreover we find that $||L,S\>\!\>=||k-L,S+3\>\!\>$. The operator $||L,S\>\!\>\to||L,S+2\>\!\>$ only reverses the sign of the RR coupling, thus it corresponds to the transitions from a brane to its anti-brane.

In order to identify CFT boundary states with matrix factorisations, it is useful to compare certain invariants on both sides. On the one hand, the number of branes/matrix factorisations for each minimal model match. 
It is also possible to calculate the overlap between two boundary states $||L,S\>\!\>$ and $||\hat L,\hat S\>\!\>$ (see eg \cite{Kapustin:2002bi, Brunner:2005fv}). The number of propagating states between the branes are also found to match the number of states in the cohomology of the associated two matrix factorisation. Therefore the identification in the case of minimal models seems to be clear, given by
\eqn{
	q_n \sim ||n-1,0\>\!\>\ .
}

\subsection{Permutation branes}

One can continue with these checks for models with more dimensions and also for the Gepner models. In this case, the central charges of the minimal models add up and the resulting theory differs in certain aspects from the original theories. In particular new branes can appear in the spectrum. An important class of them are the so-called permutation branes \cite{Recknagel:2002qq}  (see also \cite{Gaberdiel:2002jr}). They mix contributions from the currents-multiplet from the different minimal models (denoted by 1 and 2 here) at the boundary and are characterised by
\eqn{
	\left(L_n^{(1)}-\bar L_{-n}^{(2)}\right)||B\>\!\> =
		\left(L_n^{(2)}-\bar L_{-n}^{(1)}\right)||B\>\!\> &= 0\\
	\left(J_n^{(1)}-\bar J_{-n}^{(2)}\right)||B\>\!\> =
		\left(J_n^{(2)}-\bar J_{-n}^{(1)}\right)||B\>\!\> &= 0\\		
	\left(G_r^{\pm(1)}+i\eta_1\bar G_{-r}^{\pm(2)}\right)||B\>\!\> =
		\left(G_r^{\pm(2)}+i\eta_1\bar G_{-r}^{\pm(1)}\right)||B\>\!\> &= 0
}
with $\eta_1=\eta_2$. The corresponding boundary states
are given by
\eqn{
	&||[L,M,S_1,S_2]\>\!\>\\
	&= \frac{1}{2\sqrt{2}}\sum_{l,m,s_1,s_2}\frac{S_{Ll}}{S_{0l}}e^{i\pi Mm/(k+2)}e^{-i\pi(S_1s_1-S_2s_2)/2}
		||[l,m,s_1]\otimes[l,-m,-s_2]\>\!\>^\s\ ,
}
where the quantum state $||[l,m,s_1]\otimes[l,-m,-s_2]\>\!\>^\s$ denotes the Ishibashi state for the permutation $\s$.
The sum runs over all indices for which
\eqn{
	l+m+s_1 \qquad \text{and}\qquad s_1-s_2\qquad \text{even},
}
so that
\eqn{
	L+M+S_1-S_2\qquad\text{even}.
}
These states are identified under the equivalence relation
\eqn{
	[L,M,S_1,S_2] \sim [k-L, M+k+2, S_1+2, S_2]\ .
}

The identification of these states on the side of the Landau-Ginzburg model has been done in \cite{Brunner:2005fv, Enger:2005jk}. It was found that
\eqn{
	||[L,M,0,0]\>\!\> \leftrightarrow
		J=\prod_{m=\frac{M-L}{2}}^{\frac{M+L}{2}}\left(x_1-\hat\eta_mx_2\right)\ ,
}
where $\hat\eta_m$ denotes the $(k+2)$-th roots of -1. This identification is supported by a matching of the number of states in the corresponding Hilbert space and by the correct symmetry properties. It has also been shown that these permutation factorisations together with graded tensor products of factorisations are the basic building blocks of D-branes in the Gepner model, including the D0 and D2 branes.

\subsection{Geometry of branes}

It is not clear how the geometry of the branes constructed via matrix factorisations can be read-off directly from the factorisation. But there are indirect ways viable, relying on the identification of certain topological invariants. These can be intersection numbers or also bundle data.

Let us consider a concrete example. On the quintic
\eqn{
	W=x_1^5+x_2^5+x_3^5+x_4^5+x_5^5
}
one can find a matrix factorisation
\eqn{
	q=q_1\odot q_2 \odot q_3\odot q_4\ ,
}
determined by
\eqn{
	J_1=x_1\qquad J_2=x_2\qquad J_3=x_3\qquad J_4=x_4-\hat\eta x_5\ ,
}
where $\hat\eta$ is one of the 5th roots of -1.

It is very suggestive to give the common locus $J_i=0$ a geometrical meaning, in particular since this corresponds to a point in projective space. Indeed, an identification of this matrix factorisation with a D0 brane on $W=0$ has been achieved in  \cite{Brunner:2005fv, Enger:2005jk}.

A similar construction can be applied to construct D2 branes. An example for such a case is the factorisation
\eqn{
	q=q_1\odot q_2 \odot q_3\ ,
}
with
\eqn{
	J_1=x_1\qquad J_2=x_2-\hat\eta x_3\qquad J_3=x_4-\hat\eta' x_5\ .
}
Again, the common locus is geometrically a complex line, and the identification is supported by calculations of intersection numbers and charges of the brane.

It seems to be a general property that $\bigcap_i J_i$ gives the geometry of the brane, as long as all $J_i$ are linear in $x_j$ \cite{Aspinwall:2006ib, Orlov:2003yp, Ezhuthachan:2005jr}. When this is not the case, the identification is not clear anymore.

\section{Topological correlators}

The computation of correlators of between elements of the chiral ring can be done completely in the topological theory. This represents a clear simplification of many calculations, since a closed formula is available. Not unlike the case of the closed string theory,
scaling invariance is used to derive a residue-formula for boundary- and bulk-boundary-correlators. Our arguments follow \cite{Kapustin:2003ga} and \cite{Herbst:2004ax}.

Starting in the path-integral formalism, correlators can be written as
\eqn{
	\<{\cal O}\> &= \int DX D\bar X D\Pi D\bar\Pi\;{\cal O}\; e^{-S^{\rm bulk}} e^{-S^{\rm bdry}}\ .
}
Invariance under rescaling of the worldsheet metric $h\to\l^2h$ shows that the path-integral localises like in the bulk case to constant field maps, i.e.\ instantons, and $D=[q,\cdot]_\pm=0$. Only contributions from zero modes survive, as the contributions of the non-zero modes cancel each other. 
For the B-twisted LG theory on a Riemann surface without boundaries, the zero modes come from constant 
scalars $x^i$, $\bar x^i$, $\bar\psi^i_\pm$ and closed 1-forms $\psi^i_\pm$. For the disk topology, there are 
no $\psi^i_\pm$ zero modes. Furthermore, although the boundary condition does not 
affect the bosonic zero modes, it leads to the following relation among the 
fermionic zero modes: 
\eqn{
	\bar\psi^i_- = \bar\psi^i_+ \equiv \bar\psi^i\ .
}
The path-integral now reduces to an ordinary integral where the measure is given by $d^nxd^n\bar x d^n\bar\psi$.

For the boundary field we have already noted that they can be represented as Clifford matrices. It is easy to convince oneself, either by direct calculation or by the arguments presented e.g.\ in \cite{Takayanagi:2000rz}, that the integral over $\pi$ and $\bar\pi$ yields a supertrace in the matrix notation. Path ordering, which usually must be taken into account, does not appear here, because the integral is already reduced to constant modes.

Like in the bulk case there is a localisation on the critical points of $W$ and in addition on $D=0$. Thus $D$ can be expanded around the critical points in the bosonic variables and explicitly expressed in terms of $q$. The resulting integral can be evaluated, as has been demonstrated in \cite{Kapustin:2003ga}. The resulting formula is
\eqn{\label{BBcorr}
	\<\a\phi\>_{\rm disk} &=
	\frac{1}{(2\pi i)^n} \int d^nx \frac{\a\cdot {\rm STr} \left[\P_1q\P_2q\cdots\P_nq \phi\right]}{\P_1W\P_2W\cdots\P_nW}\ .
}
Here $\a$ is a bulk insertions given by a polynomial in the bulk ring, and $\phi$ is an element of the boundary ring.

Note also that the distinction between bulk and boundary operators in the topological theory is not completely clear, since there is a natural map $\one\;$id, which maps any bulk operator to a boundary operator.

As in the bulk case, this formula for the correlators has its limitations. In particular it is not possible to apply it to calculate expectation values of unintegrated operators. That means that only three-point functions can be evaluated with it.

We also remark that there is a generalisation to worldsheets with genus $g$ and handles $h$, which results basically in a furnishing with powers of the Hessian $H^g$ and products over the contributions from each handle \cite{Kapustin:2003ga}.

The correlators obtained by (\ref{BBcorr}) can be identified with the analogous correlators obtained from CFT. This is a powerful statement, since it allows to make contact between the two descriptions. This will be exploited in the following chapter to obtain expressions for effective superpotentials.

%
%
%

\chapter{Open-closed superpotential}
\label{ch-superpot}

\label{model}

In this chapter we will derive an expression for a bulk induced superpotential by investigating D2-branes wrapping holomorphic 2-cycles of
the quintic. Our starting point is the Fermat quintic given by the
following hypersurface in $\IP_4$
\be
\label{Fermat}
x_1^5+ x_2^5 + x_3^5 + x_4^5 + x_5^5=0 \subset \IP_4 \ .
\ee
We are interested in a special family of branes wrapping rational
curves, which has been studied from a mathematical point of view in
\cite{AK} and from a physics point of view in \cite{Ashok:2004xq}, see
\cite{Brunner:1999jq} for earlier work. More concretely, the family of
curves we have in mind is given by 
\be
\label{Katzslines}
(x_1,x_2,x_3,x_4,x_5) =  (u,\eta u,av,bv,cv) \ ,
\quad  \hbox{where} \quad
a^5+b^5+c^5=0 \ .
\ee
Here $a,b,c \in \CC$, $\eta$ is a $5^{th}$ root of $-1$, and $(u,v)$
parametrise a $\IP_1$. The three complex parameters $a,b,c$ are
subject to projective equivalence and the complex equation
in (\ref{Katzslines}), so that the above equations describe a one
parameter family of $\IP_1$'s. In fact there are $50$ such families
since there are $10$ possibilities to pick a pair of coordinates 
that are proportional to $u$, and $5$ choices for $\eta$. These
families intersect along the lines
\be
x_i - \eta x_j =0\ , \quad x_k - \eta' x_l =0\ , \quad x_m =0 \ ,
\ee
where $i,j,k,l,m$ are all disjoint and $\eta$ and $\eta'$ are 
$5^{th}$ roots of $-1$. For example, the set 
\be
x_1-\eta x_2 =0\ , \quad x_3-\eta' x_4=0\ , \quad x_5=0
\ee
describes a particular $\IP_1$ in (\ref{Katzslines}) with $c=0$, 
$a=\eta'$, $b=1$. Likewise, it describes a $\IP_1$ in the
family
\be
(x_1,x_2,x_3,x_4,x_5) = (av, bv, u,\eta' u, cv)
\ee
with $a=1$, $b=\eta$ and $c=0$. Starting from such a configuration,
one can thus move along either of the two families 
of which this $\IP_1$ is part. However, once one has started to move
away in one direction, the other becomes obstructed \cite{AK}. For 
concreteness we shall mainly consider in the following the family of
curves associated to (\ref{Katzslines}) although everything we say
can be easily generalised to the other classes of branes.
 
{}From a conformal field theory point of view, the existence of the
above families of $\IP_1$'s implies that the open string spectrum of
every corresponding brane contains an exactly marginal boundary
operator which we shall denote by $\psi_1$. At the above intersection
points there will be a second exactly marginal operator which we
shall call $\psi_2$ \cite{Ashok:2004xq}. The fact that moving away in
one direction obstructs the other should imply that the effective 
superpotential contains a term of the form  
\be\label{bdlr}
{\cal W}(\psi_1, \psi_2)= \psi_1^3 \psi_2^3 \ . 
\ee
This was argued on physical grounds in \cite{Brunner:1999jq} and later  
confirmed in \cite{Ashok:2004xq}. Recently it was shown in
\cite{Aspinwall:2007cs} that (\ref{bdlr}) is already the full
superpotential for the fields $\psi_1$ and $\psi_2$. We shall 
reproduce this result, using somewhat different methods, at the end of
section~2.
\medskip

The above discussion applies to the Gepner point of the quintic, where
the hypersurface is described by equation (\ref{Fermat}). It is well
known that at a generic point in the complex structure moduli space of
the quintic, there are only discretely many ($2875$) rational
2-cycles; in particular there are therefore no continuous families of
$\IP_1$'s if we perturb the theory away from the Gepner
point. Geometrically, this means that at a generic point in the above
moduli space of branes, the complex structure deformations are
obstructed, as has already been discussed 
in \cite{AK}. From a worldsheet point of view this should therefore
mean that the effective superpotential contains a term of the form  
\be\label{bdlr1}
{\cal W}(\psi_1,\psi_2,\Phi_i)= \psi_1^3 \psi_2^3 + 
\sum_i \Phi_i \, F_i(\psi_1,\psi_2) + \cdots \ , 
\ee
where the $\Phi_i$ describe the different complex structure deformations.  

In the following we shall mainly consider the special deformations of  
the quintic described by 
\be\label{special}
x_1^5 + x_2^5 + x_3^5 + x_4^5 + x_5^5 + x_1^3 s^{(2)}(x_3,x_4,x_5) 
=0 \ , 
\ee
where $s^{(2)}$ is a polynomial of degree $2$ in $x_3,x_4$ and $x_5$,  
The only curves that survive this deformation are those for which
\be\label{gconstraint}
a^5 + b^5 + c^5 = 0 \qquad \hbox{and} \qquad
s^{(2)}(a,b,c)=0 \ .
\ee
These equations determine a discrete set of points; in fact, counting
multiplicities there are precisely $10$ solutions, as follows from 
Bezout's theorem. 

The deformations (\ref{special}) are special in that the term linear
in $\Phi$ in (\ref{bdlr1}) is independent of $\psi_2$. In this case we
can then determine the function $F(\psi_1,\psi_2)$ exactly, and thus
give a complete description for how the system behaves under the
corresponding bulk perturbation; this will be described in detail in
section~3. As we shall see, the bulk perturbation induces a boundary
RG flow that is the gradient flow of the function $F$; in
particular the solutions to (\ref{gconstraint}) 
are precisely the critical points of $F$.

\section{2-branes on the quintic}

The starting point of our construction of bulk induced superpotentials will be the investigation of families of D2-branes
(\ref{Katzslines}) at the Fermat point in the Landau-Ginzburg model description. At this point in moduli space the corresponding conformal field theory is known. As soon as bulk perturbations are switched on, the boundary moduli space changes. Most of the configurations which appeared in D2-families in the unperturbed backgrouns are then found to break supersymmetry. The relationship to conformal field theory will be used in order to derive renormalisation group equations for this case. Finally this will enable us to find explicit expressions for the bulk induced superpotential.

\subsection{The matrix factorisations description}

At the Fermat point the quintic is described by the Gepner model
corresponding to five copies of the $N=2$ minimal model at $k=3$. 
The branes of interest are B-type branes of this superconformal field
theory. As we shall see, isolated D-branes can be constructed as 
permutation branes in conformal field theory 
\cite{Recknagel:2002qq},
but in order to understand the full moduli space of branes 
a treatment in the formalism of matrix factorisations is more adequate.

\noindent At the Gepner point the relevant LG superpotential is    
\be\label{2.1}
W_0 = x_1^5 + x_2^5 + x_3^5 + x_4^5 + x_5^5 \ . 
\ee
The first step consistes in the construction of a matrix factorisation $q$ with
\be
q^2=W_0 \one \ .
\ee
$q$ is the boundary part of  the BRST operator $Q$, and together with the
bulk BRST charge squares to $0$. In particular, $q$ is fermionic and 
can be expressed as a linear combination of (non-BRST closed)
fermionic operators $\pi^i$ and their conjugates $\bar\pi^i$,
$i=1,\ldots,n$, that live at the boundary, 
\be\label{Qcliff}
q= \sum_{i=1}^n \big( \pi^i J_i + \bar\pi^i E_i) \ .
\ee
These fermions form a $2^n$ dimensional representation of the Clifford
algebra 
\be
\{\pi^i,\bar\pi^j\} = \delta^{ij} \ , \qquad 
\{\pi^i,\pi^j\} = \{\bar\pi^i,\bar\pi^j\} = 0  \ . 
\ee
The square of $q$ is given by
\be
q^2 = \Bigl( \sum_i E_i J_i\Bigr) \cdot \one
\ee
and hence $q$ defines a matrix factorisation if
\be\label{factorize}
W= \sum_i E_i J_i \ .
\ee
Turning the argument around, whenever $W$ can be written in the form 
(\ref{factorize}) a suitable matrix factorisation is given by
(\ref{Qcliff}). 
The matrix factorisation description captures all topological aspects
of the corresponding D-branes. For example, one can determine from
$q$ the topological part of the open string spectrum and the
topological RR charges, {\it etc}. What will be most important for our
purposes is the Kapustin-Li formula \cite{Kapustin:2003ga} that allows
one to calculate bulk-boundary correlators (or boundery three point
functions) exactly. If we denote a topological bulk field by $\Phi$
and the boundary field by $\psi$, then the disk correlator is 
\be\label{KL}
\langle\Phi\, \psi\rangle = 
{\rm Res}\; \Phi 
\frac{ {\rm STr} \left [\,\partial_{x_1}q \dots \partial_{x_5} q \,\psi
\right]}{\partial_{x_1}W\dots\partial_{x_5} W}\ ,
\ee
where the residue is taken at the critical points of LG superpotential
$W$. More details about this formula can be found in
\cite{Kapustin:2003ga,Herbst:2004ax}. 

It has been noted before that strictly speaking, to find an LG description of the quintic one has to
consider an orbifold of the theory (\ref{2.1}). This ${\mathbb Z}_5$ orbifold 
projects onto states with integer $U(1)$ charge in the closed string 
sector. As usual, the consequence for the open string sector is
\cite{Ashok:2004zb,Hori:2004ja,Walcher:2004tx} 
that we need to specify 
in addition a representation of the orbifold
group on the Chan-Paton labels. The open string spectrum is then given
by the ${\mathbb Z}_5$ invariant part of the cohomology of the BRST
operator. In the following, the additional representation label will
play no further role, since we will only consider a single D-brane
with an arbitrary but fixed representation label. 
\medskip

The D2-branes of interest correspond to a family of matrix
factorisations that can be constructed as follows, using ideas similar 
to what was done in 
\cite{Brunner:2006tc,Brunner:2004mt} (see also \cite{Hori:2004ja}).
We define 
\be\label{ansatz}
J_1 = x_1 - \eta x_2 \ , \qquad J_4 = a x_4 - b x_3 \ , \qquad
J_5 = cx_3 - a x_5  \ , 
\ee
and look for common solutions of $J_1=J_4=J_5=0$ and $W=0$. 
If $\eta$ is a fifth root of $-1$ and $a\neq 0$, we get a solution if  
\be\label{moduli}
a^5 + b^5 + c^5 = 0 \ .
\ee
If this is the case we can use the Nullstellensatz to
write 
\be
W_0 = J_1 \cdot E_1 + J_4 \cdot E_4 + J_5 \cdot E_5 \ ,
\ee
where $E_i$ are polynomials in $x_j$. We then obtain a matrix
factorisation by the procedure outlined above. More specifically, we
introduce $8\times 8$ matrices $\pi^i$ and $\bar{\pi}^i$, $i=1,4,5$,
that form a representation of the Clifford algebra, and obtain a
family  of matrix factorisations $q(a,b,c)$ 
\be\label{Qref}
q(a,b,c) = \sum_{j=1,4,5} \left( J_j \pi^j + E_j \bar\pi^j \right) \ . 
\ee
By construction $q(a,b,c)$ satisfies then 
$q(a,b,c)^2 = W_0 \cdot \one$. 

Following the geometrical interpretation of matrix factorisations
elaborated in 
\cite{Orlov:2003yp,Aspinwall:2006ib} (see also \cite{Ezhuthachan:2005jr})
 these matrix factorisations provide 
the LG-description of the D2-branes described in section
\ref{model}. Indeed, read as equations in 
${\mathbb P}_4$, the equations $J_1=J_3=J_5=0$ describe precisely 
the geometrical lines (\ref{Katzslines}).

The moduli space of such branes has complex dimension one. Indeed, it
is straightforward to see that rescaling $(a,b,c)$ by a common factor 
results in an equivalent factorisation; thus (\ref{moduli}) can be
thought of as an equation in $\mathbb{C}\mathbb{P}^2$, and hence
describes a one-complex-dimensional curve.\footnote{In the above
description we have not treated the three variables $a$, $b$ and $c$
on an equal footing, and hence $a$ could not be zero. It should
be clear, however, that we can also use a different chart in which
$a=0$ is possible. In this way we can obtain a matrix factorisation
associated to $(a,b,c)$ provided that not all three $a$, $b$ and $c$
are simultaneously zero and that (\ref{moduli}) holds. See 
\cite{Hori:2004ja}
for an explicit change of coordinates in a
different example.} Furthermore we note that special points on this
curve correspond to standard permutation branes 
\cite{Recknagel:2002qq}: for example for $a\neq 0$ and $b=0$ we may
use the projective equivalence to set $a=1$. Then $c$ must be a fifth
root of $-1$, leading  precisely to a permutation factorisation of the
form discussed in 
\cite{Brunner:2005fv,Enger:2005jk}. This
identification is also in agreement with the analysis of
\cite{Ashok:2004zb,Brunner:2005fv} where it was shown that one of these matrix
factorisations carries indeed the charge of a D2-brane.

\begin{SCfigure}
  \centering
  \includegraphics[width=.55\textwidth]{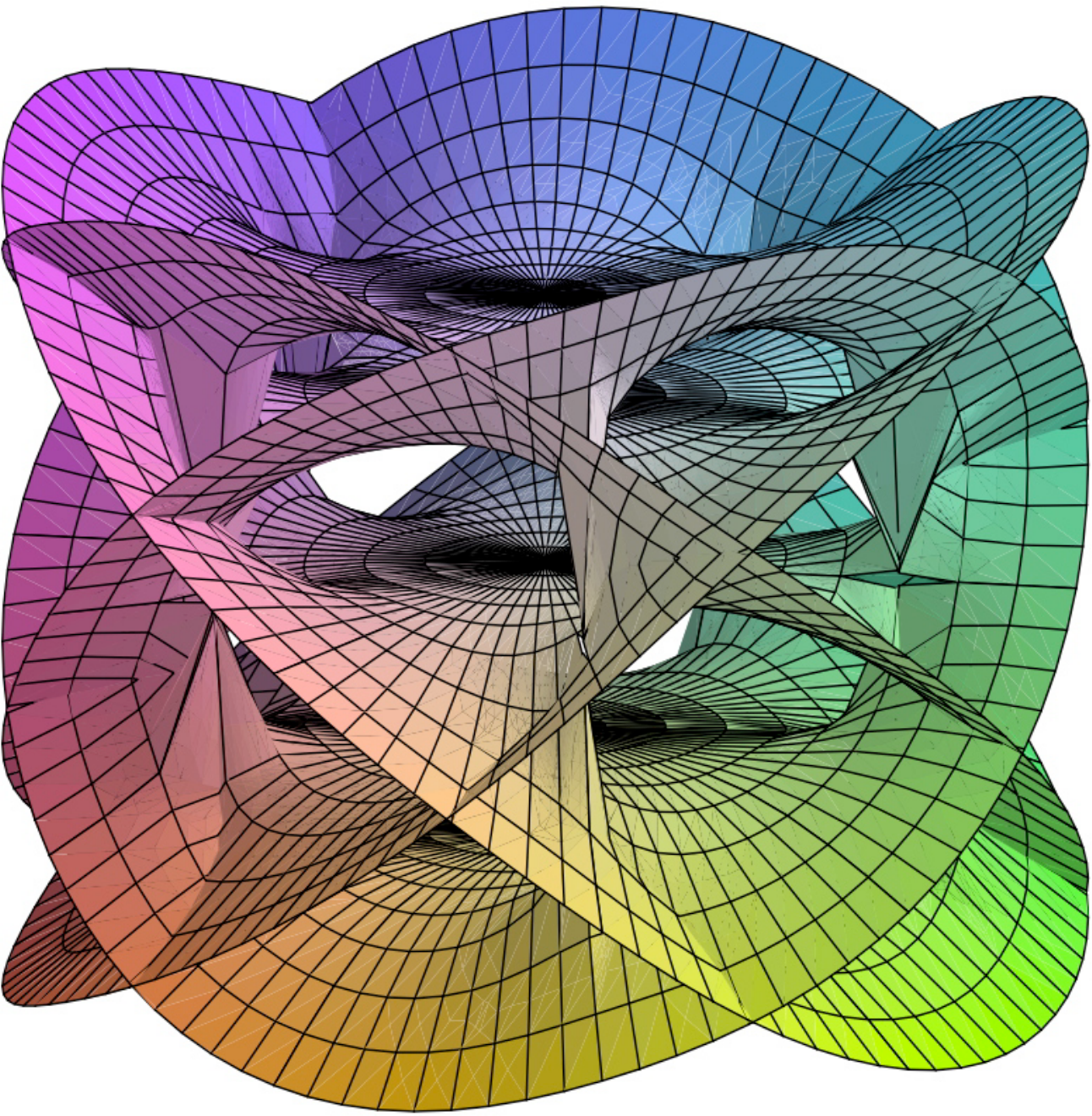}
  \caption{Riemann surface associated to the D-brane moduli space,
consisting of five copies of the complex plane. The real and imaginary
part of $b$ have been plotted horizontally, the vertical axis is
the imaginary part of $c$. The five sheets arranged vertically at
$b=0$ reflect the five possibilities for $c^5=-1$.} 
\end{SCfigure}

\subsection{The fermionic spectrum}

The fact that these matrix factorisations form a 1-complex dimensional
moduli space means that at every point in the moduli space the open
string cohomology contains at least one fermion of $U(1)$-charge
one. Indeed, this is just the matrix factorisation analogue of the
fact that each such D-brane must have an exactly marginal boundary
operator in its spectrum. From a matrix factorisation point of view,
the corresponding fermion can be easily constructed. Since by 
assumption $a\neq 0$, we may always rescale the parameters so that $a=1$. 
Let us first consider a generic point in moduli space where 
$bc\neq 0$. We then have a family of factorisations parametrised by
$(b,c)$ subject to $b^5 + c^5 = -1$. As long as $c\neq 0$, we can
locally solve this equation for $c$, {\it i.e.}\ we can express
$c\equiv c(b)$, and thus obtain a matrix factorisation $q(b)$. Since
$W_0$ does not depend on $(a,b,c)$, it then follows that   
\be
\{ q(b), \partial_b q(b) \} = 0 
\ee
which is precisely the condition for $\psi = \partial_b q(b)$ to
define a fermion of the cohomology defined by $q(b)$. For the case
under consideration, we find explicitly
\be\label{fermion1}
\psi_b \equiv \partial_b q(b) = - x_3\, \pi^4 
- \frac{b^4}{c^4} x_3 \, \pi^5 
+ \left(\partial_b E_4\right)\, \bar\pi^4 
+ \left(\partial_b E_5\right)\,  \bar\pi^5 \ ,
\ee
where we have used that 
\be
\left. \frac{\partial b}{\partial c} \right|_a = - \frac{c^4}{b^4} \ . 
\ee
In the next section, it is shown by explicit computation that $\psi_b$ is
non-trivial in cohomology. Obviously, we could have equally expressed 
$b\equiv b(c)$ (for $b\neq 0$) and written 
$q(a,b,c) \equiv q(c)$. Then the derivation with respect to $c$ also
defines a fermion 
\be\label{fermion}
\psi_c \equiv 
\partial_c q(c) = \frac{c^4}{b^4} x_3 \, \pi^4
+ x_3\, \pi^5 
+ \left(\partial_c E_4\right)\, \bar\pi^4 
+ \left(\partial_c E_5\right)\,  \bar\pi^5 \ .
\ee
It is easy to see that for $bc\neq 0$ so that both $\psi_b$ and
$\psi_c$ are well defined, $\psi_b\cong \psi_c$ in cohomology. In the
following we shall denote the equivalence class to which $\psi_b$ and
$\psi_c$ belong by $\psi_1$. More specifically, we shall usually
take $\psi_1\equiv \psi_b$ and assume that $c\neq 0$. 
\smallskip

\noindent
The full fermionic cohomology of $q(a,b,c)$ at $U(1)$-charge
$1$ is however bigger: in addition to $\psi_1$ it also contains a
second fermion that we shall call $\psi_2$. This is explained in
appendix~A, where $\psi_2$ is explicitly constructed (for 
$c\neq 0$). In general, however, $\psi_2$ does not define a
modulus. In fact, using the Kapustin-Li formula \cite{Kapustin:2003ga}
one easily finds that  
\be\label{kap-li}
B_{\psi_2\psi_2\psi_2} = -\frac{2}{5}\eta^4 \frac{b^3}{c^9} \ . 
\ee
Unless $b=0$ the three-point function of $\psi_2$ on the boundary does
not vanish, and hence $\psi_2$ is not an exactly marginal boundary 
field 
\cite{Recknagel:1998ih}. This shows that at generic points 
in the moduli space (\ref{moduli}) there is only one exactly marginal 
operator, whereas at the special point $b=0$ an additional marginal 
operator appears, indicating an additional  branch of the moduli
space. This is in nice agreement with the geometric analysis of
the previous section, since at $b=0$ the above moduli space intersects with the
branch where the roles of $J_1$ and $J_5$ can be interchanged.
In fact, this can also be seen from the explicit formula for $\psi_2$,
see (\ref{psi2ex}).

The three-point function (\ref{kap-li}) verifies 
the superpotential term (\ref{bdlr}) that was already obtained in  
\cite{Ashok:2004xq} by other means. Furthermore, 
after rescaling $\psi_2 \mapsto \hat\psi_2 = c^3 \psi_2$, 
the $b$-dependence of the three-point function for $\hat\psi_2$ is
simply proportional to $b^3$. (Recall that $c\equiv c(b)$.) Using the
arguments of section 4.1 this then implies that, with respect to this
normalisation, the effective superpotential does not contain any 
higher order contributions (in $\psi_1$) to the term 
$\psi_1^3 \, \hat\psi_2^3$ in (\ref{bdlr}). This is in agreement with
the \cite{Aspinwall:2007cs}.

\section{The cohomology of the factorisations}

In this section we want to determine the full fermionic cohomology of
$U(1)$-charge $1$ for the factorisations $q(a,b,c)$ (\ref{ansatz})
with $a\neq 0$. First we observe that the coordinates
involved in $J_1$ and $E_1$ (namely $x_1$ and $x_2$) do not appear in
$J_4, E_4$ or $J_5, E_5$. Therefore the cohomology $H$ of $Q$
separates into 
\be
H(q) = H(q_1)\odot H(q_2)\ ,
\ee  
where $q_1$ and $q_2$ are the separate factorisations
\ba
q_1 &=& \pi^1 J_1+\bar\pi^1 E_1\ ,\nonumber\\
q_2 &=& \pi^4 J_4 + \pi^5 J_5 + \bar\pi^4 E_4 +\bar\pi^5 E_5 \ . 
\ea

\noindent The explicit polynomials are
\eqn{
J_1 &= x_1-\eta x_2 \\ E_1 &
       = \prod_{\eta'^5=-1,\eta'\ne\eta}(x_1-\eta'x_2) \\
J_4 &= ax_4-b x_3 \\ E_4 &= \hphantom{-}\frac{1}{a^5}\left(
	b^4x_3^4 + ab^3x_3^3x_4 + a^2b^2x_3^2x_4^2
        +a^3bx_3x_4^3+a^4x_4^4\right) \\
J_5 &= cx_3-ax_5 \\ E_5 &=-\frac{1}{a^5}\left(
	c^4x_3^4+ac^3x_3^3x_5+a^2c^2x_3^2x_5^2
        +a^3cx_3x_5^3+a^4x_5^4\right)\ . 
}
The cohomology of $q_1$ has been calculated in
\cite{Ashok:2004zb,Brunner:2005fv,Enger:2005jk}, and consists of four  
bosonic elements of $U(1)$-charge $0$, $\tfrac{2}{5}$,
$\tfrac{4}{5}$ and $\tfrac{6}{5}$, respectively; it does not
contain any fermions at all. Thus in order to obtain a fermion of the 
full factorisation, we need to tensor one of these bosons with a
fermion from $q_2$. We are only interested in fermions of $q$ of total 
$U(1)$-charge $1$. Since the $U(1)$-charge of the fermions in $q_2$ is
always positive, there are three cases to consider:
the fermions in the cohomology of $q_2$ can have $U(1)$-charges $1$,
$\frac{3}{5}$ or $\frac{1}{5}$ which together with the boson of $q_1$
of $U(1)$-charges $0$, $\frac{2}{5}$ or $\frac{4}{5}$, respectively,
then produce a fermion of total $U(1)$-charge $1$. Thus it is
sufficient to analyse the fermionic cohomology of $q_2$ for these
three $U(1)$-charges separately.  

\subsection{The $q_2$-fermions of charge $1$}

The general $q_2$-closed fermion has an expansion (the closure
conditions force the absence of any higher powers of boundary
fermions)  
\be
\psi=\pi^4 p_4 + \bar\pi^4 m_4 + \pi^5 p_5 + \bar\pi^5 m_5  \ ,
\ee
where we have dropped some exact terms --- see (\ref{exact}) below. 
The requirement that $\psi$ has $U(1)$-charge $1$ implies that 
$p_4$ and $p_5$ are polynomials of degree $1$ (thus each $p_i$ has $3$
parameters)  while $m_4$ and $m_5$ are polynomials of degree $4$ (with
$15$ parameters each), giving in total $36$ parameters. The condition
that $\psi$ is closed implies further that 
\be
J_4 \, m_4 + J_5 \, m_5 + E_4 \, p_4 + E_5 \, p_5 = 0 \ .
\ee
The left hand side is a homogeneous polynomial of degree $5$, and hence
represents $21$ conditions. We have checked (using standard matrix
techniques) that these $21$ conditions are independent. This implies
that the space of closed fermions of the $U(1)$-charge $1$ is 
$15$-dimensional. 

It remains to determine how many of them are exact. To see this we
make the following ansatz for the most general boson, 
\eqn{
\Lambda = \hat a &+ \hat b \pi^4\bar\pi^4 + \hat c\pi^4\pi^5 
		+ \hat d\pi^4\bar\pi^5  
		+ \hat e\bar\pi^4\pi^5\\
		 &+\hat f\bar\pi^4\bar\pi^5 
		+ \hat g\pi^5\bar\pi^5 
		+ \hat h\pi^4\bar\pi^4\pi^5\bar\pi^5 \ . 
}
Then
\begin{align}
[Q,\Lambda]=
&\pi^4\left(-\hat bJ_4-\hat dJ_5-\hat cE_5\right) 
+ \pi^5\left(\hat eJ_4-\hat gJ_5+\hat cE_4\right) \nonumber \\
&+\bar\pi^4\left(\hat bE_4-\hat eE_5-\hat fJ_5\right) 
	+ \bar\pi^4\left(\hat dE_4+\hat gE_5+\hat eJ_4\right) 
\nonumber \\
&-\pi^4\bar\pi^4\pi^5 \hat hJ_5 + \pi^4\bar\pi^4\bar\pi^5 \hat hE_5 
-\pi^4\pi^5\bar\pi^5 \hat hJ_4 + \bar\pi^4\pi^5\bar\pi^5 \hat hE_4 
\ . \label{exact}
\end{align}
Consistency with the ansatz for $\psi$ requires $\hat h=0$ and 
$\hat c=0$. Moreover $\hat a$ can be set to zero, too. The other
parameters must be polynomials of degree $0$, except for $\hat f$
which has to have degree $3$ (and therefore $10$ parameters). In total 
the space of exact fermionis is described by $14$ parameters. Again,
using standard matrix methods, we have shown that these $14$
parameters are linearly independent. This implies that 
the fermionic cohomology of $q_2$ of $U(1)$-charge $1$ is
$1$-dimensional. A representative of the corresponding cohomology
class for $q$ is (for $c\neq 0$)  
\be
	\psi_1 = \partial_b q
\ee
or explicitly
\eqn{
\psi_1 = &-x_3 \pi^4 
	+\frac{1}{a^5}\left[
	4b^3x_3^4 + 3ab^2x_3^3x_4 
        + 2a^2bx_3^2x_4^2+a^3x_3x_4^3\right]\bar\pi^4 \\
	&- \frac{b^4}{c^4} x_3 \pi^5
		+\frac{b^4}{a^5c^4}\left[
	4c^3 x_3^4+3ac^2x_3^3x_5 
		+ 2a^2cx_3^2x_5^2+a^3x_3x_5^3\right]\bar\pi^5 \ .
}

\subsection{The $q_2$-fermions of charge $\frac{3}{5}$}

The same arguments can be used to determine the fermions of $U(1)$-charge
$\tfrac{3}{5}$. In this case, $p_4$ and $p_5$ have both degree $0$ 
({\it i.e.}\ are constants) while $m_4$ and $m_5$ have both degree
$3$ (with 10 parameters each), giving rise to $22$ parameters.
The closure condition is now given by a polynomial of degree $4$,
leading to $15$ (independent) equations. Thus the space of closed
fermions is in this case $7$-dimensional.

For exact fermions we find that they are described by bosons $\Lambda$
with $\hat a=0$, $\hat b=0$, $\hat d=0$, $\hat c=0$, $\hat e=0$, 
$\hat g=0$, $\hat h=0$ and $\hat f$ a polynomial of degree $2$
(with $6$ parameters). Thus there are $6$ different exact fermions,
and we have checked that they are in fact linearly independent. This
implies that there is precisely one fermion of charge $\tfrac{3}{5}$
in the cohomology of $q_2$. A representative of the corresponding
cohomology class for $q$ is given by (for $c\neq 0$)  
\begin{align}
\psi_2 = x_1\, \partial_b \, \Bigl[
b\pi^4 - c\pi^5
- (b^4x_3^3+b^3x_3^2x_4+b^2x_3x_4^2+bx_4^3)&\bar\pi^4 \nonumber \\
+ (c^4x_3^3 + c^3x_3^2x_5 + c^2x_3x_5^2 + cx_5^3)&\bar\pi^5 
		\Bigr] \ , 
\end{align}
or, since $\psi_1$ is proportional to $x_3$, 
\be\label{psi2ex}
\psi_2 = \frac{x_1}{x_3} \psi_1 \ .
\ee

\subsection{The $q_2$-fermions of charge $\frac{1}{5}$}

For fermions of charge $\frac{1}{5}$, our ansatz has $12$ parameters,
and the closure condition leads to $9$ linearly independent
conditions. Thus there are $3$ different closed fermions. In
$\Lambda$, all parameters are zero except $\hat f$, which is a
polynomial of degree $1$ with $3$ independent parameters. This implies
that all $3$ closed fermions are in fact exact, and hence that the
cohomology is trivial.

\section{Bulk induced renormalisation group flow}

Now we want to consider the bulk perturbation of the above Gepner
model by the bulk operator $\Phi$, {\it i.e.} we consider the
perturbed superpotential 
\be\label{bper}
W = W_0 + \lambda\, \Phi  \ , \qquad
\Phi = x_1^3 \, s^{(2)}(x_3,x_4,x_5) \ , 
\ee
where $s^{(2)}$ is the polynomial of section~1.1 that we expand 
as\footnote{Everything we are going to say is essentially unchanged if
we were to replace $x_1^3$ by an arbitrary third order polynomial in
$x_1$ and $x_2$.}  
\be\label{defo}
s^{(2)}(x_3,x_4,x_5) = \sum_{q+r+s=2} s^{(2)}_{qrs}\, \,
x_3^q \, x_4^r \, x_5^s \ .
\ee
{}From a conformal field theory point of view the perturbation is
generated by an exactly marginal bulk field in the $cc$ ring.
We want to understand what happens to the D-branes
described by the moduli space (\ref{moduli}) under this
perturbation. We shall be able to give a fairly complete description
of this problem by combining the ideas of \cite{Fredenhagen:2006dn}
with matrix factorisation techniques. In particular, this will allow
us to calculate the effective superpotential for the boundary
parameters $(a,b,c)$ exactly.   

One way to address this problem is to study the deformation theory of
matrix factorisations, following 
\cite{Hori:2004ja} (see also \cite{Ashok:2004xq}). Suppose that $q_0$ is a factorisation of
$W_0$. Then we ask whether we can find a deformation $q$ of $q_0$,
{\it i.e.}
\be
q= q_0 + \lambda q_{1} + \lambda^2 q_{2} + \cdots
\ee
such that $q^2 = W_0 + \lambda \Phi$. Expanding this equation to first
order in $\lambda$, we find the necessary condition that $\Phi$ must be
exact with respect to $q_0$, {\it i.e.}\  of the form 
$\Phi = \{ q_0, \chi \}$ for some $\chi$. In general this condition
will not be met; for example for the case at hand where 
$q_0\equiv q(a,b,c)$ and $\Phi$ is given by (\ref{bper}), we find that
$\Phi$ is exact if and only if 
\be\label{constraint}
a^5 + b^5 + c^5 = 0 \qquad \hbox{and} \qquad
s^{(2)}(a,b,c)=0 \ .
\ee
On the other hand, if this condition {\it is met}, it is easy to see
that we {\it can} in fact extend the matrix factorisation for
arbitrary (finite) values of $\lambda$. Indeed, if we consider the
same ansatz as in (\ref{ansatz}), it is clear 
that we can find a joint solution to $J_1=J_4=J_5=0$ and
$W=W_0+\lambda \Phi=0$ if $(a,b,c)$ satisfies (\ref{constraint}). It
then follows by the same arguments as above that there exists a
matrix factorisation for all values of $\lambda$ (that is by
construction a deformation of $Q(a,b,c)$). 

Unless $s^{(2)}\equiv 0$, the set of constraints (\ref{constraint})
has only finitely many discrete solutions; in fact, counting
multiplicities, there are precisely $10$ solutions, as follows from 
Bezout's theorem. This ties in nicely with our geometric expectations
since at a generic point in the complex structure moduli space only
finitely many holomorphic 2-cycles exist.

\subsection{Combining with conformal field theory}

As we have just seen, for $\lambda=0$ we have a one-parameter family of
superconformal D2-branes, while for $\lambda\neq 0$ only discrete
possibilities remain. The situation is therefore very analogeous to
the example studied in  
\cite{Fredenhagen:2006dn}. There a general
conformal field theory analysis of this problem was suggested that we
now want to apply to the case at hand.

In \cite{Fredenhagen:2006dn}
 the coupled bulk and boundary
deformations of a boundary conformal field theory were studied, and
the resulting renormalisation group identities were derived. It
was found that an exactly marginal bulk operator may cease to be
exactly marginal in the presence of a boundary. If this is the case it
will induce a renormalisation group flow on the boundary that will
drive the boundary condition to one that is again conformal with
respect to the deformed bulk theory. If we denote the boundary
coupling constant corresponding to the boundary field $\psi_j$ 
of conformal weight $h_j$ by $\mu_j$, then the perturbation by the
exactly marginal bulk operator $\lambda \Phi$ will induce the RG
equation  
\be\label{RG}
\dot\mu_j = (1- h _j)\mu_j +  \frac{\lambda}{2} B_{\Phi \psi_j} 
+ {\cal O}(\mu\lambda, \lambda^2, \mu^2)\ , 
\ee
where $B_{\Phi \psi_j}$ is the bulk-boundary operator product
coefficient. Since the first term in (\ref{RG}) damps the flow of any
irrelevant operators, it is sufficient to study this equation only for
the marginal or relevant boundary fields, {\it i.e.} for those that
satisfy $h_j\leq 1$. 

For the case at hand, we do not have an explicit conformal field
theory description of the D-branes away from the specific points where
$abc=0$. On the other hand, we know (based on supersymmetry) that the
open string spectrum will not contain any relevant (tachyonic)
operators. Furthermore, the above discussion suggests that
everywhere in moduli space each brane has precisely two marginal 
operators in its spectrum, namely the operators corresponding to the
open string fermions described by $\psi_1$ and $\psi_2$ --- see
appendix~A for details. The two boundary operators $\psi_1$ and
$\psi_2$ are topological, and so is the bulk perturbation $\Phi$. In
particular, this implies that we can determine the coefficients
$B_{\Phi \psi_1}$ and $B_{\Phi \psi_2}$ that are important for the RG
equations using {\it topological methods}, without having to solve the 
full conformal field theory (which would be impossibly difficult)! 

Using the Kapustin-Li formula (\ref{KL}) we find (we are
working in a patch where $a=1$) 
\be\label{psi2}
B_{\Phi\psi_2} = 0 
\ee
for all $(a,b,c)$, as well as 
\be
B_{\Phi \psi_b} = 
	\frac{\eta^4}{25} c^{-4} s^{(2)}(1,b,c) \ ,
\ee
and similarly for 
\be
B_{\Phi \psi_c} =  
	- \frac{\eta^4}{25} b^{-4} s^{(2)}(1,b,c) \ .
\ee
All of these calculations were performed in the unperturbed bulk
theory. Since the bulk-boundary coupling between $\Phi$ and $\psi_2$
vanishes (\ref{psi2}), this field is not switched on by $\Phi$. The RG
flow  will therefore only involve $\psi_1$, and for this we find
\be\label{bdot}
\dot{b} = \lambda\frac{\eta^4}{50} c^{-4} s^{(2)}(1,b,c) \ ,
\ee
or 
\be\label{cdot}
\dot{c} = - \lambda\frac{\eta^4}{50} b^{-4} s^{(2)}(1,b,c) \ .
\ee
In particular, we see that the solutions to (\ref{constraint}) are
precisely the fixed points under the RG equation. Thus any brane
described by $(a,b,c)$, will flow to one of these $10$ fixed points
under the RG flow.

\subsection{Differentials on the Fermat curve and their integrals} 

Let us consider the Fermat curve defined by 
\be\label{fermat}
	\hat{b}^5+\hat{c}^5=1 \ .
\ee
For $a=1$ this is the curve that describes the brane moduli space 
$1+b^5+c^5=0$ provided we identify $\hat{b}=-b$ and $\hat{c}=-c$. The
general theory of globally defined differentials is described in
\cite{Lang}. The simplest class of differentials, the differentials of
the first kind, are those that are holomorphic on the full curve. They
are of the form
\be\label{first}
\omega_{rs} = \hb^r \hc^s\,\frac{\tfrac{1}{5}d(\hb^5)}{\hb^5 \hc^5} 
            = \hb^{r-1} \hc^{s-1}\frac{d\hb}{\hc^4} \ ,
\ee
where $r,s,\geq 1$. Since $\hb^4 d\hb =-\hc^4 d\hc$ this is equivalent
to  
\be\label{second}
\omega_{rs} = -\hb^{r-1} \hc^{s-1}\frac{d\hc}{\hb^4} \ .
\ee
The first formula (\ref{first}) is defined on the patch of the moduli
space where $\hc\neq 0$, while the second (\ref{second}) is defined
for $\hb\neq 0$. Since on (\ref{fermat}) $\hb \hc\neq 0$ at least one
of these two expressions is everywhere well-defined. In particular,
this therefore proves that the differentials $\omega_{r,s}$ are
holomorphic for finite $\hb$ and $\hc$. The only potential poles may
thus appear at $\hb,\hc=\infty$. Expanding around $\hb=\infty$ shows
that the differentials are finite as long as $r+s\le 4$. Therefore we
find the holomorphic differentials (for $\hc\neq 0$)
\be
	\frac{1}{\hc^4}d\hb,\;
	\frac{1}{\hc^3}d\hb,\;
	\frac{1}{\hc^2}d\hb,\;
	\frac{b}{\hc^4}d\hb,\;
	\frac{\hb^2}{\hc^4}d\hb,\;
	\frac{\hb}{\hc^3}d\hb \ .
\ee
In fact this is a basis for the holomorphic differentials on the
curve. Its number is equal to the genus of the curve.

\subsubsection{Integrating the holomorphic differentials}

In order to calculate the effective superpotential we need to
integrate these holomorphic differentials. For all of them the answer
can be expressed in terms of a hypergeometric function. In fact in the
chart where $\hc\neq 0$ we have 
\be
\label{EQHypergeoInt1}
\int_0^{\hb} \omega_{rs}
=\int_0^{\hb} d \tilde{b}\,\frac{\tilde{b}^{r-1}\hc(\tilde{b})^{s-1}}{
\hc(\tilde{b})^4} 
= \frac{1}{r}\, \hb^r\,
{}_2{\rm F}_1(\tfrac{r}{5},1-\tfrac{s}{5};1+\tfrac{r}{5}; \hb^5) \ .
\ee
On the other hand in the chart with $\hb\neq 0$ we get instead 
\be
\label{EQHypergeoInt2}
\int_0^{\hc} \omega_{rs}
=-\int_0^{\hc} d\tilde{c}\,\frac{\hb(\tilde{c})^{r-1} 
\tilde{c}^{s-1}}{\hb(\tilde{c})^4} 
= -\frac{1}{s} \hc^s\,
{}_2{\rm F}_1(\tfrac{s}{5},1-\tfrac{r}{5};1+\tfrac{s}{5}; \hc^5) \ .
\ee
In particular, the formula for the effective superpotential 
(\ref{superex}) follows directly from (\ref{EQHypergeoInt1}). Note
that the reference point $\hat{b}_0=0$ corresponds to $\hat{c}_0^5=1$,
and vice versa.

\subsection{Renormalisation group flow as gradient flow}

Actually, the above renormalisation group flow is a gradient flow, as was also the case
in the example studied in \cite{Fredenhagen:2006dn}.\footnote{For
exactly marginal bulk deformations this may in fact follow from the
analysis of \cite{Friedan:2003yc}.} In fact, we can integrate the RG
equation for $b$ in (\ref{bdot}) to   
\be\label{Wb}
\dot{b} = \partial_b {\cal W}(a,b,c) \ ,
\ee
where ${\cal W}(a,b,c)$ is evaluated on the moduli space
(\ref{moduli}) with  $a^5+b^5+c^5=0$ and we have rescaled
$a=1$. Similarly, the same function ${\cal W}(a,b,c)$ also controls
the RG equation for $c$ in (\ref{cdot}) 
\be\label{Wc}
\dot{c} = \partial_c {\cal W}(a,b,c) \ , 
\ee
where again $a=1$ and we regard $b$ as a function of $c$ via the
constraint $a^5+b^5+c^5=0$. To determine ${\cal W}(a,b,c)$ explicitly
we need to integrate
\be
\int_{b_0}^b db' B_{\Phi \psi_b} = \frac{\eta^5}{25} 
\int_{b_0}^b db' c^{-4} s^{(2)}(a,b',c) \ .
\ee
The integral is along a line on the Riemann surface starting at a fixed
reference point $b_0$ that we take to be $0$ and ending at $b$. Since
$b$ parametrises the brane moduli space, it has a natural physical
interpretation as the position of the brane.  The integrand is a
holomorphic one-form on the Riemann surface parametrising the moduli
space, see appendix~B for more details. The potential therefore has a
natural geometric interpretation as the Abel-Jacobi map associated 
to a one-form on the Riemann surface whose points label the brane
positions. Which particular one-form is to be integrated is determined
by the bulk deformation under consideration. 

Since the integrals of such forms are known, we can give explicit
formulae for ${\cal W}(a,b,c)$ in each patch. As explained in
appendix~B, in the patch where $a=1$ and $c\neq 0$ (so that
$c\equiv c(b)$ is well defined) one obtains
\be\label{superex}
{\cal W}(1,b,c)
= \lambda\frac{\eta^4}{50}\sum_{q+r+s=2}\frac{1}{r}s^{(2)}_{qrs}
  (-b)^r \,
  _2{\rm F}_1(\tfrac{r}{5},1-\tfrac{s}{5};1+\tfrac{r}{5};-b^5)
\ . 
\ee
It is also checked there that this function satisfies both (\ref{Wb})
and (\ref{Wc}). 

By combining abstract conformal field theory arguments with
topological methods we can thus give a complete description of the RG
flow: the D2-brane simply follows the gradient flow of ${\cal W}$ to
arrive at one of its local minima, which are precisely the points
characterised by (\ref{constraint}). As in \cite{Fredenhagen:2006dn},
in the RG scheme in which we always remain in the original moduli
space, this analysis is exact in the boundary moduli, and first order
in the bulk coupling constant. Obviously the picture we have found
ties in very nicely with the geometric expectations of section~1.1.

We should note that it is crucial in this analysis that the bulk
perturbation by $\Phi$ does not switch on $\psi_2$, {\it i.e.}\ that 
$B_{\Phi \psi_2}=0$. Otherwise the bulk perturbation would switch on a
boundary field that would lead us out of the original moduli
space and we would not be able to iterate the RG equations. This is
the reason why we restricted our analysis to the bulk perturbations of
the form described in (\ref{defo}).

\subsubsection{Comparing different charts}

Since the differentials we have integrated are globally defined, the 
two expressions we obtain in different charts, namely
(\ref{EQHypergeoInt1}) and (\ref{EQHypergeoInt2}), must agree, once we
have taken into account that the lower bound of the integrals are
different. This can also be checked explicitly. In order to see this
we use the identity
\eqn{
{}_2{\rm F}_1&(\mfa,\mfb;\mfc;1-z) \\&=
\frac{\Gamma(\mfc)\Gamma(\mfa+\mfb-\mfc)}{\Gamma(\mfa)\Gamma(\mfb)}
{}_2{\rm   F}_1(\mfc-\mfa,\mfc-\mfb;
\mfc-\mfa-\mfb+1;z)\, z^{\mfc-\mfa-\mfb} \\ 
&\quad\quad + \frac{\Gamma(\mfc)\Gamma(\mfc-\mfa-\mfb)}
      {\Gamma(\mfc-\mfa)\Gamma(\mfc-\mfb)}
{}_2{\rm F}_1(\mfa,\mfb;\mfa+\mfb-\mfc+1;z) \ .
}
This allows us to rewrite the right hand side of
(\ref{EQHypergeoInt1}) as 
\eqn{ \label{zwischen}
&\frac{1}{r}\hb^r \hc^s\,
\frac{\Gamma(1+\tfrac{r}{5})\Gamma(-\tfrac{s}{5})}
{\Gamma(\tfrac{r}{5})\Gamma(1-\tfrac{s}{5})}\,
{}_2{\rm F}_1(1,\tfrac{r+s}{5};1+\tfrac{s}{5};\hc^5) \\
&\qquad+ \frac{1}{r} \hb^r\,
\frac{\Gamma(1+\tfrac{r}{5})\Gamma(\tfrac{s}{5})}
{\Gamma(\tfrac{r+s}{5})}\,
{}_2{\rm F}_1(\tfrac{r}{5};1-\tfrac{s}{5};1-\tfrac{s}{5};\hc^5) \ .
}
With the help of the identities
\begin{align}
{}_2{\rm F}_1(\mfa, \mfc;\mfc;z) &= (1-z)^{-\mfa}\\
{}_2{\rm F}_1(\mfa,\mfb;\mfc;z) &= 
(1-z)^{\mfc-\mfa-\mfb}\, {}_2{\rm F}_1(\mfc-\mfa,\mfc-\mfb;\mfc;z)
\end{align}
as well as properties of the $\Gamma$-function, (\ref{zwischen}) then 
becomes 
\be
-\frac{1}{s} \hc^s\,
{}_2{\rm F}_1(\tfrac{s}{5},1-\tfrac{r}{5};1+\tfrac{s}{5};\hc^5)
+ \frac{1}{r}\frac{\Gamma(1+\tfrac{r}{5})\Gamma(\tfrac{s}{5})}
{\Gamma(\tfrac{r+s}{5})} \ .
\ee
By the Gauss hypergeometric theorem the second term is precisely the 
value of the right hand side of (\ref{EQHypergeoInt1}) for
$\hat{b}=1$, while the first term agrees with
(\ref{EQHypergeoInt2}). Since $\hat{b}=1$ corresponds to $\hat{c}=0$,
the second term just accounts for the fact that the reference points
in the two line integrals (\ref{EQHypergeoInt1}) and
(\ref{EQHypergeoInt2}) are different, and we have therefore proven our
claim. In particular, this then implies that the function ${\cal W}$ 
defined by (\ref{superex}) solves both (\ref{Wb}) and (\ref{Wc}).

\section{Superpotentials}

As has been indicated already in previous chapters,
the function ${\cal W}$ has actually an interpretation in terms of the
effective spacetime superpotential. It therefore encodes information about the different vacua of the model (at least in the part of the moduli space under consideration). 

In the above we have seen explicitly that the RG flow is a gradient  
flow of a potential. This potential is precisely the contribution to
the effective superpotential ${\cal W}$ that is first order in the
bulk field  $\Phi$ and exact in the boundary field $\psi_1$. To see
this we simply note that the term that appears on the right hand side
of (\ref{RG}) is the bulk-boundary coefficient that involves one
insertion of the bulk field $\Phi$ and one insertion of the boundary
field $\psi_1$ (that couples to $\mu$).  This bulk-boundary correlator
was evaluated at an {\it arbitrary} point in the brane moduli space;
if we start around any given point of the brane moduli space, the
above expression  therefore involves an arbitrary number of insertions
of $\psi_1$ (that allow one to move around this brane moduli
space). Thus the right-hand-side of (\ref{RG}) is the generating
function describing symmetrised correlators involving an arbitrary
number of boundary fields $\psi_1$, together with one insertion of the
boundary field $\psi_1$ and one insertion of the bulk field $\Phi$. We
can produce the insertion of the boundary field $\psi_1$ by taking a 
derivative with respect to the corresponding boundary coupling
constant. It thus follows that the function ${\cal W}$ (that we
obtained by integrating up the right hand side of (\ref{RG}))
is precisely the generating function of one bulk field $\Phi$ with an
arbitrary number of boundary fields. It therefore defines the
corresponding contribution of the effective superpotential.

It is also clear from this argument that this method can be applied to
calculate the corresponding terms of the effective superpotential for 
an arbitrary bulk deformation, not just one of the form
(\ref{defo}). For the other cases, the result is however trivial: the
complex structure deformations (\ref{defo}) are the only monomials 
(instead of $x_1^3$ we may also allow for an arbitrary third order
polynomial in $x_1$ and $x_2$) for
which the bulk-boundary OPE coefficient with $\psi_1$ is
non-zero. Thus to first order in the bulk perturbation the
above terms are the only terms that appear in the effective
superpotential. It should also be obvious how to perform the same
analysis for the other ($45$) families of D2-branes.

It should be noted that a priori only the minima of the effective superpotential do have a physical interpretation. The function itself may be subject to field redefinitions, so that it not clear if ${\mathcal W}$ contains definite off-shell information. Nevertheless it is interesting to learn how the open string moduli spaces change under bulk deformations, and to see how they are connected. 

Note also that for the investigation of the open-closed moduli space it was only necessary to deal with the {\it boundary} part of the BRST operator. This is very reminiscent of bulk deformations which have been considered in chapter \ref{ch-ren} for bosonic BSFT. In fact, the restriction to $q$ instead of $Q$ is possible because in the topological theory the bulk-boundary map is almost trivial. It is given by the trivialisation
$\Phi \mapsto \Phi \; \one$,
where $\Phi$ is a bulk field. 
This map involves no other modes than the constant one. This has the effect that OPEs between any fields are always regular and do not contain singularities. Therefore there is a trivial map from bulk to boundary fields which makes it possible to view the bulk fields naturally as a subset of the boundary fields. Once the projection on constant modes is abandoned, extra singularities will appear when moving bulk operators to the boundary, as explained in previous sections. These were seen to lead finally to non-local excitations, thus in a non-topological setting one would again expect the appearance of non-local terms.

%
%
%

\chapter{Conclusions}

The main goal of the work presented in this thesis was to develop a version of BSFT which is valid in curved backgrounds, to find a way to isolate closed string deformations in this language and to find support for the idea that the open string field theory is indeed capable of describing closed string deformations. 

In the bosonic case it has been possible to achieve these goals and fortify the approach by concrete calculations.
The paths taken rested on an extension of BSFT suitable for curved target spaces. While the factorisation conjecture, which enabled this extensions, has only been proven for WZW targets, and therefore for a large and important class of target spaces, it does not seem too farfetched to put this forward also for general targets. 

Further tests of the constructed BSFT action have been provided. In fact it has been shown that tachyon condensation on D-branes yields the expected results.
For this we have applied the open-closed string correspondence developed in chapter \ref{ch-fact} \cite{Baumgartl:2004iy} to a specific example, where the qualitative features observed should be rather generic. Apart from the numerical values not much depended on the details of the group manifold in question. Given the highly symmetric set-up one might hope that some of the phenomena discussed in chapter \ref{ch-ren} and in \cite{Baumgartl:2006xb} within perturbation theory could be established exactly at least for some simple processes. 

In particular within the perturbative approximation utilised here we are not able to see all D2-branes corresponding to conjugacy classes of the group. Rather we only see the `biggest' 2-brane. This should be related to the fact that we worked in the large radius regime. Pushing the perturbation in $\l$ further it is conceivable that additional fixed points appear which describe `smaller' conjugacy classes, but in order to see these much more powerful methods are needed. More interestingly it would be worthwhile investigating, if non-symmetry preserving branes exist in these models. 
Also, although we have observed the absence of divergences in the 2-brane theory by brute force computation, there may well be symmetry arguments that imply finiteness of the loop correction. It would be interesting to know if such a symmetry exists, in particular in view of a non-perturbative approach to these models.

With chapter \ref{ch-superpot} a first step has been taken to repeat the bosonic approach in a supersymmetric setting. The starting point for this investigation has been a spacetime much more complicated than flat space, namely the quintic. Most remarkably, it has been possible to derive exact results on the open string moduli space of this Calabi-Yau and its behaviour under closed string deformations. The observations made are in agreement with the philosophy put forward in chapters  \ref{ch-fact} and \ref{ch-ren}. The fact that closed string deformations can be treated completely by looking only at the boundary BRST operator is, from this point of view, to be expected. If this is a relict of the immense simplification achieved by projecting on the topological sector of the theory can only be decided once the factorisation conjecture (\ref{EQNConjecture}) has been proven for supersymmetric theories on Calabi-Yau.
In addition we have worked exclusively in the B-model, because there a description of D-branes which is close to the worldsheet formalism is available. It is not clear how to conduct similar calculations in the A-model, or how to consider even situations, where A- and B-branes are considered simultaneously. 

While a supersymmetric version of (\ref{EQNConjecture}) for arbitrary target spaces is an important further step in the investigation of open-closed correspondence, the results of chapter  \ref{ch-superpot} do have immediate application to other problems, too. Prominent among them is the existence of open-closed Picard-Fuchs equations 
\cite{Lerche:2002yw, Walcher:2006rs}. This in turn opens up the door for an investigation of mirror symmetry when both, open and closed string moduli are included. This is an interesting question in itself, but it should be investigated with the far aim of shedding light on a general understanding of open-closed correspondence in string field theory.

\backmatter

\markboth{}{}

\end{document}